\documentclass[aps,twocolumn,prx,preprintnumbers, nofootinbib,10pt]{revtex4-2}
\usepackage{xpatch}
\makeatletter
\patchcmd{\@ssect@ltx}
    {\addcontentsline{toc}{#1}{\protect\numberline{}#8}}
    {}
    {}
    {}
\makeatother
\pdfoutput=1
\usepackage{color}
\usepackage{enumitem}
\usepackage{mathtools}
\usepackage[dvipsnames]{xcolor}

\usepackage{hyperref}
\definecolor{amaranth}{rgb}{0.9, 0.17, 0.31}
\definecolor{forestForestGreen(web)}{rgb}{0.13, 0.55, 0.13}
\definecolor{blue(munsell)}{HTML}{005567}
\definecolor{bblue}{rgb}{0.0, 0.58, 0.71}
\hypersetup{
pdfstartview={FitH}, 
pdftitle={}, 
colorlinks=true, 
linkcolor=blue(munsell), 
citecolor=blue(munsell), 
filecolor=magenta, 
urlcolor=blue(munsell)
}
\usepackage{pgfplots}
\pgfplotsset{compat=1.18}
\usepackage{amssymb}
\usepackage{amsfonts}
\usepackage{graphicx}
\usepackage{epstopdf}
\usepackage{dcolumn}
\usepackage{amsmath}
\usepackage{latexsym,bm}
\usepackage{amsthm}
\usepackage{slashed}
\usepackage{float}
\usepackage{color}
\usepackage{url}
\usepackage{rotating}
\usepackage[normalem]{ulem}
\usepackage{faktor}
\usepackage{tikz-cd}
\usepackage{tensor}
\usepackage{array}

\usepackage{tabularx}
\usepackage{makecell}
\usepackage[normalem]{ulem}
\usepackage{changepage}
\numberwithin{equation}{section}

\newcommand{\ssn}[1]{{#1}}

\newcommand{\green}[1]{{\color{ForestGreen}  #1}}

\usepackage{cancel}

\newcommand{\prx}[1]{{#1}}

\usepackage{makecell}
\newtheorem{theorem}{Theorem}

\newtheorem{corollary}{Corollary}

\newcommand{\bin}{\text{bin}}

\usepackage{tikz}
\usepackage{diagbox}
\usetikzlibrary{positioning}
\usetikzlibrary{calc}
\usetikzlibrary{decorations.pathreplacing,calligraphy}

\usepackage{xstring}
\usetikzlibrary{decorations.pathmorphing} 
\usetikzlibrary{decorations.markings} 
\usetikzlibrary{arrows} 
\usetikzlibrary{shapes} 
\usetikzlibrary{matrix} 
\usetikzlibrary{positioning} 
\usepackage[english]{babel} 
\usepackage[autostyle]{csquotes}
\usepackage{pifont}
\usetikzlibrary{shapes.multipart}

\tikzset{->-/.style={decoration={
  markings,
  mark=at position .6 with {\arrow{Latex[length=1.5mm,width=1.5mm]}}},postaction={decorate}}}

\usepackage{quantikz}

\newcommand{\bea}{\begin{eqnarray}}
\newcommand{\eea}{\end{eqnarray}}
\newcommand{\be}{\begin{equation}}
\newcommand{\ee}{\end{equation}}
\newcommand{\ba}{\begin{aligned}}
\newcommand{\ea}{\end{aligned}}
\newcommand{\bit}{\begin{itemize}}
\newcommand{\eit}{\end{itemize}}
\newcommand{\ben}{\begin{enumerate}}
\newcommand{\een}{\end{enumerate}}
\newcommand{\nn}{\nonumber}

\newcommand{\id}{\text{id}}

\newcommand{\lb}{\left(}
\newcommand{\rb}{\right)}

\newcommand{\Z}{{\mathbb Z}}

\newcommand{\bC}{{\mathbb C}}

\newcommand{\Auto}{K}
\newcommand{\auto}{\alpha}

\newcommand{\cA}{\mathcal{A}}

\newcommand{\cC}{\mathcal{C}}

\newcommand{\cH}{\mathcal{H}}

\newcommand{\cL}{\mathcal{L}}

\newcommand{\cO}{\mathcal{O}}
\newcommand{\cP}{\mathcal{P}}

\newcommand{\cS}{\mathcal{S}}

\newcommand{\cX}{\mathcal{X}}

\newcommand{\cZ}{\mathcal{Z}}

\newcommand{\fB}{\mathfrak{B}}

\renewcommand{\dim}{\text{dim}}

\def\thirdcolor{orange}
\def\BlueColor{MidnightBlue!50}

\def\GreenColor{ForestGreen}

\makeatother

\usepackage{physics}
\renewcommand{\ol}{\overline}

\newcommand{\bbI}{\mathbb I}

\newcommand{\diag}{\text{diag}}

\usepackage{float}

\usepackage{makecell}

\makeatletter
\newcommand\xlabel[2][]{\phantomsection\def\@currentlabelname{#1}\label{#2}}
\makeatother

\renewcommand{\r}{{\color{red}  r}}
\newcommand{\g}{{\color{ForestGreen}  g}}
\renewcommand{\b}{{\color{blue}  b}}

\makeatletter
\def\l@subsubsection#1#2{}
\makeatother

\begin{document}

\title{Constant-Depth Clifford-Hierarchy Gates via Non-Abelian Surface Codes}

\author{Alison Warman}
\author{Sakura Sch\"afer-Nameki}

\affiliation{Mathematical Institute, University of Oxford, Woodstock Road, Oxford, OX2 6GG, United Kingdom}

\begin{abstract} 
\noindent 
We present an entirely 2D constant-depth realization of topologically protected phase gates at any level of the Clifford hierarchy, and beyond, using non-Abelian surface codes. Our construction encodes a logical qubit in the quantum double $D(G)$ of a non-Abelian group $G$ on a triangular spatial patch. The logical gate is implemented by a constant-depth circuit constructed from stacking on the spatial region a symmetry-protected topological (SPT) phase specified by a group 2-cocycle and boundary counter-terms.
The  Bravyi--K\"onig theorem limits the unitary gates implementable by constant-depth quantum circuits on Pauli stabilizer codes in $D$ dimensions to the $D$-th level of the Clifford hierarchy. We bypass this limitation, by constructing constant-depth unitary gates at arbitrary levels of the Clifford hierarchy purely in 2D, without sacrificing locality or fault tolerance, at the cost of using the quantum double of a non-Abelian group $G$. Specifically, for $G = D_{4N}$, the dihedral group of order $8N$, we realize the phase gate
$T^{1/N} = \mathrm{diag}(1, e^{i\pi/(4N)})$
in the logical $\overline{Z}$ basis. In this context, we propose a non-abelian stabilizer group formalism, which we work out for dihedral groups.
For $8N = 2^n$, the logical gate lies at the $n$-th level of the Clifford hierarchy and, importantly, has a qubit-only realization: we show that it can be constructed in terms of Clifford-hierarchy stabilizers for a code with $n$ physical qubits on each edge of the lattice.
We also discuss code-switching to the double surface-code $D(\mathbb{Z}_2\times\mathbb{Z}_2)$, to complete a universal gate-set in this setup. 

\end{abstract}

\maketitle

\tableofcontents


\section{Introduction}


\begin{figure}
\centering
\begin{tikzpicture}
\begin{scope}[shift={(0,0)}]
\draw [thick, fill= \BlueColor, opacity=0.5]  
(0,0) -- (4,0) -- (2.7,1.5)  --(0,0); 
 \draw [very thick] (0,0) -- (4,0) -- (2.7,1.5)  --(0,0); 
\draw[thick] (0,0) -- (0, -4) ;
\draw[thick] (4,0) -- (4, -4) ;
\draw[thick, dashed] (2.7,1.5) -- (2.7, 1.5-4) ;
\draw[fill= \BlueColor, opacity=0.4] (0,0) -- (0, -4)  -- (4,-4) -- (4,0) -- (0,0) ;
\begin{scope}[shift={(0,-4)}]
\draw [thick, fill= \BlueColor, opacity=0.3]  
(0,0) -- (4,0) -- (2.7,1.5)  --(0,0); 
 \draw [thick] (0,0) -- (4,0) -- (2.7,1.5)  --(0,0); 
 \node at (1.7,1.5) {$D(G)$}; 
  \node at (1.7,3.5) {$D(G)$}; 
 \draw[very thick, red] (2.5, 0.5) --  (1.8, 1) ; 
\draw[very thick, blue] (2.5, 0.5) --  (3.3, 0.8) ; 
\draw[very thick, \thirdcolor] (2.5, 0.5) -- (2.5, 0)  ; 
\node[red, left] at (2.55, 0.9){$a_1$} ;
\node[blue, right] at (2.7, 0.9) {$a_2$} ;
\node[right, \thirdcolor] at (2, 0.3) {$a_3$} ;
 \end{scope}
\begin{scope}[shift={(0,-2)}]
\draw [thick, fill= \GreenColor, opacity=0.5]  
(0,0) -- (4,0) -- (2.7,1.5)  --(0,0); 
 \draw [very thick] (0,0) -- (4,0) -- (2.7,1.5)  --(0,0); 
\node[right]  at (1.7,0.5) {$U_{\alpha,\beta}$};
 \end{scope}
\node[\thirdcolor, above] at (2,0) {$\cL_3$};
\node[red] at (1.2,1) {$\cL_1$};
\node[blue, right] at (3.3,1.0) {$\cL_2$};
\end{scope}
\end{tikzpicture}
\caption{Quantum double $D(G)$ for a finite (non-Abelian) group $G$ on a space-time prism. The vertical faces of the prism are the three topological boundary conditions specified by $\cL_i$ for $i\in\{1,2,3\}$. Along a triangular spatial slice, we insert $U_{\alpha,\beta}$. For $G=D_{4N}$ it is specified by the non-trivial 2-cocycle $\alpha \in H^2 (D_{4N}, U(1)) \cong \Z_2$, which labels an SPT phase in the 2D spatial slice. A logical state  is specified by anyons forming a trivalent junction: the red and blue fuse to the \thirdcolor\ anyon, which can end on the third boundary. With suitable choices of boundary conditions $\cL_i$, we show that this realizes a single logical qubit {on which} the $T^{1/N}$ phase gate is implemented by a constant-depth circuit. 
  \label{fig:Prism} }
\end{figure}

\noindent
{Fault tolerant} realization of non-Clifford gates is a key bottleneck for scalable, universal quantum computation. In two spatial dimensions (2D), Pauli stabilizer codes with geometrically local checks are constrained via the Bravyi-K\"onig (BK) theorem: 
any locality-preserving logical unitary on a $D$-spatial dimensional topological Pauli stabilizer code is contained in the $D$-th level of the Clifford hierarchy \cite{Bravyi:2012rnv}. In particular, for $D=2$ such logical gates are therefore at most Clifford, precluding constant-depth implementations of non-Clifford gates such as the $T=\diag(1,e^{i\pi/4})$ gate. This has underpinned, for example, the  reliance on magic-state distillation in 2D Pauli stabilizer codes \cite{Bravyi2005magicstate, Bravyi:2012lxn}, which comes with substantial overheads. 

In this work we show that, by moving beyond Pauli stabilizer codes to 2D non-Abelian surface codes (in particular quantum doubles of the dihedral groups $D(D_{4N})$), one can nevertheless realize topologically protected constant-depth phase gates at arbitrarily high levels of the Clifford hierarchy, and even beyond. Concretely, we construct a family of 2D codes in which a single logical qubit admits a constant-depth implementation of the phase gate $T^{1/N}$, acting as $\diag(1,e^{i\pi/(4N)})$ in the logical $\overline{Z}$ basis. For special values of $N=2^{n-3}$, these are in the $n$-th level of the Clifford hierarchy and have a realization purely in terms of $n$ physical qubits on each {edge of the lattice}.

\vspace{1mm}
\noindent\textbf{Prior work using non-Abelian Surface Codes.}
Recent approaches have suggested alternatives to magic state distillation, relying on non-Abelian topological orders, in particular quantum doubles $D(G)$ of non-Abelian groups $G$. 
One method utilizes code-switching protocols involving non-Abelian quantum doubles $D(G)$ for magic state preparation \cite{Laubscher:2019rss,Davydova:2025ylx, Huang:2025cvt}. Another one, known as ``hybrid lattice surgery'' \cite{Huang:2025ump} generalizes the standard lattice surgery involving the same code patches \cite{Horsman:2011hyt,Cowtan:2022csx,Cowtan:2025vok} to merge and split operations between Abelian and non-Abelian phases. 
This enables injection or teleportation of non-Clifford operations (including higher-level gates) back into the standard surface code, with a TQFT description in terms of interfaces between 2D topological orders. 
It was shown in \cite{Huang:2025ump}, that these codes can realize $T^{1/N}$ gates for any integer $N$, going even beyond the Clifford hierarchy. 
However, in these architectures the gates are not implemented {\it by a constant-depth circuit} on a single code patch. 
Another approach \cite{Hsin:2024nwc, Kobayashi:2025cfh, Hsin:2025zgn} uses automorphisms of topological orders to realize logical gates with a constant-depth circuit. 
In particular \cite{Kobayashi:2025cfh} implemented the $T^\dagger$-gate in the 2D surface code $D^\omega(\Z_2^{3})$ with a constant-depth circuit. 
This was extended to higher dimensions, and conjectured to lower the BK bound by one. It should be noted that in this context, the $T^\dagger$-gate requires a non-Clifford gate on the physical qubits.

\vspace{1mm}
\noindent{\bf Proposed 2D Code.} 
In this paper, we present a purely 2D, topologically protected \prx{constant-depth} realization of a family of non-Clifford gates, by utilizing non-Abelian surface codes. In particular, we realize the  phase gates 
\be\label{eq:T1n}
T^{1/N} \equiv P\lb \frac{\pi}{4N}\rb 
= \diag \left(1, e^{i \pi/(4N)} \right)\,,
\quad N\in \mathbb{N} \,.
\ee
We implement them with a constant-depth circuit in a 2D surface-code constructed from non-Abelian quantum doubles $D(D_{4N})$ for dihedral groups $D_{4N}$ of order $8N$.

Concretely, we consider a triangular spatial patch (which is a cross-section of a spacetime prism), with three gapped boundaries chosen so that \prx{single qubit logical states are encoded by trivalent junction configurations of anyons, as shown in Fig.~\ref{fig:Prism}. We illustrate our setup with the familiar example of $G=\Z_2\times\Z_2$ in Sec.~\ref{sec:Z2Z2_setup}, before turning to non-Abelian codes, which will be our main focus, in Sec. \ref{sec:code_space}.} 
Along a spatial slice we insert a 2D purely spatial SPT-layer determined by a group cocycle $\alpha\in H^2(G,U(1))$. This is required to trivialize on each of the boundaries, where it is written in terms of a function 
$\beta^{(i)}$ for $i\in\{1,2,3\}$. The induced automorphism implements a diagonal, constant-depth logical unitary gate acting on a state $|g_1,g_2\rangle$, where $g_1,g_2\in G$ label a configuration of magnetic anyons, by the phase
\be \label{eq:automorphism}
U_{\alpha,\beta}(g_1,g_2)
=\frac{\alpha(g_1,g_2)\,\beta^{(3)}(g_1 g_2)}{\beta^{(1)}(g_1)\,\beta^{(2)}(g_2)} \,.
\ee
By choosing $G$ and $\alpha$ so that the action of $U_{\alpha,\beta}$ on the trivial state is the identity, while on the non-trivial $\ol{Z}$ eigenstate it is the phase $e^{i \pi/(4N)}$, we obtain a constant-depth implementation of $T^{1/N}=P(\pi/(4N))$ as in Eq.~\eqref{eq:T1n}. We first recover, in this SPT-stacking language, the familiar constant-depth $T=P(\pi/4)$ in $D(D_4)$, a non-Abelian presentation equivalent to $D^\omega(\Z_2^3)$, and then exhibit a systematic generalization to the family of dihedral groups $D_{4N}$ where the same mechanism yields $T^{1/N}$.

\vspace{1mm}
\noindent{\bf Implementation with Qubits.} 
For any $N\geq1$ these realize non-Clifford gates. While for generic $N$ these are beyond the Clifford hierarchy, for the special values
\be \label{Nnmadness}
8N= 2^n
\ee
they are in the $n$-th level of the Clifford hierarchy (and therefore non-Clifford for $n\geq 3$). For instance
\be\ba 
n=3 &\Rightarrow P(2\pi/2^3)=\diag(1,e^{i\pi/4})=T\,,\\
n=4 &\Rightarrow P(2\pi/2^4)=\diag(1,e^{i\pi/8})=T^{1/2}\,,\quad\text{etc.}
\ea\ee
The values (\ref{Nnmadness}) are also special as they allow implementation in terms of $n$ physical {\it qubits} on each {edge of the lattice}.\footnote{The total number of physical qubits is given by $n\times N_{\text{edges}}$, where $N_{\text{edges}}$ denotes the total number of lattice edges.} 

If $N$ has prime factors other than 2, the gates are beyond the Clifford hierarchy, and implementation requires qudits.

\vspace{1mm}
\noindent{\bf Conceptual Advance.}
 Firstly, non-Abelian surface-codes with gapped boundaries enlarge the landscape of topologically protected logicals available in 2D, permitting diagonal gates at \textbf{any} Clifford-hierarchy level without resorting to non-local operations or higher-dimensional layouts.
 
 Secondly, group-SPT stacking automorphisms provide a simple route to such gates: they are implemented by constant-depth circuits of local unitaries arranged along a single spatial slice,  suggesting extensions to controlled-phase gates (e.g., CS) by coupling patches and to qudit operations via other finite groups. 

From a fault-tolerance perspective, constant-depth and topologically protected logical operations are fundamentally constrained: the Eastin-Knill theorem rules out universal constant-depth gate sets on any code with single site error detection \cite{Eastin:2008aa}. Within 2D, further restrictions on topologically protected logical gates (constant-depth, geometrically local circuits) are captured by TQFTs and stabilizer-codes \cite{Beverland:2014aa,Pastawski:2014aa}. Our construction uses topological orders and  gapped boundaries of these \cite{Kitaev:2011aa}, together with SPT “stacking” ideas \cite{Chen:2011aa, Levin:2012aa}. Most importantly, we make key use of insights from condensations of anyons, that have been obtained in various different areas of physics \cite{Barkeshli:2014aa, Bhardwaj:2023idu,Bhardwaj:2024qrf, Bhardwaj:2024igy, Gai:2026hjk}.

The logical non-Clifford gates that we construct are \emph{topologically protected}, as defined in \cite{Bravyi:2012rnv}, since they can be
realized by applying a constant-depth quantum circuit on
the physical qudits (qubits for $8N=2^n$). Of course, our result does not contradict the BK theorem: it provides an alternative to the increase in space-time dimensions for non-Clifford gates, by instead remaining in 2D but increasing the local Hilbert space dimension and using \emph{\prx{non-Abelian stabilizers}}. 
We will clarify  this in the main text of the paper, but let us briefly explain the setting here.

The BK no-go theorem applies to constant-depth quantum circuits in Pauli stabilizer codes with commuting checks. By contrast, our codes originate from non-Abelian topological order. 
The code space is the ground state space of the non-Abelian surface code for $D(G)$. Unlike the case of Pauli stabilizer codes, here, the stabilizer group is  not necessarily Abelian. 
Previous work on non-abelian stabilizer groups has appeared in \cite{Ni:2014clx, Webster:2022kdn, Schotte:2020lnz}, for qubit stabilizers, as well as permutation group stabilizers in \cite{Pollatsek:2004kja}. Examples of stabilizers for $D(S_3)$ and $D(D_4)$ have appeared in \cite{Verresen:2021wdv, Huang:2025ump}. We will determine the stabilizer groups explicitly for $D(D_{4N})$.

Concretely, the stabilizers are written in terms of vertex and plaquette terms of a non-Abelian quantum double, that, while still being geometrically local, do not all commute. The logical subspace is characterized by the inequivalent ground states of the quantum double Hamiltonian \cite{Kitaev:1997wr} with boundaries \cite{Beigi:2010htr}. Each ground state is specified by condensed anyons ending only on the boundaries and not in the bulk (in order to not generate an excitation). 
In this generalized setup, logicals are not restricted by the BK bound for Pauli stabilizers, and topologically protected gates at all levels of the Clifford hierarchy (and beyond) can occur purely in 2D. We prove that one can realize a qubit logical gate at level $n$ of the Clifford hierarchy using the quantum double of the dihedral group of order $2^n$ on a physical Hilbert space of $n$ qubits on each edge of the lattice.

\vspace{1mm}
\noindent{\bf Error Correction.} 
Quantum error correction (QEC) for non-Abelian surface codes is less well explored, with some advances using the just-in-time decoder \cite{Bombin:2018wjx, Brown:2020xxo, Scruby:2020pvw, Davydova:2025ylx,Huang:2025ump}.
In particular, as discussed in \cite{Davydova:2025ylx}, the just-in-time decoder is key for non-Abelian quantum doubles, as in these settings, one cannot easily reconstruct the full history of flux-lines \prx{after many rounds of
noisy measurements}. This is due to the fact that charge operators can be absorbed by flux operators and thus such errors cannot be corrected. From a categorical perspective, this means that some of the anyons of the non-Abelian quantum double have non-invertible fusion and can absorb some of the other anyons. The just-in-time decoder is built to correct errors as they occur, thus preempting the creation of such irreversible errors. The proposal in \cite{Davydova:2025ylx} addresses this in a way that should be applicable to our setting, and we hope to return to this in the future with more detailed numerical studies.

\prx{
\noindent{\bf Code Switching.}
To obtain a universal gate set, we discuss code-switching to the double surface code (SC) $D(\Z_2\times \Z_2)$ in Sec.~\ref{sec:CodeSwitch}. 
Although we are able to generate non-Clifford gates at any level of the Clifford hierarchy in the non-abelian codes,
we do not have the full set of Clifford gates. This is of course compatible with  the Eastin-Knill theorem \cite{Eastin:2009tem}. To obtain a universal constant-depth gate-set, one can generate the Clifford gates by constant-depth circuits within the double SC. The code-switching we present allows incorporation of constant-depth Clifford and non-Clifford gates into a single computational framework.
}

\subsection{Main Results}

We will now state our main results. First, we show how group cocycles can give rise to constant-depth diagonal gates in group surface codes (Theorem \ref{thm:SPT}). We then apply it to the dihedral groups $D_{4N}$ and construct in 2D constant-depth $T^{1/N}$ gates  (Corollary \ref{coro:D4N}). Finally, we demonstrate that for $8N=2^n$ these can be realized purely in terms of qubits (Theorem \ref{thm:qubit-only}).

\begin{theorem} \textbf{Constant-depth Gates from SPT-Stacking.}
\label{thm:SPT}
Consider the quantum double $D(G)$ of a finite group $G$ on a spatial triangle. Let the three gapped boundaries be labelled by subgroups 
$K_1,K_2,K_3\subseteq G$, chosen to encode a single logical qubit.
Let $\alpha\in H^2(G,U(1))$ be a group 2-cocycle whose restriction on each boundary is trivial in group cohomology, i.e. there exist $\beta^{(i)}:K_i\to U(1)$ such that
\be
\label{eq:boundary-trivialization}
 \alpha|_{K_i} = \delta\beta^{(i)} \qquad \text{for } i\in\{1,2,3\} \,.
\ee
Define the constant-depth unitary $U_{\alpha,\beta}$ by stacking the 2D purely spatial SPT circuit for $\alpha$ on a single time-slice, with the boundary counterterm $\beta^{(i)}$ on each 1D boundary.

Then  the induced logical gate $U_{\alpha,\beta}$ is diagonal in the $\overline{Z}$  basis $\{\ol{\ket{0}},\ol{\ket{1}}\}$, with 
\be\ba\label{Ualphabeta}   
U_{\alpha,\beta}\ol{\ket{0}}&=\ol{\ket{0}}\,,\\
U_{\alpha,\beta}\ol{\ket{1}}&=\frac{\alpha(g_1,g_2)\beta^{(3)}(g_1g_2)}{\beta^{(1)}(g_1)\,\beta^{(2)}(g_2)}\,\ol{\ket{1}}  \,,
\ea\ee
for any state labeled by $g_1\in K_1$, $g_2\in K_2$ representing $\ol{\ket{1}}$. 

$U_{\alpha,\beta}$ preserves the logical codespace (since it is constructed from an automorphism of $D(G)$). Furthermore, its lattice realization involves only operators acting on $O(1)$ lattice sites: a local error can thus enlarge its support only by $O(1)$ sites. 
\end{theorem}

Applied to the dihedral groups this theorem allows us to realize non-Clifford gates at any level of the Clifford hierarchy and beyond:

\begin{corollary}\textbf{Constant-Depth Non-Clifford Gates from $\bm{D(D_{4N})}$.}
\label{coro:D4N}
For any integer $N\geq 1$, consider the order-$8N$ dihedral group (of symmetries of a $4N$-gon)
\be
G=D_{4N}=\langle r,s \mid r^{4N}=s^2=\id,\ srs=r^{-1}\rangle\,.
\ee
We consider the setup of Theorem \ref{thm:SPT} for this group $G$ and choose the three gapped boundaries to be determined by the subgroups 
\be
K_1=\langle rs\rangle \cong \Z_2\,,\quad 
K_2=\langle s\rangle \cong \Z_2 \,,\quad 
K_3=\langle r\rangle \cong \Z_{4N}\,.
\ee
Consider the non-trivial cocycle 
\be
\alpha_N\in H^2(D_{4N},U(1)) = \Z_2 \,.
\ee
On the state labeled by $g_1=rs,\; g_2=s,\; g_1g_2= r$, Eq. \eqref{Ualphabeta} evaluates to the unitary 
\be\label{Unj}
U_{\alpha,\beta}=T^{1/N}=P\left({\pi \over 4N}\right)=\mathrm{diag}\left(1,e^{i\pi \over 4N}\right)\,.
\ee
\end{corollary}
In this case we construct the non-Abelian stabilizer group explicitly determining all operator commutators. For $N=1$ it is a Clifford stabilizer, for higher $N$ non-Clifford stabilizer.

For a possible implementation, in particular of the higher-Clifford hierarchy gates,  the following theorem will be crucial, as it allows a realization of the gates in terms of purely qubit architectures:
\begin{theorem}\textbf{Qubit-only realization for $\bm{8N=2^n}$.}
\label{thm:qubit-only}
When $8N=2^n$, the quantum double $D(D_{4N})=D(D_{2^{n-1}})$ can be implemented on a qubit-only physical Hilbert space in 2D, with $n$ qubits on each lattice edge. The unitary $U_{\alpha,\beta}$ in (\ref{Unj}) becomes a geometrically local, topologically protected qubit circuit implementing the logical
$T^{2^{3-n}}= P(2\pi/2^{n})$ gate, in the $n$-th level of the Clifford hierarchy. In particular, for $n\ge 3$ it is non-Clifford.
\end{theorem}

A sample set  of the resources required for the realization in terms of qubits is given for the first few values of $n$ in Tab.~\ref{tab:Resources}.

\begin{table}
$$
\begin{array}{|c|c|c|c|c|}
\hline
N & n &  D_{4N} & \text{Gate} & \makecell[c]{\text{Total}\\ \text{phys. qbts}}  \\
\hline
1 & 3 &  D_4=\Z_{4}\rtimes\Z_2 & T=P(\pi/4) & 3\times N_{\text{edges}} \\
2 & 4 &   D_{8}=\Z_{8}\rtimes\Z_2 & T^{1/2}=P(\pi/8) & 4\times N_{\text{edges}} \\
4 & 5 &  D_{16}=\Z_{16}\rtimes\Z_2 & T^{1/4}=P(\pi/16) & 5\times N_{\text{edges}} \\
2^{n-3} & n & D_{2^{n-1}} = \Z_{2^{n-1}} \rtimes\Z_2 & T^{2^{3-n}} = P\left({\pi\over 2^{n-1}}\right) &  n \times N_{\text{edges}}\cr 
\hline
\end{array}
$$
\caption{The resources required to realize the non-Clifford gate $T^{1/N}$, which is at level $n$ of the Clifford hierarchy. Our protocol employs the quantum double of the dihedral group $D_{4N}$ of order $8N$ realized in terms of $n$ physical qubits on each lattice edge. The total number of lattice edges is denoted by $N_{\text{edges}}$. \label{tab:Resources}}
\end{table}

\medskip
\noindent\textbf{Plan of the Paper.}
We start in Sec.~\ref{sec:SPTStack} with a discussion of surface codes based on the Kitaev quantum double models $D(G)$ for finite groups $G$. We consider the spatial triangle configuration and discuss boundary conditions as well as automorphisms of the double that are obtained from stacking with SPTs. This is substantiated with a concrete lattice description. We  then prove Theorem \ref{thm:SPT} in Sec.~\ref{sec:Constant-DepthGates}.
In Sec.~\ref{sec:CliffHier} we apply this formalism to {construct} the full single qubit diagonal Clifford hierarchy by realizing constant-depth $T^{1/N}$ gates, thus proving Corollary \ref{coro:D4N}. Finally, in Sec.~\ref{sec:Qubits} we show that for $8N=2^n$ these codes can be realized in a purely qubit setup, proving Theorem \ref{thm:qubit-only}.
Finally, we discuss code-switching to surface codes in Sec.~\ref{sec:CodeSwitch}. We conclude with discussions and outlook in Sec.~\ref{sec:Dis}. App. \ref{app:GroupTheory} contains a review of quantum doubles and group 2-cocycles, App.~\ref{app:StabilizerComms} the details on our $D(D_{4N})$ stabilizer groups and App. \ref{app:T_from_rgb} a summary of the $T$-gate construction of \cite{Kobayashi:2025cfh}.

\section{Constant-Depth Gates from SPT-Stacking}
\label{sec:SPTStack}

The general setup of this paper is the quantum double $D(G)$ of a finite group $G$, described in terms of a topological code in 2D space (and 2+1 space-time dimensions). The {space-time} geometry is that of a prism, with triangular spatial cross-section. \prx{After illustrating our setup for the familiar case of $G=\Z_2\times\Z_2$ in Sec.~\ref{sec:Z2Z2_setup}, in Sec.~\ref{sec:code_space} we describe the general framework for encoding logical states in the geometry of Fig.~\ref{fig:Prism} for any $D(G)$, whose lattice realization we provide in Sec. \ref{sec:lattice}.}
In Sec. \ref{sec:Constant-DepthGates}, we show how to construct constant-depth gates by stacking a purely spatial SPT on a time-slice. Operationally, this stacking is implemented by a constant-depth layer of local diagonal unitaries, whose phases are determined by a group 2-cocycle $\alpha$ together with boundary  counter-terms/functions $\beta^{(i)}$ on the three edges of the triangular patch.

\subsection{\prx{$\Z_2\times\Z_2$ Code on a Triangle}} \label{sec:Z2Z2_setup}
\prx{Before discussing non-abelian codes, as a warm-up we illustrate our setup for $G=\Z_2\times\Z_2$ whose group generators we denote by $m_1,m_2$ and elementary irreducible representations (irreps) by $e_1,e_2$, such that $m_i$ is charged under $e_i$ for $i=1,2$. We place the code on a surface patch with the following three boundary conditions, given by Lagrangian algebras of anyons that can consistently end on each boundary\footnote{\prx{See \cite{Davydova:2025ylx, Kobayashi:2025cfh} for setups with a similar geometry. Another way to think about this is to consider the doubled picture, which corresponds to a $\Z_2\times\Z_2$ color code, see \cite{Davydova:2023mnz}.}}
\be\ba
   \cL_1 & = (1\oplus m_1 m_2)(1 \oplus e_1 e_2) \cr 
    \cL_2 &= (1\oplus m_2)(1 \oplus e_1) \cr 
    \cL_3 &= (1\oplus m_1) (1 \oplus e_2) \,.
\ea\ee
The logical Pauli $\ol{Z}$ and $\ol{X}$ operators are shown in Fig.~\ref{fig:Z2Z2_LatticeAnyons}.
\begin{figure}
\begin{tikzpicture}
\begin{scope}[shift={(0,0)}]
\draw[step=0.5,black] (0,0) grid (3,3);
\draw[black, very thick] (0,3) -- (3,3);
\draw[black, very thick] (3,3) -- (3,0);
\draw[black, very thick] (0,3) -- (0,0);
\draw[black, very thick] (0,0) -- (3,0);
\node[red,  left] at (0,1.5) {$\cL_1$};
\node[blue,  above] at (1.5,3) {$\cL_2$};
\node[\thirdcolor,  right] at (3,1.5) {$\cL_3$};
\node[\thirdcolor,  below] at (1.5,0) {$\cL_3$};
\draw[red, very thick] (0,1.5) -- (1.5,1.5);
\draw[blue, very thick] (1.5,3) -- (1.5,1.5);
\draw[\thirdcolor, very thick] (3,1.5) -- (1.5,1.5);
\node[blue, right] at (1.5,2.25) {$e_1$};
\node[red, above] at (0.75,1.5) {$e_1e_2$};
\node[\thirdcolor, above] at (2.25,1.5) {$e_2$};
\end{scope}
\begin{scope}[shift={(4.5,0)}]
\draw[step=0.5,black] (0,0) grid (3,3);
\draw[black, very thick] (0,3) -- (3,3);
\draw[black, very thick] (3,3) -- (3,0);
\draw[black, very thick] (0,3) -- (0,0);
\draw[black, very thick] (0,0) -- (3,0);
\node[red,  left] at (0,1.5) {$\cL_1$};
\node[blue,  above] at (1.5,3) {$\cL_2$};
\node[\thirdcolor,  right] at (3,1.5) {$\cL_3$};
\node[\thirdcolor,  below] at (1.5,0) {$\cL_3$};
\draw[red, very thick, dashed] (0,1.75) -- (1.75,1.75);
\draw[red, very thick] (0,1.5) -- (0,2);
\draw[red, very thick] (0.5,1.5) -- (0.5,2);
\draw[red, very thick] (1,1.5) -- (1,2);
\draw[red, very thick] (1.5,1.5) -- (1.5,2);
\draw[blue, very thick, dashed] (1.75,3) -- (1.75,1.75);
\draw[blue, very thick] (2,3) -- (1.5,3);
\draw[blue, very thick] (2,2.5) -- (1.5,2.5);
\draw[blue, very thick] (2,2) -- (1.5,2);
\draw[\thirdcolor, very thick] (2,1.5) -- (1.5,1.5);
\draw[\thirdcolor, very thick] (2,1) -- (1.5,1);
\draw[\thirdcolor, very thick] (2,0.5) -- (1.5,0.5);
\draw[\thirdcolor, very thick] (2,0) -- (1.5,0);
\draw[\thirdcolor, very thick, dashed] (1.75,0) -- (1.75,1.75);
\node[red, above] at (0.8,1.95) {$m_1m_2$};
\node[blue, right] at (1.9,2.25) {$m_2$};
\node[\thirdcolor, right] at (1.9,0.75) {$m_1$};
\end{scope}
\end{tikzpicture}
\caption{\prx{Lattice model realization of logical operators from electric (left) and magnetic (right) anyons in $D(\Z_2\times\Z_2)$. The code is placed on a 2D square lattice with BCs $\cL_1, \cL_2$ and two adjacent boundaries with $\cL_3$: this is equivalent to the triangular configuration in the continuum and encodes a single logical qubit.}
\label{fig:Z2Z2_LatticeAnyons}}
\end{figure}
Explicitly, denoting by $Z_1,X_1$ and $Z_2,X_2$ the physical Pauli operators for each $\Z_2\times\Z_2$ generator and by $\gamma_i$ and $\xi_i$ for $i\in\{1,2,3\}$ the ribbons on the direct and dual lattices shown respectively on the left and right sides of Fig.~\ref{fig:Z2Z2_LatticeAnyons}, the logical Pauli operators are written as: 
\be\ba \label{eq:Z2Z2_Pauli}
    \ol{Z}&=\prod_{l\in\gamma_1}(Z_1Z_2)_l \prod_{j\in\gamma_2}(Z_1)_j \prod_{k\in\gamma_3}(Z_2)_k\,,\\
    \ol{X}&=\prod_{l\in\xi_1}(X_1X_2)_l \prod_{j\in\xi_2}(X_2)_j \prod_{k\in\xi_3}(X_1)_k\,.
\ea\ee
They obey the Pauli algebra
\be \label{eq:-1_e1m1}
    \ol{Z}^2=\bbI\,,\quad \ol{X}^2=\bbI\,, \quad \ol{Z}\ol{X}=-\ol{X}\ol{Z}\,,
\ee
where the last relation follows from $(Z_1X_1X_2)_l=-(X_1X_2Z_1)_l$ at the left edge $l=\xi_1\cap\gamma_2$ of the plaquette\- where the three magnetic anyons end, which is to the north-east of the vertex on which the electric anyons end, as shown in Fig.~\ref{fig:Z2Z2_LatticeAnyons}. From the continuum anyon description, the $-1$ phase in \eqref{eq:-1_e1m1} corresponds to the braiding phase of $m_1m_2\in\cL_1$ with $e_1\in\cL_2$.
}

\subsection{Non-Abelian Surface Code on a Triangle} 
\label{sec:code_space}
We consider the quantum double $D(G)$ for a finite group $G$, as shown in Fig.~\ref{fig:Prism}. The code-space is obtained as follows:
each of the three edges labeled by $i\in\{1,2,3\}$ is a gapped boundary condition of $D(G)$, which mathematically is specified by $(K_i,\varphi_i)$ for $K_i\subseteq G$ a subgroup and $\varphi_i\in H^2(K_i,U(1))$ a group 2-cocycle, up to conjugacy~\cite{Ostrikmodule,Davydov2009ModularIF,davydov2017lagrangian,Natale2017}. See e.g.~\cite{Beigi:2010htr,delaFuente:2023whm,Gai:2026hjk} for maps from the mathematical characterization $(K_i,\varphi_i)$ to the physical description in terms of condensed anyons. 
Each gapped boundary is specified by a Lagrangian algebra, which describes a maximal set of anyons that can be simultaneously consistently condensed
\be
\cL_i = \bigoplus_{a} n_a \,a \,,
\ee
where $n_a$ are non-negative integers and $a$ are the anyons, which are representations of the quantum double $D(G)$. In particular, the anyons in a given Lagrangian algebra need to be mutually local.

\prx{To define the code space}, 
consider a set of three boundary conditions $\cL_1, \cL_2, \cL_3$ with two triplets of anyons, $(a_1, a_2, a_3)$ and $(b_1, b_2, b_3)$, such that:
\begin{itemize}
    \item $a_i,b_i\in\cL_i$ for $i\in\{1,2,3\}$,
    \item $a_1\otimes a_2 \supseteq a_3$ and $b_1\otimes b_2 \supseteq b_3$ so that the anyons in each triplet meet at a tri-valent junction,
    \item the anyons in each triplet are mutually local among each other,
    \item the $(a_1, a_2, a_3)$ triplet braids non-trivially with the $(b_1, b_2, b_3)$ one. 
\end{itemize}
\prx{We will take the $(b_1, b_2, b_3)$ triplet to be comprised of abelian anyons of order two, to encode the logical Pauli $\ol{Z}$, whose eigenstates we denote by $\ol{\ket{m}}$, for $m\in\{0,1\}$. We will take the braiding of the $(a_1, a_2, a_3)$ anyon triplet with the $(b_1, b_2, b_3)$ one to be $-1$, so that the logical operator encoded by $(a_1,a_2,a_3)$ will anti-commute with $\ol{Z}$. The detailed requirements on the anyons in each triplet will be presented after discussing the lattice model in Sec.~\ref{sec:lattice}.\footnote{A single logical qubit will be our main focus in this work, however, the general theory we discuss is also applicable to different logical encodings: for example if the braiding is a phase $e^{2\pi i/d}$ the logical state will be a qudit, and if we have $k$ disjoint  pairs of anyon triplets ($(a_1, a_2, a_3)$,\, $(b_1, b_2, b_3)$), the system will describe $k$ logical qudits (with a possibly different $d$ for each pair).}}

\subsection{Lattice Model} \label{sec:lattice}
We will use Kitaev's quantum double lattice model $D(G)$~\cite{Kitaev:1997wr, Beigi:2010htr, Albert:2021vts,Li2025_QDboundary}, which has on each edge a local Hilbert space $\cH=\bC[G]$ with an orthonormal basis labeled by the group elements $\{\ket{h}\;:\;h\in G\}$. These are acted upon by left and right multiplication operators
\be \label{eq:LgRg}
    L^g\ket{h}=\ket{gh}\,,\qquad   R^g\ket{h}=\ket{hg^{-1}}\,,
\ee
and diagonal operators for each irrep $\bm{R}$ of $G$ and $j,k\in\{1,\cdots ,\dim(\bm{R})\}$
\be \label{eq:irrep_ops_G}
    Z_{\bm{R}}^{j,k}\ket{g}=M_{\bm{R}}^{j,k}(g)\ket{g}\,,
\ee
where $M_{\bm{R}}(g)$ is the matrix representation of $g$ in the irrep $\bm{R}$. The set 
\be \label{eq:basis_Z_ops}
\{ Z_{\bm{R}}^{j,k}\,:\,\bm{R}\in\text{Irreps}(G)\,,\,   j,k=1,\cdots,\dim(\bm{R})\}
\ee 
is comprised of $\sum_{\bm{R} \in \text{Irreps}}\dim(\bm{R})^2=|G|$ operators, which are all linearly independent due to the Schur orthogonality relations. They therefore provide a basis of diagonal operators acting on $\cH$.

One can write projectors onto each group element, acting as: 
\be \label{eq:Tops}
    T_+^g\ket{h} =\delta_{g,h}\ket{h}\,,\qquad  
    T_-^g\ket{h}=\delta_{g^{-1},h}\ket{h}\,.
\ee
The quantum double Hamiltonian is written in terms of vertex and plaquette operators, which, on a square lattice, take the form:
\be\ba \label{eq:AvBp}
    A_v^{g}= \, & \begin{tikzpicture}[baseline]
\begin{scope}[shift={(0,0)}]
\draw[thick, ->-] (-1,0) to (0,0);
\draw[thick, ->-] (0,0) to (1,0);
\draw[thick, ->-] (0,-1) to (0,0);
\draw[thick, ->-] (0,0) to (0,1);
\node[above] at (0.2, 0) {$_v$};
\node[above] at (-0.7,0.05) {$R^g$};
\node[above] at (0.7,0.05) {$L^g$};
\node[right] at (0.05,0.7) {$L^g$};
\node[right] at (0.05,-0.7) {$R^g$};
\draw[fill=black] (0,0) ellipse (0.05 and 0.05);
\end{scope}
\end{tikzpicture},\\
    B_p^{g}= & \sum\limits_{g_1,g_2,g_3,g_4}\delta_{g,\,g_1g_2g_3^{-1}g_4^{-1}} \times \cr 
&
\qquad 
\times  \begin{tikzpicture}[baseline]
\begin{scope}[shift={(0,0)}]
    \draw[->-] (-1,-0.5) to (-1,0.5);
    \node at (-1.5,0) {$\Bigg|$};
    \node at (-1.3,0) {$_{g_1}$};
    \draw[->-] (-1,0.5) to (0,0.5);
    \node at (-0.5,0.8) {$_{g_2}$};
    \draw[->-] (0,-0.5) to (0,0.5);
    \node at (0.3,0) {$_{g_3}$};
    \draw[->-] (-1,-0.5) to (0,-0.5);
    \node at (-0.5,-0.8) {$_{g_4}$};
    \node at (0.5,0) {$\Bigg>$};
    \node at (-0.5,0) {$_p$};
    \begin{scope}[shift={(2.2,0)}]
    \draw[->-] (-1,-0.5) to (-1,0.5);
    \node at (-1.5,0) {$\Bigg<$};
    \node at (-1.3,0) {$_{g_1}$};
    \draw[->-] (-1,0.5) to (0,0.5);
    \node at (-0.5,0.8) {$_{g_2}$};
    \draw[->-] (0,-0.5) to (0,0.5);
    \node at (0.3,0) {$_{g_3}$};
    \draw[->-] (-1,-0.5) to (0,-0.5);
    \node at (-0.5,-0.8) {$_{g_4}$};
    \node at (0.5,0) {$\Bigg|$};
    \node at (-0.5,0) {$_p$};        
    \end{scope}
\end{scope}    
\end{tikzpicture}\,.
\ea\ee
A pair $s=(v, p)$ of an adjacent vertex $v$ and plaquette $p$ is called a site. For different sites $s=(v,p)$ and $s'=(v',p')$ with $v\neq v'$ and $p\neq p'$, the vertex and plaquette terms commute, while on the same site they satisfy \cite{Kitaev:1997wr,Beigi:2010htr}
\be\ba \label{eq:non-comm_A_B}
    A^g_s A^h_s&=A^{gh}_s\,,&\;\; (A^g_s)^\dagger&=A^{g^{-1}}_s\,,\\
    B^g_s B^h_s&=\delta_{g,h} B^h_s\,, &\;\; (B^g_s)^\dagger&=B^g_s\,,\\
    A^g_s B^h_s &= B^{ghg^{-1}}_s A^g_s\,,
\ea\ee
for a non-Abelian group $G$, they do not all commute.

\vspace{1mm}
\noindent\textbf{Lattice realization of logical states.} We will take the $(b_1,b_2,b_3)$ anyon triplet configuration to be purely electric (pure irreps) placed on the edges of the direct lattice (solid line) and the $(a_1,a_2,a_3)$ triplet to be purely magnetic (pure conjugacy classes) \prx{supported on cycles of the dual lattice} (dashed lines), as shown in Fig.~\ref{fig:LatticeAnyons}. 

We will chose the orientations of the horizontal edges to point to the right and the vertical edges upwards.
\begin{figure}
\begin{tikzpicture}
\begin{scope}[shift={(0,0)}]
\draw[step=0.5,black] (0,0) grid (3,3);
\draw[black, very thick] (0,3) -- (3,3);
\draw[black, very thick] (3,3) -- (3,0);
\draw[black, very thick] (0,3) -- (0,0);
\draw[black, very thick] (0,0) -- (3,0);
\node[red,  left] at (0,1.5) {$\cL_1$};
\node[blue,  above] at (1.5,3) {$\cL_2$};
\node[\thirdcolor,  right] at (3,1.5) {$\cL_3$};
\node[\thirdcolor,  below] at (1.5,0) {$\cL_3$};
\draw[red, very thick] (0,1.5) -- (1.5,1.5);
\draw[blue, very thick] (1.5,3) -- (1.5,1.5);
\draw[\thirdcolor, very thick] (3,1.5) -- (1.5,1.5);
\node[blue, right] at (1.5,2.25) {$b_2$};
\node[red, above] at (0.75,1.5) {$b_1$};
\node[\thirdcolor, above] at (2.25,1.5) {$b_3$};
\end{scope}
\begin{scope}[shift={(4.5,0)}]
\draw[step=0.5,black] (0,0) grid (3,3);
\draw[black, very thick] (0,3) -- (3,3);
\draw[black, very thick] (3,3) -- (3,0);
\draw[black, very thick] (0,3) -- (0,0);
\draw[black, very thick] (0,0) -- (3,0);
\node[red,  left] at (0,1.5) {$\cL_1$};
\node[blue,  above] at (1.5,3) {$\cL_2$};
\node[\thirdcolor,  right] at (3,1.5) {$\cL_3$};
\node[\thirdcolor,  below] at (1.5,0) {$\cL_3$};
\draw[red, very thick, dashed] (0,1.75) -- (1.75,1.75);
\draw[red, very thick] (0,1.5) -- (0,2);
\draw[red, very thick] (0.5,1.5) -- (0.5,2);
\draw[red, very thick] (1,1.5) -- (1,2);
\draw[red, very thick] (1.5,1.5) -- (1.5,2);
\draw[blue, very thick, dashed] (1.75,3) -- (1.75,1.75);
\draw[blue, very thick] (2,3) -- (1.5,3);
\draw[blue, very thick] (2,2.5) -- (1.5,2.5);
\draw[blue, very thick] (2,2) -- (1.5,2);
\draw[\thirdcolor, very thick] (2,1.5) -- (1.5,1.5);
\draw[\thirdcolor, very thick] (2,1) -- (1.5,1);
\draw[\thirdcolor, very thick] (2,0.5) -- (1.5,0.5);
\draw[\thirdcolor, very thick] (2,0) -- (1.5,0);
\draw[\thirdcolor, very thick, dashed] (1.75,0) -- (1.75,1.75);
\node[red, above] at (0.75,1.9) {$a_1$};
\node[blue, right] at (2,2.25) {$a_2$};
\node[\thirdcolor, right] at (2,0.75) {$a_3$};
\end{scope}
\end{tikzpicture}
\caption{Lattice model realization of logical states from  electric $b_i$ (left) and magnetic $a_i$ (right)  anyons. The $D(G)$ surface code is placed on a square lattice with BCs $\cL_1, \cL_2$ and two adjacent boundaries with $\cL_3$: this is equivalent to the triangular configuration in the continuum.
\label{fig:LatticeAnyons}}
\end{figure}

Note that there are multiple possible choices for resolving the intersection site of the electric and magnetic triangles, given by the plaquette where the magnetic anyons $a_1,a_2,a_3$ end and the vertex where the electric anyons $b_1,b_2,b_3$ end. In Fig.~\ref{fig:LatticeAnyons} we have chosen it to be given by a site connecting the vertex and the plaquette to its north-west.\footnote{This is consistent with the choice of end-site $s_1$ for ribbon operators stretching from a boundary to the bulk made in \cite{Beigi:2010htr}.} With this convention the linking is $S_{a_1,b_2}$ (with $S$ the modular $S$-matrix). We remark that in our examples below, the linking phase will not depend on how we resolve the junction.

The Hamiltonian for the quantum double \cite{Kitaev:1997wr}, was extended to a setup with boundaries  in \cite{Beigi:2010htr}. Our setup comprises of a bulk and three boundaries indexed by $i\in\{1,2,3\}$, each corresponding to a subgroup $K_i$, with trivial 2-cocycle. The Hamiltonian is:
\be\label{eq:Hamiltonian}
    H=-\sum_v A_v -\sum_p B_p -\sum_{i=1}^3\sum_{s_i}(A_{s_i}^{K_i}+B_{s_i}^{K_i}) \,.
\ee
Here, the bulk operators are defined as
\be\ba \label{eq:Av_Bp}
    A_v:=\frac{1}{|G|}\sum_g A_v^g\,,\qquad B_p:=B_p^{\id}
\ea\ee
and the boundary ones as
\be
    A_{s_i}^{K_i}:=\frac{1}{|K_i|}\sum_{k\in K_i} A_{s_i}^{k}\,,\qquad 
    B_{s_i}^{K_i}:=\sum_{k\in K_i} B_{s_i}^{k} \,,
\ee
where $\{s_i\}$ are sites along the boundary $i$ and $A^k_{s_i},B^k_{s_i}$ are truncated at the boundary. \prx{There can also be contributions from the corners $s_{ij}$, at the pairwise intersections of boundaries $i$ and $j$. These contributions add vertex term acting on the two edges forming the corner $s_{ij}$ 
\be \label{eq:A_corner}
A_{s_{ij}}^{K_i\cap K_j} = {1\over |K_i \cap K_j|} \sum_{k\in K_i \cap K_j} A_{s_{ij}}^{k} \,.
\ee
In the codes we consider $K_i \cap K_j=\{\id\}$ for all pairs $i\neq j$, therefore these corner contributions will all be trivial. 
}

\vspace{1mm}
\noindent{\bf Non-Abelian Stabilizer Code.}
The Hamiltonian operators $A_v$ and $B_p$ are projectors and commute, therefore the quantum double Hamiltonian is commuting projector \cite{Kitaev:1997wr,Beigi:2010htr}. However, the set of all the $A_v^g$ and $B_p^g$ operators is non-commutative, recall \eqref{eq:non-comm_A_B}. We define the multiplicative commutator of invertible operators as
\be\label{commutdef}
  [\cO_1,\cO_2]:=\cO_1^{-1} \cO_2^{-1} \, \cO_1 \, \cO_2 \,.
\ee
The $A_v^g$ operators are unitary (since $L^g$ and $R^g$ are) with $(A_v^g)^{-1}=A_v^{g^{-1}}$, and obey the commutation relation
\be
    [A_v^g , A_v^h]=A_v^{g^{-1}h^{-1}gh}=A_v^{[g,h]}\,,
\ee
where, 
for a non-Abelian group $[g,h]:=g^{-1}h^{-1}gh$ can be non-trivial. 

The $B_p^g$ terms are instead projectors and are therefore non-invertible. We can however write diagonal unitary operators acting on each plaquette. 
For the codes relevant to this work, we will provide a concrete set of operators in Sec. \ref{sec:Qubits} that, together with the $A_v^g$ terms, form the {non-Abelian} stabilizer group, as we prove in App. \ref{app:StabilizerComms}.

The logical states are the inequivalent ground states of \eqref{eq:Hamiltonian}. Since the Hamiltonian terms are commuting projectors, each ground state $\ket{\psi}$ has to satisfy:
\be\ba
    A_v\ket{\psi}=B_p\ket{\psi}=A_{s_i}^{K_i}\ket{\psi}=B_{s_i}^{K_i}\ket{\psi}=\ket{\psi}\,,
\ea\ee
i.e. $\ket{\psi}$ is stabilized by the Hamiltonian terms.
Let us define the trivial ground state as:
\be \label{eq:idid_state_lattice}
    \ol{\ket{\id,\id}}:=\prod_v A_v\; \bigotimes_l \ket{\id}_l
\ee
where $v$ runs over all (bulk and boundary) vertices, and $l$ runs over all edges, while for the state labeled by the magnetic anyons $a_1,a_2,a_3$ we will denote:
\be\ba \label{eq:a_anyons}
    a_1:=([g_1],1)\,,\quad a_2:=([g_2],1)\,,\quad  a_3:=([g_1g_2],1)\,,
\ea\ee
where $1$ denotes the identity irrep and we have chosen the conventions to match \eqref{eq:triangle_g1g2}. The state corresponding to $(a_1,a_2,a_3)$ is:
\be \label{eq:g1g2_state_lattice}
    \ol{\ket{g_1,g_2}}:=\prod_v A_v\; L_{\xi_1}^{g_1} L_{\xi_2}^{g_2} L_{\xi_3}^{g_1g_2}\bigotimes_l \ket{\id}_l\,.
\ee
Here, $v$ runs over all (bulk and boundary) vertices, $l$ runs over all edges and $\xi_i$ for $i\in\{1,2,3\}$ are ribbons on the dual lattice, shown as dashed lines in Fig.~\ref{fig:LatticeAnyons}, with $L_\xi^g$ denoting the tensor product of left-multiplication operators by $g$ along the ribbon $\xi$, oriented as in \eqref{eq:triangle_g1g2}.\footnote{Note that we can use this simplified expression since we are acting on $\bigotimes_l \ket{\id}_l$. For ribbon operators acting on a general state, see \cite{Kitaev:1997wr,Beigi:2010htr,Bombin:2007qv}.}
The states
\be \label{eq:repr_states}
    \bigotimes_l \ket{\id}_l\,,\quad \text{and}\quad L_{\xi_1}^{g_1} L_{\xi_2}^{g_2} L_{\xi_3}^{g_1g_2}\bigotimes_l \ket{\id}_l
\ee
are $+1$ eigenstates of the plaquette term $B_p$. For the first state this is obvious, while for the second we note that the non-trivial plaquettes are of the type:
\be \label{eq:easy_plaquettes}
    \begin{tikzpicture}
        \begin{scope}[shift={(-2.5,0)}]
    \draw[->-] (-1,-0.5) to (-1,0.5);
    \node at (-1.3,0) {$_{g_1}$};
    \draw[->-] (-1,0.5) to (0,0.5);
    \draw[->-] (0,-0.5) to (0,0.5);
    \node at (0.3,0) {$_{g_1}$};
    \draw[->-] (-1,-0.5) to (0,-0.5);
    \end{scope}
    \begin{scope}[shift={(0,0)}]
    \draw[->-] (-1,-0.5) to (-1,0.5);
    \draw[->-] (-1,0.5) to (0,0.5);
    \node at (-0.5,0.8) {$_{g_2}$};
    \draw[->-] (0,-0.5) to (0,0.5);
    \draw[->-] (-1,-0.5) to (0,-0.5);
    \node at (-0.5,-0.8) {$_{g_2}$};
    \end{scope}
    \begin{scope}[shift={(2.5,0)}]
    \draw[->-] (-1,-0.5) to (-1,0.5);
    \draw[->-] (-1,0.5) to (0,0.5);
    \node at (-0.5,0.8) {$_{g_1g_2}$};
    \draw[->-] (0,-0.5) to (0,0.5);
    \draw[->-] (-1,-0.5) to (0,-0.5);
    \node at (-0.5,-0.8) {$_{g_1g_2}$};
    \end{scope}
    \end{tikzpicture}
\ee
along the ribbons and
\be \label{eq:hard_plaquette}
\begin{tikzpicture}
        \begin{scope}[shift={(4.4,0)}]
    \draw[->-] (-1,-0.5) to (-1,0.5);
    \node at (-1.3,0) {$_{g_1}$};
    \draw[->-] (-1,0.5) to (0,0.5);
    \node at (-0.5,0.8) {$_{g_2}$};
    \draw[->-] (0,-0.5) to (0,0.5);
    \draw[->-] (-1,-0.5) to (0,-0.5);
    \node at (-0.5,-0.8) {$_{g_1g_2}$};
    \end{scope}
\end{tikzpicture}
\ee
on the plaquette in which the three ribbons end.
All these states are $+1$ eigenstates of $B_p$ since the product of group elements taking into account orientation (as in \eqref{eq:AvBp}) is the identity group element. Since $B_p$ and $A_v$ are commuting projectors, the states \eqref{eq:idid_state_lattice} and \eqref{eq:g1g2_state_lattice} are therefore ground states of \eqref{eq:Hamiltonian}. They are orthogonal since they have different commutation relations with the electric anyon triangle (since, by assumption the $(a_1,a_2,a_3)$ and $(b_1,b_2,b_3)$ anyon configurations braid non-trivially), hence \eqref{eq:idid_state_lattice} and \eqref{eq:g1g2_state_lattice} correspond to distinct logical states.

\prx{\noindent\textbf{Logical Operators.} In this work, we will focus on boundary conditions that encode a single logical qubit, i.e. we will require only one pair of magnetic and electric anyon triplets $(a_1,a_2,a_3)$ and $(b_1,b_2,b_3)$.

We will take the $(b_1,b_2,b_3)$ anyon triplet to be comprised by 1-dimensional and order-two irreps $(\bm{R_1,R_2,R_1R_2})$ of $G$ so that
the logical Pauli $\ol{Z}$ is realized as
\be \label{eq:triangle_R1R2}
\begin{tikzpicture}[baseline]
\begin{scope}[shift={(0,0)}]
\draw [thick, fill= \BlueColor, opacity=0.5]  
(0,0) -- (4,0) -- (2,2.5)  --(0,0); 
 \draw [very thick] (0,0) -- (4,0) -- (2,2.5)  --(0,0); 
\draw[very thick, red, ->-] (1, 1.25) -- (2, 0.8)  ; 
\draw[very thick, blue, ->-] (3, 1.25) -- (2, 0.8)   ; 
\draw[very thick, \thirdcolor, ->-] (2, 0.8)-- (2, 0)   ; 
\node[red] at (1.55, 1.3) {$\bm{R_1}$} ;
\node[blue] at (2.45, 1.3) {$\bm{R_2}$} ;
\node[right, \thirdcolor] at (2, 0.35) {$\bm{R_1R_2}$} ;
\draw[fill=black] (0,0) ellipse (0.05 and 0.05);
\draw[fill=black] (4,0) ellipse (0.05 and 0.05);
\draw[fill=black] (2,2.5) ellipse (0.05 and 0.05);
\node[\thirdcolor, below] at (2,0) {$\cL_3$};
\node[left] at (-0.3,1.3) {$\ol{Z}\quad\mapsto$};
\node[red, left] at (0.9,1.2) {$\cL_1$};
\node[blue, right] at (3.1,1.2) {$\cL_2$};
\end{scope}
\end{tikzpicture}
\ee
On the lattice, it is concretely written as
\be \label{eq:logical_Z_R1R2}
\qquad\ol{Z}=\prod_{l\in\gamma_1}(Z_{\bm{R_1}})_l \prod_{j\in\gamma_2}(Z_{\bm{R_2}})_j \prod_{k\in\gamma_3}(Z_{\bm{R_1R_2}})_k\,,\\
\ee
where $\gamma_i$ for $i\in\{1,2,3\}$ are the ribbons on the lattice that support the $b_i$ anyons as shown in Fig.~\ref{fig:LatticeAnyons}. Since we require these irreps to be of order-two, it follows immediately that
\be
    \ol{Z}^2=\bbI\,.
\ee

The $(a_1,a_2,a_3)$ triplet is instead comprised of non-abelian anyons \eqref{eq:a_anyons} with non-invertible bulk fusions. To define an invertible logical $\ol{X}$, we will therefore require
\be
    [g_1]\cap K_1=\{g_1\}\,,\quad [g_2]\cap K_2=\{g_2\}\,.
\ee
This ensures that the boundary conditions labeled by $K_1$ and $K_2$ will select a 1-dimensional (invertible) subspace of the anyons $a_1,a_2$ corresponding to the group elements $g_1,g_2$ respectively. The fusion constraints will then select the group element $g_1g_2\in[g_1g_2]$ as an invertible subspace of the anyon $a_3$. In order for the logical operator to be of order-two we will also require:
\be \label{eq:g1_g1_order2}
    g_1^2=\id\,,\quad g_2^2=\id\,, 
\ee
which furthermore imply that $(g_1g_2)(g_2g_1)=\id$\,. Finally, we will require that 
\be \label{eq:K3_conj_class}
    [g_1g_2]\cap K_3=\{g_1g_2,g_2g_1\}\,.
\ee
The logical $\ol{X}$ operator is then defined as:
\be \label{eq:triangle_g1g2}
\begin{tikzpicture}[baseline]
\begin{scope}[shift={(0,0)}]
\draw [thick, fill= \BlueColor, opacity=0.5]  
(0,0) -- (4,0) -- (2,2.5)  --(0,0); 
 \draw [very thick] (0,0) -- (4,0) -- (2,2.5)  --(0,0); 
\draw[very thick, red, ->-] (1, 1.25) -- (2, 0.8)  ; 
\draw[very thick, blue, ->-] (3, 1.25) -- (2, 0.8)   ; 
\draw[very thick, \thirdcolor, ->-] (2, 0.8)-- (2, 0)   ; 
\node[red] at (1.55, 1.25) {$g_1$} ;
\node[blue] at (2.45, 1.25) {$g_2$} ;
\node[right, \thirdcolor] at (2, 0.35) {$g_1g_2$} ;
\draw[fill=black] (0,0) ellipse (0.05 and 0.05);
\draw[fill=black] (4,0) ellipse (0.05 and 0.05);
\draw[fill=black] (2,2.5) ellipse (0.05 and 0.05);
\node[\thirdcolor, below] at (2,0) {$\cL_3$};
\node[left] at (-0.3,1.3) {$\ol{X}\quad\mapsto$};
\node[red, left] at (0.9,1.2) {$\cL_1$};
\node[blue, right] at (3.1,1.2) {$\cL_2$};
\end{scope}
\end{tikzpicture}
\ee
When computing the product of this operator with itself, we will use the fact that when $\textcolor{\thirdcolor}{g_1g_2}$ passes through $\textcolor{blue}{g_2}$, it gets conjugated (see e.g. \cite[Fig.~7]{Barkeshli:2014cna}) to 
\be \label{eq:g2_cong}
    \textcolor{blue}{g_2}\textcolor{\thirdcolor}{g_1g_2}\textcolor{blue}{g_2^{-1}}=\textcolor{\thirdcolor}{g_2g_1} \,,
\ee
where we write both $\textcolor{\thirdcolor}{g_1g_2}$ and $\textcolor{\thirdcolor}{g_2g_1}$ in \textcolor{\thirdcolor}{orange} since they both belong to subgroup $K_3$, labeling the third boundary, recall \eqref{eq:K3_conj_class}. Using \eqref{eq:g2_cong} and \eqref{eq:g1_g1_order2}, we conclude that the $\ol{X}$ operator squares to $\bbI$:
\be \label{eq:XX}
\begin{tikzpicture}[baseline]
\begin{scope}[shift={(0,0)}]
\draw [thick, fill= \BlueColor, opacity=0.5]  
(0,0) -- (4,0) -- (2,2.5)  --(0,0); 
 \draw [very thick] (0,0) -- (4,0) -- (2,2.5)  --(0,0); 
\end{scope}
\begin{scope}[shift={(0.15,0.45)}]
\draw[very thick, red, ->-] (1.15 , 1.15 ) -- (2, 0.8);  
\draw[very thick, blue, ->-] (2.62, 1.1) -- (2, 0.8)   ; 
\draw[very thick, \thirdcolor, ->-] (2, 0.8)-- (2, -0.45)   ; 
\node[red] at (1.55, 1.25) {$g_1$} ;
\node[blue] at (2.3, 1.25) {$g_2$} ;
\node[right, \thirdcolor] at (2, -0.1) {$g_2g_1$} ;
\end{scope}
\begin{scope}[shift={(0,0)}]
\draw[very thick, red, ->-] (1, 1.25) -- (2, 0.8)  ; 
\draw[very thick, blue, ->-] (3, 1.25) -- (2, 0.8)   ; 
\draw[very thick, \thirdcolor, ->-] (2, 0.8)-- (2, 0)   ; 
\node[red] at (1.55, 1.25) {$g_1$} ;
\node[blue] at (2.45, 1.2) {$g_2$} ;
\node[right, \thirdcolor] at (1, 0.35) {$g_1g_2$} ;
\draw[fill=black] (0,0) ellipse (0.05 and 0.05);
\draw[fill=black] (4,0) ellipse (0.05 and 0.05);
\draw[fill=black] (2,2.5) ellipse (0.05 and 0.05);
\node[\thirdcolor, below] at (2,0) {$\cL_3$};
\node[left] at (-0.3,1.3) {$\ol{X}\ol{X}\quad\mapsto$};
\node[right] at (4.5,1.3) {$=\quad \bbI$};
\node[red, left] at (0.9,1.2) {$\cL_1$};
\node[blue, right] at (3.1,1.2) {$\cL_2$};
\end{scope}
\end{tikzpicture}
\ee

With the conventions of Fig.~\ref{fig:LatticeAnyons} and the choice of irreps explained above, the property of $-1$ braiding between the magnetic $(a_1,a_2,a_3)$ and electric $(b_1,b_2,b_3)$ anyon triplets translates to requiring $\bm{R_2}(g_1)=-1$.\footnote{\prx{This follows from the modular $S$-matrix element $S_{a_1,b_2}$, whose general expression can be found in \cite{Beigi:2010htr,Coste:2000tq}.}} In terms of the lattice operators, using \eqref{eq:LgRg}-\eqref{eq:irrep_ops_G} we have that $(L^{g_1}Z_{\bm{R_2}})_l=-(Z_{\bm{R_2}}L^{g_1})_l$ at the left edge $l=\xi_1\cap\gamma_2$ of the plaquette where the three $(a_1,a_2,a_3)$ anyons end, which is to the north-east of the vertex on which the $(b_1,b_2,b_3)$ anyons end, as shown in Fig.~\ref{fig:LatticeAnyons}.\footnote{\prx{This is analogous to the $\Z_2\times\Z_2$ case discussed below \eqref{eq:Z2Z2_Pauli}.}}
At the level of the logical operators, this implies the anti-commutation relation between $\ol{Z}$ and $\ol{X}$:
\be
    \ol{ZX}=-\ol{XZ}\,.
\ee
Finally, note that the logical states \eqref{eq:idid_state_lattice} and \eqref{eq:g1g2_state_lattice} form a complete qubit logical basis for the non-abelian code $D(G)$ on a triangle: they are respectively the $+1$ and $-1$ eigenstates of the logical $\ol{Z}$ operator \eqref{eq:logical_Z_R1R2}. $\ol{X}$ anti-commutes with $\ol{Z}$ and thus exchanges $\ol{\ket{0}}$ and $\ol{\ket{1}}$.}

\subsection{Constant-Depth Gates}
\label{sec:Constant-DepthGates}

We now prove Theorem \ref{thm:SPT}. 
Given a surface code we can construct unitary operators $U$ that implement automorphisms of $D(G)$ and are topologically protected. We will apply these along a spatial slice as shown in Fig.~\ref{fig:Prism}. 

\vspace{1mm}
\noindent{\bf SPTs as Automorphisms of $D(G)$.}
The mathematical classification of automorphisms of $D(G)$  was provided in~\cite{Davydov2009ModularIF}:
they are specified by a subgroup 
\be
\Auto\subseteq G\times G\,,
\ee
and 2-cocycle 
\be
\auto\in H^2(\Auto,U(1))
\ee
(see App. \ref{app:2coc} for a brief summary) such that $p_1(K)=p_2(\Auto)=G$, where $p_k: G\times G \to G$ are the projections onto the $k$-th factor and the cocycle needs to be such that 
\be
\epsilon(g,h):=\frac{\auto(g,h)}{\auto(ghg^{-1},\,g)} 
\ee
on $(\Auto\cap(G\times\{\id\}))\times (\Auto\cap(\{\id\}\times G))$ is non-degenerate.

Intuitively one can understand this as follows: an automorphism is an interface between $D(G)$ with itself, which in turn can be classified by gapped boundary conditions on the folded theory 
$D(G) \boxtimes \overline{D(G)} \cong D(G \times G)$, which are indeed classified by subgroups and cocycles (see ~\cite{Ostrikmodule,Davydov2009ModularIF,davydov2017lagrangian,Natale2017} and App. \ref{sec:GappedBC}). The conditions on $\Auto$ and $\auto$ ensure that the resulting gapped boundary of $D(G\times G)$ corresponds to an automorphism of $D(G)$: i.e. each anyon is mapped uniquely to an anyon.

\vspace{1mm}
\noindent{\bf Constant-Depth Gates from SPT-stacking.}
We will consider automorphisms of $D(G)$ where $\Auto = G_{\diag} \subset G\times G$. Since $G^\diag\cong G$, the automorphism is thus specified by a group 2-cocycle $\auto\in H^2(G,U(1))$.\footnote{\prx{We will require that the 2-cocycle representative $\alpha$ satisfies the properties of \cite[Lemma 5.1]{Beigi:2010htr}, which we summarize in \eqref{eq:lemma5.1}.}}
This automorphism will be realized by inserting the 2D surface along {a fixed time slice} into the prism, {i.e. on a 2D spatial surface, as shown in Fig.~\ref{fig:Prism}}. On the anyons labeled by the identity group element (electric anyons) this acts trivially, but it acts non-trivially on the magnetic and dyonic anyons.

For simplicity, in the following we will choose boundary conditions $\cL_i(K_i,\varphi_i)$ of $D(G)$ with trivial 2-cocycles $\varphi_i\equiv1$. In order for the $\auto$ surface to end on the boundaries of the prism, we must require that for each boundary condition $\cL (K_i, 1)$, there are 1-cochains $\beta^{(i)}: K_i\to U(1)$ that satisfy 
\be \label{eq:alphaK_delta_beta}
    \auto(g,h)\big\vert_{K_i}=\frac{\beta^{(i)}(g)\beta^{(i)}(h)}{\beta^{(i)}(gh)}\quad \forall\;g,h\in K_i\,.
\ee
This ensures that the 2-cocycle $\auto$ is trivialized on the boundary and that $\frac{\auto|_{K_i}}{\delta\beta^{(i)}}$ can consistently end on {it}. Note that the above equation determines $\beta^{(i)}$ up to 1-cocycles in $H^1(K_i,U(1))$, which are 1-dimensional representations of $K_i$.
At each corner $C_{ij}$, we require:
\be \label{eq:beta_i_beta_j}
    \beta^{(i)}(g)=\beta^{(j)}(g) \quad \forall\; g\in K_i\cap K_j\,.
\ee
\prx{Denote by $|g_1, g_2\rangle$ the logical state obtained by applying an operator labeled by $(g_1,g_2,g_1g_2)$ as shown in \eqref{eq:triangle_g1g2}: the trivial logical state is obtained by setting $g_1=\id,g_2=\id$ while the non-trivial state by applying the non-trivial $g_1,g_2$ appearing in the logical $\ol{X}$ operator \eqref{eq:triangle_g1g2}.}

{The automorphism stacking} gives rise to a diagonal constant-depth gate 
that acts on the state $|g_1, g_2\rangle$ by the phase 
\be \label{eq:Ualphabeta_in_proof}
    U_{\alpha,\beta}(g_1,g_2)=\frac{\auto(g_1,g_2)\beta^{(3)}(g_1g_2)}{\beta^{(1)}(g_1)\beta^{(2)}(g_2)}\,,
\ee
where $\alpha(g_1,g_2)$ is the bulk contribution coming from the 2-cocycle and each $\beta^{(i)}(g_i)$ is the boundary counter-term required for $\alpha$ to end consistently on the $i$-th boundary.
Since $U_{\alpha,\beta}$ is constructed from an automorphism of $D(G)$, it preserves the logical codespace (i.e. maps ground states of the quantum double to ground states): the encoded gate is therefore a logical operator on the codespace. Note that if $\auto=\delta\beta$ on the whole group $G$, then $U_{\alpha,\beta}$ is the trivial logical operator.

\vspace{1mm}
\noindent{\textbf{Lattice implementation of $U_{\alpha,\beta}$.}} On the lattice, 
denote by $\fB_i$ is the $i$th boundary on which the anyons in the Lagrangian algebra $\cL_i$ are condensed. The boundary 1-cochains $\beta^{(i)}$ for $i\in\{1,2,3\}$ are realized as a diagonal gate $M^{\beta^{(i)}}$ acting on each lattice edge $e$ along $\fB_i$ in the basis $\{\ket{g}:g\in G\}$ as follows:
\be \label{eq:Mbetai_def}
    M^{\beta^{(i)}}_e\ket{g}_e=\beta^{(i)}(g)\ket{g}_e \,.
\ee
The 2-cocycle $\alpha$ is evaluated on a triangulated plaquette $p$ by acting with the corresponding operator $M^\alpha_p$:  
\be \label{eq:Malpha_general}
\begin{tikzpicture}
\begin{scope}[shift={(-1,0)}, scale= 1.2]
    \draw[->-] (-1,-0.5) to (-1,0.5);
    \node at (-1.3,0) {${{g_1}}$};
    \draw[->-] (-1,0.5) to (0,0.5);
    \node at (-0.5,0.8) {${{g_2}}$};
    \draw[->-] (0,-0.5) to (0,0.5);
    \node at (0.3,0) {${g_3}$};
    \draw[->-] (-1,-0.5) to (0,-0.5);
    \node at (-0.5,-0.8) {${g_4}$};
    \draw  (-1,-0.5) -- (0,0.5) ;
    \node at (-0.7, 0.2) {$\circlearrowright$}; 
     \node at (-0.25, -0.2) {$\circlearrowleft$}; 
    \node at (-1.8,0) {$M^\alpha_p\, \Bigg|$};
    \node[right] at (0.3, 0) {$\Biggl>$};
    \end{scope}   
    \begin{scope}[shift={(2.3,-2)}, scale= 1.2]
    \draw[->-] (-1,-0.5) to (-1,0.5);
    \node at (-1.3,0) {${{g_1}}$};
    \draw[->-] (-1,0.5) to (0,0.5);
    \node at (-0.5,0.8) {${{g_2}}$};
    \draw[->-] (0,-0.5) to (0,0.5);
    \node at (0.3,0) {${g_3}$};
    \draw[->-] (-1,-0.5) to (0,-0.5);
    \node at (-0.5,-0.8) {${g_4}$};
    \draw  (-1,-0.5) -- (0,0.5) ;
    \node at (-0.7, 0.2) {$\circlearrowright$}; 
     \node at (-0.25, -0.2) {$\circlearrowleft$}; 
    \node at (-1.6,0) {$ \Bigg|$};
       \node[right] at (0.3, 0) {$\Biggl>$ \,.};
       \node at (-2.8,0) {$=\quad${\Large $\frac{\alpha(g_1,g_2)}{\alpha(g_4,g_3)}$}};
    \end{scope}  
\end{tikzpicture}
\ee
The constant-depth gate $U_{\alpha,\beta}$ will therefore be implemented as:
\be \label{eq:U_alpha_lattice}
     U_{\alpha,\beta}=\prod_{p}M_p^\alpha \, \prod_{e\in \fB_1} \lb M^{\beta^{(1)}}_e\rb^\dagger
    \prod_{e\in \fB_2} \lb M^{\beta^{(2)}}_e\rb^\dagger
    \prod_{e\in \fB_3} M^{\beta^{(3)}}_e  \,.
\ee
Each of these operators acts diagonally by phases, and thus has a standard decomposition into diagonal gates. 

\smallskip
\noindent
\prx{\textbf{Proof of Theorem~\ref{thm:SPT}.} 
We now explicitly demonstrate that the logical action implemented by the $U_{\alpha,\beta}$ circuit \eqref{eq:U_alpha_lattice} is \eqref{eq:Ualphabeta_in_proof}. Note that we choose the 2-cocycle $\alpha$ such that properties \eqref{eq:lemma5.1} hold: in particular $\alpha(\id,g)=\alpha(g,\id)=1$ and $\alpha(g,g^{-1})=1$, for all $g\in G$. These also imply that the 1-cochains $\beta^{(i)}$ can be taken to satisfy $\beta^{(i)}(\id)=1$ and $\beta^{(i)}(g^{-1})=\beta^{(i)}(g)^{-1}$ for all $g\in G$.

We will compute the action of $U_{\alpha,\beta}$  \eqref{eq:U_alpha_lattice} on the representative states \eqref{eq:repr_states} and show that it is \eqref{eq:Ualphabeta_in_proof}. We will then demonstrate in App. \ref{app:proof_logical_action_indep} that this action is the same on all physical states (related by $A_v$ operators) appearing in the logical states \eqref{eq:idid_state_lattice} and \eqref{eq:g1g2_state_lattice}, therefore $U_{\alpha,\beta}$ encodes the logical action \eqref{eq:Ualphabeta_in_proof}.

Let us first consider the trivial logical state \eqref{eq:idid_state_lattice}, with representative shown on the left of \eqref{eq:repr_states}. Since $\alpha(\id,\id)=1$ and  $\beta^{(i)}(\id)=1$, all operators in the circuit \eqref{eq:U_alpha_lattice} act on this representative state trivially with phase $1$.

We now turn to the non-trivial logical state \eqref{eq:g1g2_state_lattice}, with representative state written on the right of \eqref{eq:repr_states}. On the plaquettes shown in \eqref{eq:easy_plaquettes}, $M^\alpha_p$ \eqref{eq:Malpha_general} evaluates to $1$, since $\alpha(\id,g)=\alpha(g,\id)=1$. Instead, on the plaquette where the three magnetic anyons meet, with group elements shown in \eqref{eq:hard_plaquette}, the operator  $M^\alpha_p$ \eqref{eq:Malpha_general} produces the phase $\alpha(g_1,g_2)$. On each boundary, the representative state on the right of \eqref{eq:repr_states} has a single edge with $\ket{g_1},\ket{g_2},\ket{g_1g_2}$ respectively, while all others edges have the trivial group element $\ket{\id}$. The $M_e^\beta$ operators in \eqref{eq:U_alpha_lattice} thus evaluate to $\frac{\beta^{(3)}(g_1g_2)}{\beta^{(1)}(g_1)\,\beta^{(2)}(g_2)}$. The action of the $U_{\alpha,\beta}$ circuit \eqref{eq:U_alpha_lattice} on the representative states in \eqref{eq:repr_states} is therefore precisely \eqref{eq:Ualphabeta_in_proof}.

Importantly, the logical action of $U_{\alpha,\beta}$ does not depend on the choice of representative physical state in the superpositions \eqref{eq:idid_state_lattice} and \eqref{eq:g1g2_state_lattice}: we prove this in App. \ref{app:proof_logical_action_indep}. In conclusion, the logical action of the $U_{\alpha,\beta}$ circuit \eqref{eq:U_alpha_lattice} on the logical states \eqref{eq:idid_state_lattice} and \eqref{eq:g1g2_state_lattice} is precisely \eqref{eq:Ualphabeta_in_proof}.}
$\qedsymbol$
\smallskip

From the lattice expression \eqref{eq:U_alpha_lattice}, it is clear that the gate $U_{\alpha,\beta}$ is topologically protected (as defined in \cite{Bravyi:2012rnv}), since it is implemented by applying a constant-depth quantum circuit on the physical qudits. Importantly, the operators appearing in \eqref{eq:U_alpha_lattice} act only on a few qudits (each $M^\alpha_p$ acts on the four qudits on the edges of the plaquette and each $M_e^{\beta^{(i)}}$ acts only on a single qudit at edge $e$). The operator \eqref{eq:U_alpha_lattice} is therefore inherently \prx{fault tolerant}: an error in any single gate appearing in its expression cannot spread beyond $O(1)$ lattice sites. 

We will now construct automorphisms of specific quantum doubles, and show that they realize constant-depth non-Clifford gates on the logical qubit states.

\section{Constant-Depth Clifford-Hierarchy Gates from Non-Abelian Stabilizer Codes} \label{sec:CliffHier}

We now apply the general framework to the construction of constant-depth logical $T^{1/N}$ gates \prx{in 2D via the $D(D_{4N})$ surface code}. This proves Corollary \ref{coro:D4N}.

\subsection{Constant-Depth $T$-gate from $D(D_4)$ Clifford Stabilizer Code} 
\label{sec:T}
First, let us re-derive the constant-depth $T$-gate, which was realized in \cite{Kobayashi:2025cfh} in terms of the description of the topological order $D^\omega(\Z_2^3)$  (we have summarized this in App. \ref{app:T_from_rgb}) in terms of  the (isomorphic) topological order $D(D_4)$. This latter presentation will be central to the generalization to $T^{1/N}=P(\pi/(4N))$ constant-depth gates. 
First we summarize some aspects of the surface code for $D(D_4)$. 

\subsubsection{Clifford Stabilizer Code for $D(D_4)$}
\label{sec:CliffStabD4}

Here, the non-Abelian group is the dihedral group (of symmetries of the square) 
\be \label{eq:D4}
    D_4=\langle r,s \,|\, r^4=s^2=\id\,,\; srs=r^{-1}\rangle \,.
\ee
Its irreducible representations (irreps) are:
\begin{itemize}
    \item 4 irreps of dimension 1, generated by $1_r$ and $1_s$, where: 
    \be
    \ba
        1_r(r)&=+1\,, & 1_r(s)&=-1\,,\\
        1_s(r)&=-1\,, & 1_s(s)&=+1\,,\\
    \ea
    \ee
    \item An irrep of dimension 2, which we label as $E$ with non-zero characters
    \be
        \chi_E(\id)=2\,,\quad\chi_E(r^2)=-2\,\,.
    \ee
\end{itemize}
On the lattice, the local Hilbert space on each edge is spanned by the group elements:
\be \label{eq:D4_elements}
    \ket{r^as^j}\mapsto\ket{a}\ket{j}\,,\quad a\in\{0,1,2,3\}\,,\quad j\in\{0,1\}\,,
\ee
in terms of a $\Z_4$-qudit and a qubit. 
We will denote by $\cX,\cZ$ the generalized Pauli operators for the $\Z_4$ qudit and by $X,Z$ the standard Paulis for the qubit. Furthermore, let $
\cC$ be the charge-conjugation operator on the $\Z_4$ qubit. They satisfy, for $a\in\{0,1,2,3\}$:
\begin{align} \label{eq:Z4_X_Z_C}
    \cC\ket{r^a}&=\ket{r^{4-a}}\,,\;\;\cX\ket{r^a}=\ket{r^{a+1}}\,,\;\; \cZ\ket{r^a}=i^a\ket{r^a}\,,\nn\\
\cC \cX \cC^\dagger&=\cX^\dagger\,,\quad\;\;\; \cC \cZ \cC^\dagger=\cZ^\dagger\,.
\end{align}

On this Hilbert space, the left and right multiplication operators for $D_4$ are generated by \cite{Albert:2021vts}:
\begin{align}\label{eq:D4_LR_ops}
    L^r&=\cX\otimes\bbI\,,&\;\; R^r&=\cX^{-Z}\,,\nn\\
    L^s&=\cC\otimes X\,,&\;\; R^s&=\bbI\otimes X\,,
\end{align}
\prx{where for $R^r$ we use the notation
\be
\cX^{-Z}:=\cX^{-1}\otimes\ket{0}\bra{0}+\cX\otimes\ket{1}\bra{1}\,,
\ee
}
while the diagonal irrep operators are realized as \cite{Albert:2021vts}:
\begin{align} \label{eq:D4_rep_ops}
    Z_{1_r}&=\bbI\otimes Z\,, \quad Z_{1_s}=\cZ^{2}\otimes \bbI\,, \quad Z_{1_{rs}}=\cZ^{2}\otimes Z\,,\nn\\
    Z_E&=
    \begin{pmatrix}
        \cZ \otimes \ket{0}\bra{0} & \cZ\otimes \ket{1}\bra{1} \\
        \cZ^\dagger\!\otimes \ket{1}\bra{1} & \cZ^\dagger\!\otimes \ket{0}\bra{0} \\
    \end{pmatrix}\,.
\end{align}

We can write  a stabilizer group for \prx{the} $D(D_4)$ surface code as follows: 
\be
\cS_{D(D_4)} = \left\langle A_v^r \,,\, A_v^s\,,\, S_p^r \,,\, S_p^s  \right\rangle \,,
\ee
where 
\be
\ba
A_v^{r} &= 
\begin{tikzpicture}[baseline]
\begin{scope}[shift={(0,0.1)}]
\draw[thick, ->-] (-1,0) to (0,0);
\draw[thick, ->-] (0,-1) to (0,0);
\draw[thick, ->-] (0,0) to (0,1);
\draw[thick, ->-] (0,0) to (1,0);
\draw[fill=black] (0,0) ellipse (0.05 and 0.05);
\node[] at (0.15, 0.15) {$_v$};
\node[above] at (-0.6,0.05) {$\mathcal{X}^{-Z}$};
\node[right] at (0.04,-0.7) {$\mathcal{X}^{-Z}$};
\node[right] at (0.04,0.8) {$\mathcal{X}$};
\node[above] at (0.75,0.05) {$\mathcal{X}$};
\draw[fill=black] (0,0) ellipse (0.05 and 0.05);
\end{scope}
\end{tikzpicture}\,, \quad
A_v^{s} = 
\begin{tikzpicture}[baseline]
\begin{scope}[shift={(0,0.1)}]
\draw[thick, ->-] (-1,0) to (0,0);
\draw[thick, ->-] (0,-1) to (0,0);
\draw[thick, ->-] (0,0) to (0,1);
\draw[thick, ->-] (0,0) to (1,0);
\draw[fill=black] (0,0) ellipse (0.05 and 0.05);
\node[] at (0.15, 0.15) {$_v$};
\node[above] at (-0.7,0.05) {$X$};
\node[right] at (0.04,-0.7) {$X$};
\node[right] at (0.04,0.8) {$\mathcal{C}X$};
\node[above] at (0.75,0.05) {$\mathcal{C}X$};
\draw[fill=black] (0,0) ellipse (0.05 and 0.05);
\end{scope}
\end{tikzpicture} \,,\cr
S_p^r &=
\begin{tikzpicture}[baseline]
\begin{scope}[shift={(0,-0.4)}]
    \draw[thick, ->-] (-1,0) to (-1,1);
    \node at (-1.4,0.5) {$\mathcal{Z}_{1}$};
    \draw[thick, ->-] (-1,1) to (0,1);
    \node at (-0.5,1.4) {$\mathcal{Z}_{2}^{Z_1}$};
    \draw[thick, ->-] (0,0) to (0,1);
    \node[right] at (0.2,0.5) {$\mathcal{Z}_{3}^{-Z_1Z_2Z_3}$};
    \draw[thick, ->-] (-1,0) to (0,0);
    \node at (0.2,-0.35) {$\mathcal{Z}^{-Z_1Z_2Z_3Z_4}_{4}$};
    \node at (-0.5,0.5) {$_p$};
\end{scope}     
\end{tikzpicture} \cr 
S_p^s &=
\begin{tikzpicture}[baseline]
\begin{scope}[shift={(0,-0.4)}]
    \draw[thick, ->-] (-1,0) to (-1,1);
    \node[left] at (-1,0.5) {$S$};
    \draw[thick, ->-] (-1,1) to (0,1);
    \node at (-0.5,1.35) {$S$};
    \draw[thick, ->-] (0,0) to (0,1);
    \node[right] at (0.1,0.5) {$S$};
    \draw[thick, ->-] (-1,0) to (0,0);
    \node at (-0.5,-0.3) {$S$};
     \node at (-0.5,0.5) {$_p$};
\end{scope}    
\end{tikzpicture} \,,
\ea\ee
where $Z_i$ for $i=1,2,3,4$ denotes the qubit Pauli $Z$ operators on the $i$-th edge of the plaquette and we use the qubit gate $S:=\diag(1,i)$. 

Unlike qubit Pauli stabilizers, 
the stabilizer group $\cS_{D(D_4)}$ is non-Abelian, with non-trivial commutators as defined in (\ref{commutdef}). These are given by 
\be
\ba
    \left[ A_{v}^{r}, A_{v}^{s} \right] 
    &= (A_{v}^{r})^{-2} \cr 
    \left[ A_{v}^{r}, S_{p_{NE}}^r \right] 
    &= (S_{p_{NE}}^s)^{-2} \cr     
    \left[ A_{v}^{s}, S_{p_{NE}}^r \right] 
    &= (S_{p_{NE}}^r)^2\,.
\ea\ee
Here $p_{NE}$ is the plaquette in the North-East direction with respect to the vertex $v$. 
The code space is defined as the subspace of the Hilbert space that has $+1$ eigenvalue for all operators in $\cS_{D(D_4)}$, {as discussed in App. \ref{app:StabilizerComms}}. 

\prx{In terms of the $\Z_4$-qudit and qubit description, the stabilizer operators are Clifford since they involve the operator $\cC$ which conjugates $\cX,\cZ$ to $\cX^\dagger,\cZ^\dagger$, recall \eqref{eq:Z4_X_Z_C}, and the Clifford $S$-gate on the qubit, in addition to the (generalized) Pauli operators $\cX,\cZ,X,Z$.} We will show in Sec. \ref{sec:Qubits} that in the 3-qubit description, the stabilizer group is comprised of 3-qubit Clifford operators, which is why we refer to this as a {\bf Clifford Stabilizer Code}. {We derive these operators and prove their commutation relations in App. \ref{app:StabilizerComms}.}

\subsubsection{Constant-Depth $T$-gate} 

The quantum double has anyons labeled by $([g], \bm{R})$,
where $[g]$ is a conjugacy class $g\in G$ and $\bm{R}$ is an irrep of the centralizer of $g$. For $D(D_4)$ this is explained in detail in \cite{Bhardwaj:2024qrf}. We will label the anyons with conjugacy classes $[s],[sr]$ whose centralizer is $\Z_2 \times \Z_2$, with irreps for each $\Z_2$ labeled by $\pm$; we provide some background and notation in App. \ref{app:GroupTheory}.
Consider the gapped boundaries 
\be\ba\label{LagsD8}
\cL_{\langle rs \rangle}&=1\oplus 1_{rs} \oplus E \oplus [rs]_{++}\oplus [rs]_{+-}\cr
\cL_{\langle s \rangle}&=1\oplus 1_s \oplus E \oplus [s]_{++}\oplus [s]_{+-}\cr 
\cL_{\langle r \rangle}&=1\oplus 1_r \oplus 2[r]\oplus [r^2]\oplus [r^2]1_r
\ea\ee
labeled by the subgroups 
\be\ba
\langle rs \rangle&=\{\id,rs\}\cong\Z_2\cr
\langle s \rangle&=\{\id,s\}\cong\Z_2\cr
\langle r \rangle&=\{\id,r,r^2,r^3\}\cong\Z_4\,.
\ea\ee

We perform an automorphism $U$ of $D(D_4)$ given by inserting along a triangular spatial slice a topological surface (see Fig.~\ref{fig:Prism}), specified by a representative of the non-trivial cohomology class
\be
\alpha\in H^2(D_4,U(1))=\Z_2\,,
\ee
which we compute in App. \ref{app:2coc}. 
One has
\be
\alpha|_{\langle s \rangle}=\alpha|_{\langle rs \rangle}= 1
\ee
so we will take the 1-cochain on the corresponding boundaries to be identically 1. Note also that $\alpha(rs,s)=1$.\\ 
On $\langle r\rangle\cong \Z_4$, instead
\be\ba \label{eq:alpharbeta_D8}
    \alpha|_{\langle r \rangle}&=\delta\beta
\ea\ee
for $\beta$ given by:
\be \label{eq:beta_r}
    \beta(\id)=\beta(r^2)=1\,,\ \beta(r^\pm )=e^{\pm i\pi/4}\,.  
\ee
One can check that all equations \eqref{eq:alpharbeta_D8}, i.e.
\be
    \alpha|_{\langle r \rangle}(g_1,g_2)=\frac{\beta(g_1)\beta(g_2)}{\beta(g_1g_2)}\,, 
\ee
are satisfied $\forall\,g_1,g_2\in\langle r\rangle=\{\id,r,r^2,r^3\}$.
The map on the anyons induced by the automorphism is the following:
\be\label{Uaction}
U:\qquad 
\ba
[r^2]\phantom{1_1} & \ \longleftrightarrow \ [r^2] 1_r \cr 
[r^2]1_s & \ \longleftrightarrow \ [r^2]1_{rs} \cr 
[s]_{++} &\ \longleftrightarrow \  [s]_{+-}\cr 
[s]_{-+} &\ \longleftrightarrow \  [s]_{--}\cr 
[rs]_{++} & \ \longleftrightarrow \  [rs]_{+-}\cr 
[rs]_{-+} & \ \longleftrightarrow \ [rs]_{--}
\ea\,.
\ee
This leaves all three \prx{Lagrangian algebras} in (\ref{LagsD8}) invariant.


\vspace{1mm}
\noindent{\bf Implementation of Logical $T$-gate.} 
We initialize the system in the eigenstates of $\ol{Z}$, i.e. $\ol{\ket{m}}$ for $m=0,1$. The state $\ol{\ket{0}}$ comes from an anyon triangle with 
\be
    g_1=g_2=\id \quad\Rightarrow\quad U(\id,\id)=1\,,
\ee 
while for the non-trivial state $\ol{\ket{1}}$ we take 
\be
g_1=rs,\;g_2=s \quad\Rightarrow\quad U(rs,s)=\beta(r)=e^{i\pi/4}\,.
\ee 
The automorphism $U$ therefore acts as 
\be
U \ol{\ket{m}} = e^{i \pi m /4}  \ol{\ket{m}} = T \ol{\ket{m}} \,,
\ee
and thus encodes the diagonal  
constant-depth $T$-gate acting on the logical qubit.

\vspace{1mm}
\noindent{\textbf{Lattice Implementation.}} On the lattice, the unitary $U_{\alpha,\beta}$ is realized as the operator in \eqref{eq:U_alpha_lattice}. In this case, the only non-identity boundary operator is the one associated to the 1-cochain $\beta$ \eqref{eq:beta_r} on the $\langle r\rangle$ boundary. On the physical Hilbert space \eqref{eq:D4_elements} on each edge $e\in \fB_3$ (where $\cL_{\langle r\rangle}$ condenses), $\beta$ is implemented by the operator:
\be  
    \ba
M^\beta_e= &\diag\left(1,e^{i\pi/4},1,e^{-i\pi/4}\right)\otimes\bbI\cr 
=& \lb (CS^\dagger)(\bbI\otimes T)\rb \otimes \bbI \,.
\ea 
\ee
In the last line we used the decomposition on the 3-qubit Hilbert space that we define in Sec.~\ref{sec:Qubits} and the standard notation for qubit operators $CS=\diag(1,1,1,i)$, $T=\diag(1,e^{i\pi/4})$, $\bbI=\diag(1,1)$.
Since $\alpha(\id,\id)=\alpha(rs,s)=1$, on the states of interest to us, $M^\alpha_p$ acts trivially, so we will omit its concrete lattice implementation.

\subsection{Constant-Depth $T^{1/N}$-gates from $D(D_{4N})$ Clifford-Hierarchy Stabilizer Codes}

\begin{figure}
\begin{tikzpicture}
\begin{scope}[shift={(0,0)}, scale=0.9]
\draw [thick, fill= \BlueColor, opacity=0.5]  
(0,0) -- (4,0) -- (2,2.5)  --(0,0); 
 \draw [very thick] (0,0) -- (4,0) -- (2,2.5)  --(0,0); 
\draw[very thick, red, ->-] (1, 1.25) -- (2, 0.8)  ; 
\draw[very thick, blue, ->-] (3, 1.25) -- (2, 0.8)   ; 
\draw[very thick, \thirdcolor, ->-] (2, 0.8)-- (2, 0)   ; 
\node[red] at (1.55, 1.3) {$1_{rs}$} ;
\node[blue] at (2.45, 1.3) {$1_s$} ;
\node[right, \thirdcolor] at (2, 0.35) {$1_{r}$} ;
\draw[fill=black] (0,0) ellipse (0.05 and 0.05);
\draw[fill=black] (4,0) ellipse (0.05 and 0.05);
\draw[fill=black] (2,2.5) ellipse (0.05 and 0.05);
\node[\thirdcolor, below] at (2,0) {$\cL_{{\langle r\rangle}}$};
\node[red, left] at (0.9,1.2) {$\cL_{\langle rs\rangle}$};
\node[blue, right] at (3.1,1.2) {$\cL_{\langle s\rangle}$};
\end{scope}
\begin{scope}[shift={(4.5,0)}, scale=0.9]
\draw [thick, fill= \BlueColor, opacity=0.5]  
(0,0) -- (4,0) -- (2,2.5)  --(0,0); 
 \draw [very thick] (0,0) -- (4,0) -- (2,2.5)  --(0,0); 
\draw[very thick, red, ->-] (1, 1.25) -- (2, 0.8)  ; 
\draw[very thick, blue, ->-] (3, 1.25) -- (2, 0.8)   ; 
\draw[very thick, \thirdcolor, ->-] (2, 0.8)-- (2, 0)   ; 
\node[red] at (1.55, 1.25) {$rs$} ;
\node[blue] at (2.45, 1.25) {$s$} ;
\node[right, \thirdcolor] at (2, 0.35) {$r$} ;
\draw[fill=black] (0,0) ellipse (0.05 and 0.05);
\draw[fill=black] (4,0) ellipse (0.05 and 0.05);
\draw[fill=black] (2,2.5) ellipse (0.05 and 0.05);
\node[\thirdcolor, below] at (2,0) {$\cL_{{\langle r\rangle}}$};
\node[red, left] at (0.9,1.2) {$\cL_{\langle rs\rangle}$};
\node[blue, right] at (3.1,1.2) {$\cL_{\langle s\rangle}$};
\end{scope}
\end{tikzpicture}
\caption{Electric and  magnetic anyon configurations, realizing the code space in $D(D_{4N})$. The LHS, which are the electric anyons, that are irreps, realizes the logical $\overline{Z}$, the RHS, which are the magnetic anyons, given by {conjugacy classes for which we show the representative} group element, braids non-trivially with it, and is a dressed version of the logical $\overline{X}$. \label{fig:EandMTriangles}}
\end{figure}

\begin{figure}
\begin{tikzpicture}
\begin{scope}[shift={(0,0)}, scale= 0.9]
\draw[step=0.5,black] (0,0) grid (3,3);
\draw[black, very thick] (0,3) -- (3,3);
\draw[black, very thick] (3,3) -- (3,0);
\draw[black, very thick] (0,3) -- (0,0);
\draw[black, very thick] (0,0) -- (3,0);
\node[red,  left] at (0,1.5) {$\cL_{\langle rs\rangle}$};
\node[blue,  above] at (1.5,3) {$\cL_{\langle s\rangle}$};
\node[\thirdcolor,  right] at (3,1.5) {$\cL_{\langle r\rangle}$};
\node[\thirdcolor,  below] at (1.5,0) {$\cL_{\langle r\rangle}$};
\draw[red, very thick] (0,1.5) -- (1.5,1.5);
\draw[blue, very thick] (1.5,3) -- (1.5,1.5);
\draw[\thirdcolor, very thick] (3,1.5) -- (1.5,1.5);
\node[blue, right] at (1.5,2.25) {$1_s$};
\node[red, above] at (0.75,1.5) {$1_{rs}$};
\node[\thirdcolor, above] at (2.25,1.5) {$1_{r}$};
\end{scope}
\begin{scope}[shift={(4.5,0)}, scale=0.9]
\draw[step=0.5,black] (0,0) grid (3,3);
\draw[black, very thick] (0,3) -- (3,3);
\draw[black, very thick] (3,3) -- (3,0);
\draw[black, very thick] (0,3) -- (0,0);
\draw[black, very thick] (0,0) -- (3,0);
\node[red,  left] at (0,1.5) {$\cL_{\langle rs\rangle}$};
\node[blue,  above] at (1.5,3) {$\cL_{\langle s\rangle}$};
\node[\thirdcolor,  right] at (3,1.5) {$\cL_{\langle r\rangle}$};
\node[\thirdcolor,  below] at (1.5,0) {$\cL_{\langle r\rangle}$};
\draw[red, very thick, dashed] (0,1.75) -- (1.75,1.75);
\draw[red, very thick] (0,1.5) -- (0,2);
\draw[red, very thick] (0.5,1.5) -- (0.5,2);
\draw[red, very thick] (1,1.5) -- (1,2);
\draw[red, very thick] (1.5,1.5) -- (1.5,2);
\draw[blue, very thick, dashed] (1.75,3) -- (1.75,1.75);
\draw[blue, very thick] (2,3) -- (1.5,3);
\draw[blue, very thick] (2,2.5) -- (1.5,2.5);
\draw[blue, very thick] (2,2) -- (1.5,2);
\draw[\thirdcolor, very thick] (2,1.5) -- (1.5,1.5);
\draw[\thirdcolor, very thick] (2,1) -- (1.5,1);
\draw[\thirdcolor, very thick] (2,0.5) -- (1.5,0.5);
\draw[\thirdcolor, very thick] (2,0) -- (1.5,0);
\draw[\thirdcolor, very thick, dashed] (1.75,0) -- (1.75,1.75);
\node[red, above] at (0.75,1.95) {$rs$};
\node[blue, right] at (2,2.25) {$s$};
\node[\thirdcolor, right] at (1.95,0.75) {$r$};
\end{scope}
\end{tikzpicture}
\caption{Lattice realization of logical states from electric anyons  (left) and magnetic anyons  (right). The $D(G)$ surface code is put on a square lattice with BCs $\cL_{\langle g\rangle}$, which in the continuum is equivalent to the triangular configuration. 
\label{fig:LatticeAnyons_D4N}}
\end{figure}

We now move to the generalization to
\be
T^{1/N}=P(\pi/(4N))= \diag(1,e^{i\pi/(4N)}) \,,
\ee
that we realize with a constant-depth circuit in a 2D surface code, whose logical operators are shown in Fig.~\ref{fig:EandMTriangles}.
Note that in \cite{Huang:2025ump} {these gates were constructed from hybrid lattice surgery}, however not as a constant-depth gate within a single code patch.

\subsubsection{Clifford-Hierarchy Stabilizer Code for $D(D_{4N})$}
\label{sec:CliffHierStab}

Consider the dihedral group $D_{4N}$ (the symmetry group of the $4N$-gon) 
\be
    D_{4N}=\langle r,s \,|\, r^{4N}=s^2=\id\,,\; srs=r^{-1}\rangle \,.
\ee
Its irreducible representations (irreps) are:
\begin{itemize}
    \item 4 irreps of dimension 1, generated by $1_r$ and $1_s$, where: 
    \be
    \ba
        1_r(r)&=+1\,, & 1_r(s)&=-1\,,\\
        1_s(r)&=-1\,, & 1_s(s)&=+1\,,\\
    \ea
    \ee
    \item $2N-1$ irreps of dimension 2, which we label as $E_\ell$ for $\ell=1,\cdots ,2N-1$, where the character of $r$ in these irreps evaluates to
    \be
        \chi_{E_\ell}(r)=(e^{i\pi\ell/(2N)}+e^{-i\pi\ell/(2N)})\,,
    \ee
    and $\chi_{E_\ell}(s)=\chi_{E_\ell}(rs)=0\;\;\forall\,\ell$.
\end{itemize}

On the lattice, the local Hilbert space on each edge is spanned by the group elements:
\be
    \ket{r^as^j}\mapsto\ket{a}\ket{j}\,,\quad  a\in\{0,1,\cdots ,4N-1\}\,,\; j\in\{0,1\}\,,
\ee
which are comprised of a $\Z_{4N}$-qudit and a qubit. 
We will denote by $\cX,\cZ$ the generalized Pauli operators for the $\Z_{4N}$ qudit and by $X,Z$ the standard Paulis for the qubit. Furthermore, let $
\cC$ be the charge-conjugation operator on the $\Z_{4N}$ qudit, satisfying for $a\in\{0,1,\cdots, 4N-1\}$:
\begin{align}
    \cC\ket{r^a}&=\ket{r^{4N-a}}\,,\;\;\cX\ket{r^a}=\ket{r^{a+1}}\,,\;\; \cZ\ket{r^a}=e^{\frac{i\pi a}{2N}}\ket{r^a}\,,\nn\\
\cC \cX \cC^\dagger&=\cX^\dagger\,,\quad\;\;\; \cC \cZ \cC^\dagger=\cZ^\dagger\,.
\end{align}
The left and right multiplication operators for $D_{4N}$ are generated by \cite{Albert:2021vts}:
\be\ba \label{eq:D4N_LR}
    L^r&=\cX\otimes\bbI\,,&\;\; R^r&=\cX^{-Z}\,,\\
    L^s&=\cC\otimes X\,,&\;\; R^s&=\bbI\otimes X\,.\\
\ea\ee
\prx{where for $R^r$ we use the notation
\be
\cX^{-Z}:=\cX^{-1}\otimes\ket{0}\bra{0}+\cX\otimes\ket{1}\bra{1}\,,
\ee
}
while the diagonal irrep operators are realized as \cite{Albert:2021vts}:
\be\ba \label{eq:D4N_rep_ops}
    Z_{1_r}&=\bbI\otimes Z\,, \quad Z_{1_s}=\cZ^{2N}\otimes \bbI\,, \quad Z_{1_{rs}}=\cZ^{2N}\otimes Z\,,\\
    Z_{E_\ell}&=
    \begin{pmatrix}
        \cZ^\ell\otimes \ket{0}\bra{0} & \cZ^\ell\otimes \ket{1}\bra{1} \\
        \cZ^{\ell\dagger}\otimes \ket{1}\bra{1} & \cZ^{\ell\dagger}\otimes \ket{0}\bra{0} \\
    \end{pmatrix}\,.
\ea\ee
We can write a stabilizer group for $D(D_{4N})$ surface code as follows: 
\be\label{StabGroupsD4N}
\cS_{D(D_{4N})} = \left\langle A_v^r \,,\, A_v^s\,,\, S_p^r \,,\, S_p^s  \right\rangle \,,
\ee
realized in terms of the $\Z_{4N}$ qudit operators $\cX, \cZ,\cC$, the qubit operators $X, Z$ and the phase 
\be 
\zeta_{4N}:=e^{2\pi i/4N}=e^{i\pi/2N}\,,
\ee
as derived in App. \ref{app:StabilizerComms}
\be \label{eq:stabilizer_ops}
\ba
A_v^{r} &= 
\begin{tikzpicture}[baseline]
\begin{scope}[shift={(0,0.1)}]
\draw[thick, ->-] (-1,0) to (0,0);
\draw[thick, ->-] (0,-1) to (0,0);
\draw[thick, ->-] (0,0) to (0,1);
\draw[thick, ->-] (0,0) to (1,0);
\draw[fill=black] (0,0) ellipse (0.05 and 0.05);
\node[] at (0.15, 0.15) {$_v$};
\node[above] at (-0.6,0.05) {$\mathcal{X}^{-Z}$};
\node[right] at (0.04,-0.7) {$\mathcal{X}^{-Z}$};
\node[right] at (0.04,0.8) {$\mathcal{X}$};
\node[above] at (0.75,0.05) {$\mathcal{X}$};
\draw[fill=black] (0,0) ellipse (0.05 and 0.05);
\end{scope}
\end{tikzpicture}\,, \quad
A_v^{s} = 
\begin{tikzpicture}[baseline]
\begin{scope}[shift={(0,0.1)}]
\draw[thick, ->-] (-1,0) to (0,0);
\draw[thick, ->-] (0,-1) to (0,0);
\draw[thick, ->-] (0,0) to (0,1);
\draw[thick, ->-] (0,0) to (1,0);
\draw[fill=black] (0,0) ellipse (0.05 and 0.05);
\node[] at (0.15, 0.15) {$_v$};
\node[above] at (-0.7,0.05) {$X$};
\node[right] at (0.04,-0.7) {$X$};
\node[right] at (0.04,0.8) {$\mathcal{C}X$};
\node[above] at (0.75,0.05) {$\mathcal{C}X$};
\draw[fill=black] (0,0) ellipse (0.05 and 0.05);
\end{scope}
\end{tikzpicture} \,,\cr
S_p^r &=
\begin{tikzpicture}[baseline]
\begin{scope}[shift={(0,-0.4)}]
    \draw[thick, ->-] (-1,0) to (-1,1);
    \node at (-1.4,0.5) {$\mathcal{Z}_{1}$};
    \draw[thick, ->-] (-1,1) to (0,1);
    \node at (-0.5,1.4) {$\mathcal{Z}_{2}^{Z_1}$};
    \draw[thick, ->-] (0,0) to (0,1);
    \node[right] at (0.2,0.5) {$\mathcal{Z}_{3}^{-Z_1Z_2Z_3}$};
    \draw[thick, ->-] (-1,0) to (0,0);
    \node at (0.2,-0.35) {$\mathcal{Z}^{-Z_1Z_2Z_3Z_4}_{4}$};
    \node at (-0.5,0.5) {$_p$};
\end{scope}     
\end{tikzpicture} \cr 
S_p^s&=\zeta_{4N}^{(\bbI-Z_1Z_2Z_3Z_4)/2}\cr
&=\tfrac{1}{2}(\bbI+Z_1Z_2Z_3Z_4)+\tfrac{1}{2}(\bbI-Z_1Z_2Z_3Z_4)\zeta_{4N}\cr
(S_p^s)^{2N} &=
\begin{tikzpicture}[baseline]
\begin{scope}[shift={(0,-0.4)}]
    \draw[thick, ->-] (-1,0) to (-1,1);
    \node[left] at (-1,0.5) {$Z_1$};
    \draw[thick, ->-] (-1,1) to (0,1);
    \node at (-0.5,1.35) {$Z_2$};
    \draw[thick, ->-] (0,0) to (0,1);
    \node[right] at (0.1,0.5) {$Z_3$};
    \draw[thick, ->-] (-1,0) to (0,0);
    \node at (-0.5,-0.3) {$Z_4$};
     \node at (-0.5,0.5) {$_p$};
\end{scope}    
\end{tikzpicture} \,. 
\ea\ee
The stabilizer group is non-abelian: its non-trivial commutators as defined in (\ref{commutdef}) are given by 
\be \label{D4NStabComms}
\ba
    \left[ A_{v}^{r}, A_{v}^{s} \right] 
    &= (A_{v}^{r})^{-2} \cr 
    \left[ A_{v}^{r}, S_{p_{NE}}^r \right] 
    &= (S_{p_{NE}}^s)^{-2} \cr     
    \left[ A_{v}^{s}, S_{p_{NE}}^r \right] 
    &= (S_{p_{NE}}^r)^2\,.
\ea\ee
Here $p_{NE}$ is the plaquette in the North-East direction with respect to the vertex $v$. We prove these in appendix \ref{app:StabilizerComms}. The code space (ground states of \eqref{eq:Hamiltonian}) have $+1$ eigenvalue under the stabilizers, while the local errors (elementary excitations) have distinct non-trivial eigenvalues.

\subsubsection{Constant-Depth $T^{1/N}$-Gate}
Consider the following gapped boundaries $\fB_i,\,i\in\{1,2,3\}$ of $D(D_{4N})$ defined by the Lagrangian algebras $\cL_{{\langle g \rangle}}$ 
\be\ba\label{D4NLags}
\cL_{\langle rs \rangle}&=1\oplus 1_{rs} \oplus_{\ell=1}^{2N-1} E_\ell \oplus [rs]_{++}\oplus [rs]_{+-}\cr
\cL_{\langle s \rangle}&=1\oplus 1_s \oplus_{\ell=1}^{2N-1} E_\ell \oplus [s]_{++}\oplus [s]_{+-}\cr
\cL_{\langle r \rangle}&=1\oplus 1_r \oplus_{a=1}^{2N-1} 2[r^a]\oplus [r^{2N}]\oplus [r^{2N}]1_r\,,
\ea\ee
which are labeled by the subgroups 
\be\ba
\langle rs \rangle&=\{\id,rs\}\cong\Z_2\cr
\langle s \rangle&=\{\id,s\}\cong\Z_2\cr
\langle r \rangle&=\{\id,r,r^2,\cdots, r^{4N-1}\}\cong\Z_{4N}\,.
\ea\ee

We now consider the automorphism of $D(D_{4N})$ obtained from stacking with the non-trivial SPT
\be 
\alpha_N\in  H^2(D_{4N},U(1))=\Z_2\,.
\ee
We explain in App.~\ref{app:2coc} how we compute $\alpha_N$. 
The automorphism acts on the anyons analogously to (\ref{Uaction}): the pure irreps (electric anyons) do not map, while the anyons with non-trivial group elements and centralizer $D_{4N}$ and $\Z_2^2$, i.e. $[r^{2N}]$, $[s]$ and $[rs]$ get permuted as in \eqref{Uaction}. This leaves the all the \prx{Lagrangian algebras} (\ref{D4NLags}) invariant. 


In order for the surface specified by $\alpha_N$ to end on the boundary labeled by a subgroup $K_i$, we need 
\be \label{eq:alpha_beta}
    \alpha_N(g_1,g_2)\big|_{K_i}=\frac{\beta_{N}^{(i)}(g_1)\beta_{N}^{(i)}(g_2)}{\beta_{N}^{(i)}(g_1g_2)}\,,\quad \forall \;g_1,g_2 \in K_i\,,
\ee
and we will always fix $\beta_N^{(i)}(\id)=1$. 
We will compute $\beta^{(i)}_N$ for the subgroups labeling the boundary conditions: 
\begin{itemize}
    \item For $K={\langle s\rangle}$ or $K={\langle rs\rangle}$,  $\alpha|_K=1$ so $\beta$ on these subgroups can also be taken to be identically 1. Note also that $\alpha_N(rs,s)=1$.
    \item For $K=\langle r\rangle$, a solution to the full set of equations 
    \be
    \alpha_N|_{\langle r\rangle}=\delta\beta_N
    \ee
    can be recursively computed to be:
    \be \label{eq:beta_r_2N}
    \beta_N(r^a)=
    \begin{cases}
        +e^{i\pi a/(4N)} & \text{if } a<2N \\
        +1 & \text{if } a = 2N \\
        -e^{i\pi a/(4N)} & \text{if } a>2N\,. \\
    \end{cases}
    \ee
   Note that $\beta_N(r)=e^{i\pi /(4N)}$ is the non-trivial phase in the qubit gate
    \be
    T^{1/N}=\diag(1,e^{i\pi/(4N)})\,.
    \ee
\end{itemize}

\noindent{\bf Implementation of $T^{1/N}$-gate.} 
We again initialize the system in the eigenstates of $\ol{Z}$, i.e. $\ol{\ket{m}}$ for $m=0,1$. The state $\ol{\ket{0}}$ comes from an anyon triangle with 
\be
    g_1=g_2=\id \quad\Rightarrow\quad U(\id,\id)=1\,,
\ee 
while for the non-trivial state $\ol{\ket{1}}$ we choose 
\be
g_1=rs,\;g_2=s \quad\Rightarrow\quad U_N(rs,s)=\beta_N(r)=e^{i\pi/(4N)}\,.
\ee 
The automorphism $U_{N}$ therefore acts as 
\be\label{UnDef}
\ba
U_{N} \ol{\ket{m}} &= e^{i \pi m /(4N)}  \ol{\ket{m}} = T^{1/N} \ol{\ket{m}}\,,
\ea\ee
and thus encodes the diagonal  
constant-depth gate $T^{1/N}=P(\pi/(4N))$ acting on the logical qubit.

\vspace{1mm}
\noindent{\bf Lattice Model Realization.}
The unitary $U_{\alpha,\beta}$ is realized as the operator in \eqref{eq:U_alpha_lattice}. 
Similarly to the logical $T$-gate implementation of Sec.~\ref{sec:T}, the only non-identity boundary operator is the one associated to the 1-cochain $\beta_N$ \eqref{eq:beta_r_2N} on the $\langle r\rangle$ boundary $\fB_3$:
\be  
    \ba
M^\beta_e= &\diag\left(1,e^{i\pi/(4N)},\cdots,1,\cdots,e^{-i\pi/(4N)}\right)\otimes\bbI \,.
\ea 
\ee
For $8N=2^n$ this gate can be realized in terms of qubits as we will discuss in the next section.  Similarly to the $D_4$ case, since $\alpha_N(\id,\id)=\alpha_N(rs,s)=1$, on the states of interest to us $M^\alpha_p$ acts trivially, so we will omit its concrete lattice implementation.

\section{Qubit Realization}
\label{sec:Qubits}
We now turn to proving Theorem \ref{thm:qubit-only}.
The topological orders $D(D_{4N})$ at first seem to require an increased complexity in terms of realization -- despite the fact that the architecture remains in 2D, the group is non-Abelian. 
For generic $4N$ this is indeed true and would require qudits for all prime factors of $8N$. For general $N>1$ we can map the local Hilbert basis of $D_{4N}=\Z_{4N}\rtimes\Z_2$ to a $\Z_N$-qudit and three-qubits, as follows:
\be \label{eq:ZN_qudit}
    \ket{r^as^j}\mapsto\ket{\lfloor a/4\rfloor \text{\;mod\;} N}\ket{\lfloor a/2\rfloor \text{\;mod\;} 2}\ket{a \text{\;mod\;} 2}\ket{j}\,.
\ee

However, we will now discuss how to realize the case of 
\be 
8N=2^n, \quad n \in \mathbb{N}_{\geq 3} 
\ee
by means of $n$ physical qubits on each lattice edge. Note that this is precisely the subset of gates that realize the higher Clifford hierarchy gates (the other values of $N$ give rise to unitaries beyond the Clifford hierarchy).

\vspace*{1mm}
\noindent{\bf Properties of Dihedral Groups.}
Dihedral groups are supersolvable, i.e. they have a normal series of finite length, starting from the trivial group and ending at the whole group $G$, such that all the successive quotients are cyclic:
\be
1\leq H_1 \leq H_2 \leq \cdots \leq H_N \leq G\,,
\ee
where each $H_i\leq G$ and $H_{i+1}/H_i$ is cyclic.

For $D_{2^{n-1}}$ the above series is:
\be
1\leq \Z_2 \leq \Z_{4} \leq \cdots \leq \Z_{2^{n-1}} \leq D_{2^{n-1}}
\ee
with $H_{i+1}/H_i=\Z_2$ for all $i$. In terms of generators, the series is written as:
\be
1\leq \langle r^{2^{n-2}} \rangle \leq \langle r^{2^{n-3}} \rangle \leq \cdots \leq \langle r \rangle \leq \langle r,s \rangle\,.
\ee
We can thus express $D_{2^{n-1}}$ as the supersolvable group with generators
\be \label{eq:D2k+2}
 D_{2^{n-1}} = \langle r^{2^{n-2}}, r^{2^{n-3}}, \cdots, r, s \rangle \,, 
\ee 
and relations 
\be\ba
(r^{2^{n-2}})^2&=\id\,,\  (r^{2^{n-3}})^2=r^{2^{n-2}}\,,\cdots, (r)^2 = r^2\,, \ (s)^2 = \id \\
[r^a, s] &:= r^{-a} s^{-1} r^a s = r^{-2a}\,,\  a={2^{n-2}},{2^{n-3}},\cdots,2,1\,.
\ea\ee

\noindent{\bf Map to Qubit Hilbert space.}
We can now map the local Hilbert space 
\be
   \cH=\{\ket{g}\,,\;g\in D_{2^{n-1}}\} 
\ee
to one spanned by $n$ qubits on a Hilbert space 
\be
\cH= (\mathbb{C}^2)^{n}\,,
\ee 
in which each factor is associated to a generator in \eqref{eq:D2k+2}. The first $n-1$ qubits are those for the normal $\Z_{2^{n-1}}$ subgroup and correspond to $r^{2^{n-2}},r^{2^{n-3}},\cdots ,r$ while the last qubit is for the non-normal $\Z_2$ generated by $s$. The map is therefore: 
\be \label{eq:gs_to_bits}
    \ket{r^as^j}\mapsto \ket{\bin(a)}\ket{j}\,,
\ee
where $\bin(a)$ denotes the binary representation of $a\in\{0,1,2,..,2^{n-1}-1\}$ and $j\in\{0,1\}$.
The single-qubit Pauli operators (which act on one entry in this tensor product) will be denoted by $X$ and $Z$.

\vspace*{1mm}
\noindent\textbf{$\bm{D_4}$ operators as 3-qubit Clifford gates.}
We start with $D_4$ and map the group elements \eqref{eq:D4_elements} to a basis of 3-qubits (2-qubits for the normal $\Z_4$ and one for the non-normal $\Z_2$), corresponding to the supersolvable generators $r^2,r,s$ according to \eqref{eq:gs_to_bits}.\footnote{The decomposition of a $\Z_4$ qudit as two qubits has previously been discussed in \cite{Moussa2016fusion}. See also \cite{Warman:2024lir} for an alternative mapping of the $D_4$ operators onto a 3-qubit Hilbert space.} Explicitly:
\be\ba
    \ket{\id}&\mapsto\ket{00}\ket{0}\,,&\ket{r}&\mapsto\ket{01}\ket{0}\,,\\
    \ket{r^2}&\mapsto\ket{10}\ket{0}\,,&\ket{r^3}&\mapsto\ket{11}\ket{0}\,,\\
    \ket{s}&\mapsto\ket{00}\ket{1}\,,&\ket{rs}&\mapsto\ket{01}\ket{1}\,,\\
    \ket{r^2s}&\mapsto\ket{10}\ket{1}\,,&\ket{r^3s}&\mapsto\ket{11}\ket{1}\,.
\ea\ee
To realize the $D_4$ multiplication and representation operators \eqref{eq:D4_LR_ops}-\eqref{eq:D4_rep_ops} on this Hilbert space, we need to describe the $\Z_4$ operators $\cC,\cX,\cZ$, acting on the $\Z_4$ group elements as
\be 
    \cC\ket{r^a}=\ket{r^{4-a}}\,,\;\;\cX\ket{r^a}=\ket{r^{a+1}}\,,\;\; \cZ\ket{r^a}=i^a\ket{r^a}\,,
\ee
in terms of two-qubit Clifford gates.

We denote by $CX$ the 2-qubit Clifford control-$X$ gate in which the first qubit is the target and the second the control:
\be \label{eq:CX_def}
    CX:=\bbI\otimes \ket{0}\bra{0}+X\otimes \ket{1}\bra{1}\,,
\ee
which realizes the $\Z_4$ operator 
\be
    \cC\mapsto CX\,.
\ee
The $\Z_4$ operator $\cX$ can then be implemented on the 2-qubit Hilbert space as:\footnote{As explicit matrices in the basis $\{\ket{00},\ket{01},\ket{10},\ket{11}\}$, we have $$\cX=
    \begin{pmatrix}
    0 & 0 & 0 & 1 \\
    1 & 0 & 0 & 0 \\
    0 & 1 & 0 & 0 \\
    0 & 0 & 1 & 0
    \end{pmatrix}=
    \begin{pmatrix}
    0 & 1 & 0 & 0 \\
    1 & 0 & 0 & 0 \\
    0 & 0 & 0 & 1 \\
    0 & 0 & 1 & 0
    \end{pmatrix}
    \begin{pmatrix}
    1 & 0 & 0 & 0 \\
    0 & 0 & 0 & 1 \\
    0 & 0 & 1 & 0 \\
    0 & 1 & 0 & 0
    \end{pmatrix}=(\bbI\otimes X)CX\,.$$}
\be \label{eq:D4_cX}
    \cX
\mapsto (\bbI\otimes X)\,CX\,.
\ee
Indeed, in each step of binary addition, we flip the right-most-digit (with $\bbI\otimes X$), and, if its initial value was 1, we also have to flip the digit to its left (i.e. apply $CX$).\\
The $\Z_4$ operator $\cZ=\diag(1,i,-1,-i)$ can be written using the qubit Pauli $Z$ and Clifford gate $S:=\diag(1,i)$ as:
\be
    \cZ\mapsto Z\otimes S\,.
\ee
With these maps, the $D_4$ operators \eqref{eq:D4_LR_ops} can then be decomposed into 3-qubits gates as follows:
\begin{align} \label{eq:D4_ops_qubits}
    L^r&= [(\bbI\otimes X)\,CX] \otimes\bbI\,,\nn\\
    R^r&=[(\bbI\otimes X)\,CX]^\dagger\!\otimes \ket{0}\bra{0}+ [(\bbI\otimes X)\,CX]\otimes \ket{1}\bra{1}\,,\nn\\
    L^s&=CX\otimes X\cr 
    R^s&=\bbI\otimes \bbI\otimes X\,,
\end{align}
and are all in the 3-qubit Clifford group: 
to see this, note that $CX$ (also known as $C$NOT) is Clifford and Clifford operators form a group \cite{Gottesman:1997qd}. It is therefore clear that $L^r,L^s,R^s$ are Clifford. Let us prove that $R^r$ is also Clifford. Recall from \eqref{eq:D4_cX}, that $(\bbI\otimes X)\,CX$ is order-4, and note that
\begin{equation}
        (\bbI\otimes X)\,CX=[(\bbI\otimes X)\,CX]^\dagger(X\otimes\bbI)\,.
\end{equation}
$R^r$ in \eqref{eq:D4_ops_qubits} can therefore be written as
\be 
\ba
    R^r&
    &=\{[(\bbI\otimes X)\,CX]^\dagger\otimes\bbI\}\,CX_{1,3}\,,
\ea
\ee
which is a product of Clifford operators and thus Clifford.

The irrep operators \eqref{eq:D4_rep_ops} can be expressed as
\begin{align}
    Z_{1_r}&\!=\!\bbI\!\otimes\!\bbI\!\otimes\!Z\,, \;\;\; Z_{1_s}\!=\!\bbI\!\otimes\! Z\!\otimes\bbI\,,\;\;\; Z_{1_{rs}}\!=\!\bbI\!\otimes\! Z\!\otimes\!Z\,,  \nn\\
    Z_E&\!=\!
    \begin{pmatrix}
        Z\otimes S \otimes \ket{0}\bra{0} & Z\otimes S \otimes \ket{1}\bra{1} \\
        Z\otimes S^\dagger\!\otimes \ket{1}\bra{1} & Z\otimes S^\dagger\!\otimes \ket{0}\bra{0} \\
    \end{pmatrix}\,.
\end{align}
$Z_{1_r},Z_{1_s},Z_{1_{rs}}$ and the unitary linear combinations 
\be\ba
   Z_E^{1,\pm}&:=Z\otimes S \otimes \ket{0}\bra{0} \;\pm\; Z\otimes S^\dagger \otimes \ket{1}\bra{1}\,,\\
   Z_E^{2,\pm}&:=Z\otimes S^\dagger \otimes \ket{0}\bra{0} \pm Z\otimes S \otimes \ket{1}\bra{1}\,,
\ea\ee
are also 3-qubit Clifford operators: $S=\diag(1,i)$ and $CZ=\diag(1,1,1,-1)$ are Clifford \cite{Gottesman:1997qd,Cui:2016bxt}. Note that
\begin{align}
     Z_E^{1,+}
     &=\big[Z\otimes S\otimes\bbI\big]\big[\bbI\otimes CZ\big]\,,
\end{align} 
is a product of Cliffords and thus Clifford. Then, $Z_E^{1,-}=Z_E^{1,+}[\bbI\otimes\bbI\otimes Z]$ is also Clifford, as are $Z_E^{2,+}=(Z_E^{1,+})^\dagger$ and $Z_E^{2,-}=(Z_E^{1,-})^\dagger$.
As discussed in Sec.~\ref{sec:CliffStabD4}, 
$D(D_4)$ is thus a Clifford stabilizer code.\footnote{This observation agrees with \cite{Kobayashi:2025cfh}, in which the alternative presentation of $D(D_4)$ as $D^\omega(\Z_2^3)$ was used. \prx{Certain Clifford-hierarchy stabilizer codes have previously been discussed in \cite{Ni:2014clx,Webster:2022kdn}.}}

\vspace*{2mm}
\noindent\textbf{$\bm{D_{2^{n-1}}}$ operators on qubit Hilbert space.} 
Let us now generalize the above to $D_{2^{n-1}}$. The $\Z_{2^{n-1}}$ operators $\cC,\cX,\cZ$, act on the $\Z_{2^{n-1}}$ group elements as
\be \label{eq:ops_to_qubitize}
    \cC\ket{r^a}=\ket{r^{-a}},\;\cX\ket{r^a}=\ket{r^{a+1}},\;\cZ\ket{r^a}=e^{i \pi a/2^{n-2}}\ket{r^a}.
\ee
From the two-qubit controlled-$X$ gate \eqref{eq:CX_def}, for $\ell\geqslant0$, we can recursively define the $(\ell+2)$ qubit gate with the first qubit as target and the remaining $\ell+1$ as controls
\be \label{eq:ClX}
    C^{\ell+1} X:=\bbI^{\ell+1}\otimes \ket{0}\bra{0}+C^{\ell}X\otimes \ket{1}\bra{1}\,.
\ee
The operators \eqref{eq:ops_to_qubitize} can be realized in terms of $(n-1)$-qubit gates as follows: 
\be\ba \label{eq:cal_ops}
\cC & \mapsto  (\bbI^{\otimes^{n-2}}\!\!\otimes X)(\bbI^{\otimes^{n-3}}\!\!\otimes CX)\cdots\!(C^{n-2}X)(X^{\otimes^{n-1}})\!\! \cr 
\cX & \mapsto (\bbI^{\otimes^{n-2}}\!\!\otimes X)(\bbI^{\otimes^{n-3}}\!\!\otimes CX)\cdots\!(C^{n-2}X) 
\cr 
\cZ&\mapsto Z\otimes S\otimes\cdots\otimes T^{2^{4-n}} \,.
\ea
\ee
The expression for $\cX$ follows from the rules of binary addition, while $\cZ$ contains in each factor a phase gate with a root of unity whose order is that of the corresponding supersolvable generator $r^a$ for $a\in\{2^{n-2},2^{n-3},\cdots,2,1\}$.
These operators can then be replaced into the $D_{4N}=D_{2^{n-1}}$ expressions \eqref{eq:D4N_LR}-\eqref{eq:D4N_rep_ops} to obtain the realization for the latter on the $n$-qubit local Hilbert space. See also \cite{Bravyi:2022zcw} for adaptive constant-depth local circuit implementations of group multiplication operators for solvable groups.
We summarized the resources for each of the gates in Tab.~\ref{tab:Resources}.

\vspace{1mm}
\noindent{\bf Lattice Realization.}
We now provide the qubit realization of the constant-depth gate (\ref{eq:U_alpha_lattice}) in terms of standard universal gate sets. First consider 
the operator for \eqref{eq:beta_r_2N} acting on a single lattice edge on the $\fB_3$ boundary (the other two boundaries contribute trivially). It is given by:
\be\ba
    M^{\beta_N} \mapsto& \lb S\otimes T\otimes\cdots\otimes T^{2^{3-n}}\rb\otimes\bbI \\
   & \times\lb Z\otimes X^{\otimes^{n-2}})(C^{n-2}S)(\bbI\otimes X^{\otimes^{n-2}}\rb\otimes \bbI\,,
\ea\ee
where we denote by $C^{n-2}S$ the $n-1$ qubit controlled-$S$ gate with the last qubit as target: $C^{n-2}S=\diag(1,1,\cdots,1,i)$. The first line corresponds to $\beta'_N$ and the second to $\kappa_N$ (defined in App. \ref{app:2coc}) so the total expression is the realization of $\beta_N=\beta'_N\kappa_N$ on the $n$-qubit Hilbert space on each lattice edge. 

As discussed in the lattice section in (\ref{eq:g1g2_state_lattice}), the logical states are identified as:
\be\ba \label{eq:logical_qubits_state}
    \overline{\ket{0}}&=\prod_v A_v\; \bigotimes_l \ket{00\cdots00}_l\,,\cr 
    \overline{\ket{1}}&=\prod_v A_v\; L_{\xi_1}^{rs} L_{\xi_2}^{s} L_{\xi_3}^{r}\bigotimes_l \ket{00\cdots 00}_l \,.
\ea\ee
See Fig.~\ref{fig:LatticeAnyons_D4N} for the configuration of the non-trivial magnetic state on the lattice. In particular there is an edge on the $\langle r \rangle$ boundary with group element $\ket{r}\mapsto\ket{00\cdots001}$. Note that the actual ground states in \eqref{eq:logical_qubits_state} are the superposition of states related to this by the action of $A_v$. Each vertex term will multiply two adjacent boundary edges by $g$ and $g^{-1}$ respectively and since $\beta_N$ is such that $\beta_N(g^{-1})=\beta_N(g)^{-1}$ (as a consequence of requiring \eqref{eq:lemma5.1} for $\alpha_N$), after acting with $M^\beta$ all states in the superposition will carry the same phase.

Furthermore, note that the bulk action of $M^\alpha$ on the logical states \eqref{eq:logical_qubits_state} is trivial. 
Thus, evaluating $U_{\alpha,\beta}$ requires only to evaluate the boundary terms $M^\beta$  on the states $\ket{\id}\mapsto\ket{00\cdots 000}$ and $\ket{r}\mapsto\ket{00\cdots001}$ (at the edge where the bulk anyon terminates), where it acts as the $T^{2^{n-3}}=P\lb\frac{2\pi}{2^n}\rb$ gate. In summary we obtain a constant-depth topologically protected gate at level $n$ of the Clifford hierarchy, realized on a physical Hilbert space with $n$ qubits on each lattice edge and acting on the logical qubit as:
\be\ba
U_{\alpha,\beta} \overline{\ket{0}} &= \overline{\ket{0}} \cr 
U_{\alpha,\beta} \overline{\ket{1}}&= 
 e^{2\pi i/ 2^n} \overline{\ket{{1}}} \,.
\ea\ee

\begin{figure}
\centering
\begin{tikzpicture}
\begin{scope}[shift={(0,0)}]
\draw [thick, fill= yellow, opacity=0.5]  
(0,0) -- (4,0) -- (2.7,1.5)  --(0,0); 
 \node at (-1,-1) {$D(\Z_2\times \Z_2)$}; 
 \draw [very thick] (0,0) -- (4,0) -- (2.7,1.5)  --(0,0); 
\draw[very thick, red] (2.5, 0.5) --  (1.8, 1) ; 
\draw[very thick, blue] (2.5, 0.5) --  (3.3, 0.8) ; 
\draw[very thick, \thirdcolor] (2.5, 0.5) -- (2.5, 0)  ; 
\node at (2, 0.2) {$U_{\alpha, \beta} \times$} ;
\node[red, left] at (2.4, 0.5){$m_1 m_2$} ;
\node[blue, right] at (2.6, 0.9) {$m_2$} ;
\node[right, \thirdcolor] at (2.6, 0.3) {$m_1$} ;
\node[\thirdcolor, below] at (2.5,0) {$\cL'_3$};
\node[red] at (1.2,1) {$\cL'_1$};
\node[blue, right] at (3.3,1.0) {$\cL'_2$};
\draw[fill= yellow, opacity=0.4] (0,0) -- (0, -2)  -- (4,-2) -- (4,-0) -- (0,-0) ;
\draw[fill= \BlueColor, opacity=0.4] (0,-2) -- (0, -8)  -- (4,-8) -- (4,-2) -- (0,-2) ;
\begin{scope}[shift={(0,-2)}]
\draw [thick, fill= orange, opacity=0.3]  
(0,0) -- (4,0) -- (2.7,1.5)  --(0,0); 
 \draw [very thick] (0,0) -- (4,0) -- (2.7,1.5)  --(0,0); 
\node[right]  at (1.7,0.5) {$\cA$};
 \end{scope}
\begin{scope}[shift={(0,-5.5)}]
\draw [thick, fill= \GreenColor, opacity=0.5]  
(0,0) -- (4,0) -- (2.7,1.5)  --(0,0); 
 \draw [very thick] (0,0) -- (4,0) -- (2.7,1.5)  --(0,0); 
\node[right]  at (1.7,0.5) {$U_{\alpha,\beta}$};
 \end{scope}
\end{scope}
\draw[thick] (0,0) -- (0, -10) ;
\draw[thick] (4,0) -- (4, -10) ;
\draw[thick, dashed] (2.7,1.5) -- (2.7, 1.5-10) ;
\begin{scope}[shift={(0,-8)}]
\draw [thick, fill= orange, opacity=0.3]  
(0,0) -- (4,0) -- (2.7,1.5)  --(0,0); 
 \draw [very thick] (0,0) -- (4,0) -- (2.7,1.5)  --(0,0); 
\node[right]  at (1.7,0.5) {$\cA$};
 \end{scope}
\begin{scope}[shift={(0,-4)}]
\draw [thick, fill= \BlueColor, opacity=0.3]  
(0,0) -- (4,0) -- (2.7,1.5)  --(0,0); 
 \draw [thick] (0,0) -- (4,0) -- (2.7,1.5)  --(0,0); 
 \draw[very thick, red] (2.5, 0.5) --  (1.8, 1) ; 
\draw[very thick, blue] (2.5, 0.5) --  (3.3, 0.8) ; 
\draw[very thick, \thirdcolor] (2.5, 0.5) -- (2.5, 0)  ; 
\node[red, left] at (2.55, 0.9){$rs$} ;
\node[blue, right] at (2.7, 0.9) {$s$} ;
\node[right, \thirdcolor] at (2.1, 0.3) {$r$} ;
\node at (1.7, 0.4) {$U_{\alpha, \beta} \times$} ;
\node[\thirdcolor, below] at (2.5,0) {$\cL_3$};
\node[red] at (1.2,1) {$\cL_1$};
\node[blue, right] at (3.3,1.0) {$\cL_2$};
 \end{scope}
\begin{scope}[shift={(0,-7)}]
\draw [thick, fill= \BlueColor, opacity=0.3]  
(0,0) -- (4,0) -- (2.7,1.5)  --(0,0); 
 \draw [thick] (0,0) -- (4,0) -- (2.7,1.5)  --(0,0); 
  \node at (-1,4) {$D(D_{4N})$}; 
  \node at (-1,0.5) {$D(D_{4N})$}; 
 \draw[very thick, red] (2.5, 0.5) --  (1.8, 1) ; 
\draw[very thick, blue] (2.5, 0.5) --  (3.3, 0.8) ; 
\draw[very thick, \thirdcolor] (2.5, 0.5) -- (2.5, 0)  ; 
\node[red, left] at (2.55, 0.9){$rs$} ;
\node[blue, right] at (2.7, 0.9) {$s$} ;
\node[right, \thirdcolor] at (2.1, 0.3) {$r$} ;
\node[\thirdcolor, below] at (2.5,0) {$\cL_3$};
\node[red] at (1.2,1) {$\cL_1$};
\node[blue, right] at (3.3,1.0) {$\cL_2$};
 \end{scope}
\begin{scope}[shift={(0,-10)}]
\draw [thick, fill= yellow, opacity=0.5]  
(0,0) -- (4,0) -- (2.7,1.5)  --(0,0); 
\draw[fill= yellow, opacity=0.4] (0,0) -- (0, 2)  -- (4,2) -- (4,-0) -- (0,-0) ;
 \draw [very thick] (0,0) -- (4,0) -- (2.7,1.5)  --(0,0); 
\draw[very thick, red] (2.5, 0.5) --  (1.8, 1) ; 
\draw[very thick, blue] (2.5, 0.5) --  (3.3, 0.8) ; 
\draw[very thick, \thirdcolor] (2.5, 0.5) -- (2.5, 0)  ; 
\node[red, left] at (2.4, 0.5){$m_1 m_2$} ;
\node[blue, right] at (2.6, 0.9) {$m_2$} ;
\node[right, \thirdcolor] at (2.6, 0.3) {$m_1$} ;
\node[\thirdcolor, below] at (2.5,0) {$\cL'_3$};
\node[red] at (1.2,1) {$\cL'_1$};
\node[blue, right] at (3.3,1.0) {$\cL'_2$};
 \node at (-1,1) {$D(\Z_2\times \Z_2)$};
 \draw[thick, ->-] (-1,7.5) -- (-1, 8.5);
  \node[left] at (-1.1, 8) {$\cA$};
  \draw[thick, ->-] (-1,4) -- (-1, 6.5);
  \node[left] at (-1.2, 5) {$U_{\alpha, \beta}$};
 \draw[thick, ->-] (-1,1.2) -- (-1, 3);
  \node[left] at (-1.1, 2.2) {$\cA$};
\end{scope}
\end{tikzpicture}
\caption{Code-switching from $D(G'=\Z_2\times\Z_2)$ to $D(G=D_{4N})$ and back to $D(G'=\Z_2\times\Z_2)$: 
The constant-depth gate $U_{\alpha, \beta}$ is performed within the non-abelian patch $D(D_{4N})$. The code switching happens through an interface $\cA$, where in the topological description we condense anyons. In the concrete example $G=D_{4N}$ and $G'=\Z_2 \times \Z_2$, the anyon triangle that realizes the logical qubit is mapped to the triangle in the $\Z_2\times \Z_2$ patch that realizes the logicals there. The boundary conditions $\cL_i$ of $D(D_{4N})$ translate into those of $D(\Z_2\times\Z_2)$ given by $\cL_i'$. 
\label{fig:Xmastree}}
\end{figure}

\section{Code Switching}
\label{sec:CodeSwitch}

\prx{To obtain a universal gate set, including the logical Clifford gates, we} present code-switching. For $D_4$ this was discussed in \cite{Davydova:2025ylx, Kobayashi:2025cfh}. We provide details on the general principles that allow us to determine such interfaces between two surface codes for groups $G$ and $G'$, and will work out the concrete example of  $G=D_{4N}$ and $G'=\Z_2 \times \Z_2$.

A general discussion of interfaces between quantum doubles $D^\omega(G)$ can be found in \cite{Naidu:2007,davydov2017lagrangian, Gai:2026hjk}. 
Here we will describe the special types of interfaces that we require in the current setup: $\omega$ is trivial and all the interfaces are simply specified by two subgroups: $K\subseteq G$ a subgroup of $G$ and $N\triangleleft K$ a normal subgroup of $K$. Such interfaces are obtained by condensing an algebra (in the Drinfeld center of $G$) that we will label by 
\be\label{CondAlg}
\cA (K, N) \,.
\ee
This provides an interface between the two quantum doubles
$D (G)$ and $D(K/N)$. Writing this in terms of anyons  that condense is done in general in \cite{Gai:2026hjk} (see also \cite{Beigi:2010htr,delaFuente:2023whm} for prior work). 
The algebra $\cA(K, N)$ can be understood as providing a map between the anyons of the two topological orders, which is equivalently a gapped boundary condition of the folded quantum double $D(G \times K/N)$. For $D_4$ all these algebras were determined in \cite{Bhardwaj:2024qrf}.

\subsection{Code-Switching from $D(D_{4N})$ to $D(\Z_2 \times \Z_2)$}

For QEC applications, it will be useful to study code-switching between the patch that realizes the non-Clifford gates $D(D_{4N})$ and abelian surface codes, in particular to $D(\Z_2 \times \Z_2)$. 

\noindent\textbf{TQFT description.} In this code switching, we want to retain the anyons that appear in the electric and magnetic triangles: in particular we will generate the $\Z_2\times \Z_2$ from $rs, s, r$ in the $D_{4N}$. For this, we need to trivialize the group $\Z_{2N}$ generated by $r^2$.
In terms of the quantum doubles, this means we consider an interface given by a condensable algebra (in the notation of (\ref{CondAlg}))
\be \label{eq:cond}
\cA= \cA(D_{4N}, \Z_{2N}) =\bigoplus_{a=0}^N [r^{2a}]=1\oplus [r^2]\oplus \cdots \oplus  [r^{2N}]\,.
\ee
This corresponds to condensing the anyons in $\Z_{2N}$, trivializing this subgroup: the reduced TO is therefore $D(\Z_2\times \Z_2)$ generated by the images of $rs$ and $s$.

We will label the simple anyons in the $D(\Z_2\times\Z_2)$ by $e_i$ $m_i$, $i=1, 2$, with $e_i^2=1=m_i^2$. Then, 
the condensation \eqref{eq:cond} induces an identification on the anyons of the non-abelian patch with these abelian anyons when passing through the interface\footnote{$D_4$ corresponds to $N=1$, in which case the sums evaluate to:
\begin{align*}
    \bigoplus_{a=0}^1 [r^{2a}] &=1\oplus [r^2]\,, & 
    \bigoplus_{a=1}^1 [r^{2a-1}] &=[r]\,.
\end{align*}}
\be  \label{eq:code_switch_map}
\ba
\bigoplus_{a=0}^N [r^{2a}]  &\sim 1  \,,&
\bigoplus_{a=0}^N [r^{2a}]\,1_s &\sim e_1  \,,\cr
\bigoplus_{a=0}^N [r^{2a}]\,1_r &\sim e_2  \,,&
\bigoplus_{a=0}^N [r^{2a}]\,1_{rs} &\sim e_1 e_2  \,,\cr
\bigoplus_{a=1}^{N} [r^{2a-1}] &\sim m_1   \,,&
\bigoplus_{a=1}^{N} [r^{2a-1}] &\sim m_1 e_2  \,,\cr
\bigoplus_{a=1}^{N} [r^{2a-1}]_{-1} &\sim m_1 e_1  \,,&
\bigoplus_{a=1}^{N} [r^{2a-1}]_{-1} &\sim m_1 e_1 e_2  \,, \cr 
[s]_{++} &\sim m_2   \,,&
[s]_{++} &\sim m_2 e_1  \,,\cr
[s]_{-+} &\sim m_2 e_2  \,,&
[s]_{-+} &\sim m_2 e_1 e_2  \,,\cr
[rs]_{++} &\sim m_1 m_2   \,,&
[rs]_{++} &\sim m_1 m_2 e_1 e_2  \,,\cr
[rs]_{-+} &\sim m_1 m_2 e_1  \,,&
[rs]_{-+} &\sim m_1 m_2 e_2  \,,
\ea
\ee
i.e. the anyons in the non-abelian patch that realize the logical qubit map in this way. Note that the anyons that do not appear in the above identifications confine. 
The gapped boundary conditions $\cL_i'$ on the triangle for the $D(\Z_2\times\Z_2)$ are obtained by mapping the boundary conditions $\cL_i$ in (\ref{D4NLags}) of the $D_{4N}$ patch following the above rules, resulting in
\be
\ba
\cL_1' & = (1\oplus m_1 m_2)(1 \oplus e_1 e_2) \cr 
\cL_2' &= (1\oplus m_2)(1 \oplus e_1) \cr 
\cL_3' &= 2(1\oplus m_1) (1 \oplus e_2) \,.
\ea
\ee
In particular in 
the doubled $\Z_2$ surface code we indeed have a logical qubit given by the magnetic triangle
\be
m_1m_2\,,\ m_2\,,\ m_1  \,,
\ee
shown in Fig.~\ref{fig:Z2Z2_LatticeAnyons} and Fig.~\ref{fig:Xmastree},
and the associated electric one with 
\be
e_1e_2 \,,\  e_1 \,, \ e_2 \,.
\ee
Thus, this interface precisely maps the logical operators of $D(D_{4N})$ into those of $D(\Z_2\times\Z_2)$.

\subsection{\prx{Code Switching on the Lattice}}

We now illustrate how to prepare the logical basis for the non-abelian $D(D_{4N})$ code on the lattice, via code-switching from $D(\Z_2\times\Z_2)$, whose setup we summarized in Sec.~\ref{sec:Z2Z2_setup}. We denote its vertex projectors by 
\be\ba \label{eq:AvZ2Z2}
A_v^{\Z_2^2}&=\frac{1}{4}(\bbI+A_v^{m_1})(\bbI+A_v^{m_2})\,, &
A_v^{m_i}= \, & \begin{tikzpicture}[baseline]
\begin{scope}[shift={(0,0)}]
\draw[thick, ->-] (-1,0) to (0,0);
\draw[thick, ->-] (0,0) to (1,0);
\draw[thick, ->-] (0,-1) to (0,0);
\draw[thick, ->-] (0,0) to (0,1);
\node[above] at (0.2, 0) {$_v$};
\node[above] at (-0.7,0.05) {$X_i$};
\node[above] at (0.7,0.05) {$X_i$};
\node[right] at (0.05,0.7) {$X_i$};
\node[right] at (0.05,-0.7) {$X_i$};
\draw[fill=black] (0,0) ellipse (0.05 and 0.05);
\end{scope}
\end{tikzpicture}
\ea\ee
with $i=1,2$ labeling the physical $X$ operators for the $\Z_2\times\Z_2$ generators.
The trivial $D(\Z_2\times\Z_2)$ logical state in the $\ol{Z}$-basis is obtained from the state with all zero qubits $\bigotimes_l \ket{00}_l$ (that automatically satisfies the plaquette\- stabilizers) by \ssn{measuring} the vertex projectors $\prod_v A^{\Z_2^2}_v$ \ssn{and post-selecting the $+1$ measurement outcomes}:
\be\ba \label{eq:0Z2Z2}
     \overline{\ket{0}}_{\Z_2^2}&=\prod_v A^{\Z_2^2}_v\; \bigotimes_l \ket{00}_l\,.
\ea\ee
The non-trivial logical qubit state in $D(\Z_2\times\Z_2)$ is $\overline{\ket{1}}_{\Z_2^2}=\ol{X\ket{0}}_{\Z_2^2}$, with the logical $\ol{X}$ operator in $D(\Z_2\times\Z_2)$ defined in \eqref{eq:Z2Z2_Pauli}, explicitly:
\be\ba \label{eq:1Z2Z2}
    \overline{\ket{1}}_{\Z_2^2}&=\prod_v A^{\Z_2^2}_v\; \prod_{l\in\xi_1}(X_1X_2)_l \prod_{j\in\xi_2}(X_2)_j \prod_{k\in\xi_3}(X_1)_k\bigotimes_l \ket{00}_l \,.
\ea\ee
Note that, in the TQFT description, these two logical states differ by non-contractible configurations of anyons $(m_1m_2,m_1,m_2)$ ending on the three $D(\Z_2\times\Z_2)$ boundaries of Sec.~\ref{sec:Z2Z2_setup} and all of these anyons are mapped to non-trivial anyons $([rs],[s],[r])$ ending on the three $D(D_{4N})$ boundaries \eqref{D4NLags} by the code-switching map of anyons \eqref{eq:code_switch_map}: this implies that the two states will remain inequivalent in $D(D_{4N})$.

To code-switch to $D(D_{4N})$ on the lattice, we couple the system to a $\Z_N$ qudit and a qubit both in their $\ket{0}$ state, using the identification \eqref{eq:ZN_qudit} or, if $2N=2^{n-2}$, to $n-2$ qubits all in $\ket{0}$, using \eqref{eq:gs_to_bits}. This map ensures that the physical qubits of the $\Z_2\times\Z_2$ are the last two qubits on each edge of the lattice. 

To obtain the $D(D_{4N})$ logical states from the $D(\Z_2\times\Z_2)$ basis $\overline{\ket{m}}_{\Z_2^2}$ states with $m=0,1$, we need to \ssn{measure} the $D(D_{4N})$ flux and vertex projectors \ssn{and post-select} to ensure that the states are $+1$ eigenstates of the $D(D_{4N})$ stabilizers \eqref{eq:stabilizer_ops}.

Let us start with the flux operators. Noting that $(S_p^s)^{2N}$ is the flux stabilizer for the last qubit, it already has eigenvalue $+1$ in the $\overline{\ket{m}}_{\Z_2^2}$ states: this means that the plaquettes already have trivial $s$-flux since $s\in D_{4N}$ is identified with $m_2\in\Z_2\times\Z_2$ (i.e. the last qubit). This also implies that $S_p^s$ also has eigenvalue $+1$ (since $(S_p^s)^{2N}$ and $S_p^s$ only differ in the phase they produce for non-stabilized states). We instead need to \ssn{measure} the diagonal projector
\be\ba
    B_p^{\Z_{4N}^{\phantom{i}}}=\frac{1}{4N}(1+S_p^r+\cdots+(S_p^r)^{4N-1})\,,
\ea\ee
\ssn{and post-select the $+1$ measurement outcome,}
to ensure that each plaquette has trivial $\Z_{4N}$ flux.

Mathematically, the code-switching involves a non-trivial group extension from the first $\Z_2$ to $\Z_{4N}$. The $\Z_2\times\Z_2$ stabilizers are therefore not sufficient to ensure trivial $\Z_{4N}$ flux, which is why one has to \ssn{measure} $B_p^{\Z_{4N}^{\phantom{i}}}$ \ssn{and post-select the $+1$ outcomes}.

For the vertex projectors $\prod_vA_v$, we note that each $D(D_{4N})$ $A_v$ projector
\be
    A_v=\frac{1}{8N}\sum_{g\in D_{4N}} A_v^g\,,
\ee
can be decomposed as
\be\ba
    A_v&=A_v^{\Z_2^s}A_v^{\Z_{4N}^r}\,,\quad \text{with}\\
    A_v^{\Z_2^s}&=\frac{1}{2}(\bbI+A_v^s)\,,\quad \text{and}\\
    A_v^{\Z_{4N}^r}&=\frac{1}{4N}(\bbI+A_v^r+A_v^{r^2}+\cdots+A_v^{r^{4N-1}})\,,
\ea\ee
so we can \ssn{measure} each $A_v$ operator in two steps: first the projector $A_v^{\Z_{4N}^r}$ with the vertex stabilizers for the $\Z_{4N}$ qudit, and then the projector $A_v^{\Z_2^s}$ for the last qubit. \ssn{In both cases, we post-select the $+1$ measurement outcome.} \ssn{For a detailed implementation for $D(D_4)$, see  \cite{Tantivasadakarn:2022ceu}, where the measurement of similar operators is done by first coupling the system to ancillas, then applying entangling gates and finally measuring out the ancillas.}
Since $D_{4N}=\Z_{4N}\rtimes\Z_2$, the $\Z_{4N}$ is a normal subgroup, so the $A_v^{\Z_2^s}$ projector commutes with the $A_v^{\Z_{4N}^r}$ projector and will not change its $(+1)$ eigenvalue. (This is despite the fact that the individual $A_v^g$ operators do not commute but obey $D_{4N}$ group multiplication \eqref{D4NStabComms}).

We will now show that the thus obtained states
\be\ba \label{eq:psi0_psi1}
     \ket{\psi_0}&:=\prod_v A_v^{\Z_2^s}A_v^{\Z_{4N}^r} \prod_p B_p^{\Z_{4N}^{\phantom{i}}}\bigotimes_l \ket{0\cdots0}_l\overline{\ket{0}}_{\Z_2^2}\,,\\
    \ket{\psi_1}&:=\prod_v A_v^{\Z_2^s}A_v^{\Z_{4N}^r} \prod_p B_p^{\Z_{4N}^{\phantom{i}}}\bigotimes_l \ket{0\cdots0}_l\overline{\ket{1}}_{\Z_2^2}
\ea\ee
are in fact the $D(D_{4N})$ logical states \eqref{eq:logical_qubits_state}
\be\ba \label{eq:D4N_logical_states}
    \overline{\ket{0}}&=\prod_v A_v\; \bigotimes_l \ket{00\cdots00}_l\,,\cr 
    \overline{\ket{1}}&=\prod_v A_v\; L_{\xi_1}^{rs} L_{\xi_2}^{s} L_{\xi_3}^{r}\bigotimes_l \ket{00\cdots 00}_l \,.
\ea\ee
Indeed, by \ssn{measuring} the projectors \ssn{and post-selecting the $+1$ outcomes}, we have ensured that \eqref{eq:psi0_psi1} are $+1$ eigenstates of the $D(D_{4N})$ stabilizers \eqref{eq:stabilizer_ops}, which is equivalent to the states being ground states of the Hamiltonian \eqref{eq:Hamiltonian} (recall the definition of the stabilizers in App.~\ref{app:StabilizerComms}). Furthermore, $\ket{\psi_0}$ contains the state $\bigotimes_l\ket{00\cdots 00}$, since $\overline{\ket{0}}_{\Z_2^2}$ \eqref{eq:0Z2Z2} contains $\bigotimes_l\ket{00}_l$ for the last two sites and we have tensored with $\bigotimes_l\ket{0\cdots0}$ for the remaining sites before applying the projectors. Therefore $\ket{\psi_0}$ is the trivial logical state $\ol{\ket{0}}$ in \eqref{eq:D4N_logical_states}.

For $\ket{\psi_1}$, recall that the group elements $rs,s,r$ are supported only on the last two qubits on each edge, as in \eqref{eq:ZN_qudit} or \eqref{eq:gs_to_bits} for the case of $N$ a power of two. Explicitly
\be\ba \label{eq:last_qubits}
    \ket{rs}&\mapsto \ket{0\cdots 0}\ket{11}\,,\\
    \ket{s}&\mapsto \ket{0\cdots 0}\ket{01}\,,\\
    \ket{r}&\mapsto \ket{0\cdots 0}\ket{10}\,,
\ea\ee
where the last two qubits are identified with the $\Z_2\times\Z_2$ physical states. These $\Z_2\times\Z_2$ states are those appearing in \eqref{eq:1Z2Z2}. After tensoring with $\ket{0\cdots 0}$, we then have that $\ket{\psi_1}$ is the unique Hamiltonian ground state containing $L_{\xi_1}^{rs} L_{\xi_2}^{s} L_{\xi_3}^{r}\bigotimes_l \ket{00\cdots 00}_l$. Therefore the state that is acted upon by the vertex stabilizer in \eqref{eq:1Z2Z2} is precisely the non-trivial $D(D_{4N})$ logical state $\ol{\ket{1}}$ of \eqref{eq:D4N_logical_states}.

To code-switch from $D(D_{4N})$ back to $D(\Z_2\times\Z_2)$, we need to measure out the $\Z_{N}$ qudit and extra qubit, or the first $n-2$ qubits on each edge for $2N=2^{n-2}$, selecting all $\ket{0}$ outcomes, i.e. we apply $\bra{0\cdots0}\otimes \bbI\otimes \bbI$ which does not affect the last two qubits and projects the others to $\ket{0}$. The plaquette stabilizers for these two remaining qubits are already satisfied, since if $g_1g_2g_3^{-1}g_4^{-1}=\id$ in $D(D_{4N})$, then the group elements will fuse to identity also after taking the quotient by $\langle r^2\rangle=\Z_{2N}$. The $\Z_2\times\Z_2$ logical states are then obtained by \ssn{measuring} the $A_v^{\Z_2^2}$ projector \eqref{eq:AvZ2Z2} \ssn{and post-selecting the $+1$ outcomes}. Combining the above operations, we will show that the states are the logicals in $D(\Z_2\times\Z_2)$:
\be\ba 
    \prod_v A_v^{\Z_2^2}
    \big(\!\bigotimes_l\bra{0\cdots0}\otimes 
    \bbI\otimes \bbI\big)
    \ol{\ket{0}}&=\overline{\ket{0}}_{\Z_2^2}\,,\\
    \prod_v A_v^{\Z_2^2}
    \big(\!\bigotimes_l\bra{0\cdots0}\otimes 
    \bbI\otimes \bbI\big)
    \ol{\ket{1}}&=\overline{\ket{1}}_{\Z_2^2}\,.
\ea\ee
Indeed, they are logical states because they are $+1$ eigenstates of all $D(\Z_2\times\Z_2)$ stabilizers and they differ by a non-contractible anyon configuration, since, as we explained  around \eqref{eq:last_qubits}, the $\ket{rs},\ket{s},\ket{r}$ $D_{4N}$ states, when restricted to the last two qubits, correspond precisely to the $\Z_2\times\Z_2$ group elements $m_1m_2,m_2,m_1$, which are supported on a non-contractible ribbon configuration on the dual lattice. In particular, the first state contains the completely trivial state $\bigotimes_l\ket{00}_l$ while the second state contains 
\be
\prod_{l\in\xi_1}(X_1X_2)_l \prod_{j\in\xi_2}(X_2)_j \prod_{k\in\xi_3}(X_1)_k\bigotimes_l \ket{00}_l,
\ee
which, recalling \eqref{eq:last_qubits}, is equal to 
\be
\Big(\!\!\bra{0\cdots0}
\otimes 
    \bbI\otimes \bbI\Big) L_{\xi_1}^{rs} L_{\xi_2}^{s} L_{\xi_3}^{r}\bigotimes_l \ket{00\cdots 00}_l\,.
\ee
This concludes the lattice description of the code-switching between the logical bases of $D(\Z_2\times\Z_2)$ and $D(D_{4N})$.

\vspace{2mm}
\noindent{\textbf{Qubit-based error-correcting protocol.}} Let us now illustrate how the non-abelian code could be used in an error-correcting protocol using physical qubits, generalizing the one presented in \cite{Kobayashi:2025cfh}: 
We consider the case of $8N=2^n$, in which our setup is qubit based, as explained in Sec.~\ref{sec:Qubits}. We will denote by $d$ the code-distance: it scales linearly with the lattice size, since we are using a topological surface code.

The whole setup is depicted in Fig.~\ref{fig:Xmastree}.
We start with the double SC, $D(\Z_2\times\Z_2)$: concretely we initialize the last two qubits on each edge, that correspond to the group elements $\ket{r}$ and $\ket{s}$, to form the logical $\ol{\ket{+}}$ state. All other qubits are initialized to $\ket{0}$.

Next, we  measure the stabilizers corresponding to the remaining $n-2$ supersolvable group generators $r^a$ for  $a\in\{{2^{n-2}},{2^{n-3}},\cdots,2\}$, each requiring $O(d)$ rounds of measurements. Decoding is achieved by means of a just-in-time (JIT) decoder as in \cite{Davydova:2025ylx}, followed by the renormalization group (RG) decoder of \cite{Duclos-Cianci:2010bfc}, as in \cite{Kobayashi:2025cfh} (and similarly to the error-correction strategy of \cite{Huang:2025ump}). 
In the $D(D_{4N})$ quantum double we apply the constant-depth unitary $U_{\alpha,\beta}$, that acts as the logical $T^{1/N}$ gate on the logical state. 
We then measure the $Z$ operators for the first $n-2$ qubits on each lattice edge, corresponding to group elements $r^a$ for  $a\in\{{2^{n-2}},{2^{n-3}},\cdots,2\}$. We correct the outcomes of $\ket{1}$ by acting with the group multiplication operators of Sec. \ref{sec:Qubits}, (recall that these are written in terms of $C^\ell X$ qubit operators). \prx{The time overhead therefore scales linearly with the code distance $d$ (for fixed $n$) as in \cite{Davydova:2025ylx,Kobayashi:2025cfh} and with  the number of physical qubits on each edge of the lattice $n$ (for fixed $d$). An in-depth discussion of a generalization of the JIT decoder for these codes will appear in future work.}

We have now returned to the starting $D(\Z_2\times\Z_2)$ setup in which we perform another $O(d)$ rounds of error-correction. The final state is:
\be
T^{1/N}\ol{\ket{+}}=\ol{\ket{0}}+e^{\frac{2\pi i}{2^{n}}}\ol{\ket{1}}\,,
\ee
a magic state at the $n$-th level of the Clifford hierarchy.

\section{Discussion and Future Directions}
\label{sec:Dis}

In this paper, we have constructed constant-depth Clifford-hierarchy gates in 2D locality preserving surface codes of the quantum double $D(G)$ for non-Abelian finite groups $G$. 
The gates are realized in terms of stacking SPTs that realize automorphisms of the quantum double $D(G)$. This in itself is a general way to implement constant-depth gates. 
Applied to the dihedral groups $D_{4N}$ (of order $8N$) we show
{how to define a non-Abelian stabilizer code: determine suitable boundary conditions for the surface code on a triangular patch, that realize a single logical qubit. We then determine the full stabilizer group (which is non-Abelian). Finally we implement the constant-depth gates that span the diagonal non-Clifford gates up to any level of the Clifford hierarchy.} This can be taken to be arbitrarily large by increasing $N$.

For $D(D_{4N})$ we determine the Clifford hierarchy stabilizers and show that these codes can realize the non-Clifford gate $T^{1/N}$. Furthermore, for $8N=2^n$, the code can be implemented entirely in terms of qubit operations, and realize gates from the Clifford hierarchy up to level $n$.
The other $N$ values are beyond the Clifford hierarchy (and would require some higher qudits for their implementation).

This gives an alternative to the Bravyi-K\"onig theorem, by considering, instead of higher dimensional codes, purely 2D ones, but increasing complexity by allowing non-Abelian surface codes. The realization in terms of qubits and purely in 2D should nevertheless give this approach an edge in terms of implementation. The physical gates required in this case are non-Clifford. 

The next most important point is to explore the QEC model, investigating the feasibility of the proposed just-in-time decoder geared towards non-Abelian surface codes in  \cite{Davydova:2025ylx}.
On a theoretical level it would be interesting to apply this systematic approach to explore gates that arise from automorphisms of $D^\omega(G)$ for any finite (non-Abelian) group $G$ with non-trivial 3-cocycle $\omega\in H^3(G,U(1))$. Such automorphisms were mathematically characterized in~\cite{Naidu:2007}. These generically include both automorphisms of the group $G$ in addition to the SPT stacking developed in the current paper.

\subsection*{Acknowledgments}

\noindent
We thank Yanzhu Chen, Alex Cowtan, Yuhan Gai, Po-Shen Hsin, Sheng-Jie Huang, Chao-Ming Jian,  Ryohei Kobayashi, Nat Tantivasadakarn, Zhenghan Wang, Dominic Williamson, Guanyu Zhu  for discussions. This work is supported by the UKRI Frontier Research Grant, underwriting the ERC Advanced Grant ``Generalized Symmetries in Quantum Field Theory and Quantum Gravity''.

The authors would like to thank the Isaac Newton Institute for Mathematical Sciences for the support and hospitality during the programme "Quantum field theory with boundaries, impurities, and defects", which inspired some of this work. The INI is supported by EPSRC Grant Number EP/z000580/1.

\appendix

\section{Aspects of Quantum Doubles} 
\label{app:GroupTheory}

\subsection{Quantum Doubles $D(G)$} \label{app:anyons_review}

In this appendix we provide a few details on quantum doubles $D(G)$ for finite groups $G$ and discuss the Lagrangian algebras that label the possible gapped boundary conditions. 
Mathematically, the quantum doubles have anyons that are given in terms of the Drinfeld center of the symmetry category $G$. 

The anyons are labeled by a conjugacy class 
\be
[g] = \{ kgk^{-1}\;:\;  k\in G\}
\ee
and an irreducible representation $\bm{R}$ of the centralizer $C_G(g)$ of $g \in [g]$
\be
C_G(g) = \{h\in G:\, g h = h g \} \,.
\ee
Let us first detail this for  $D_4$ whose presentation is in \eqref{eq:D4}: 
\be
\begin{array}{|c|c|c|c|}
\hline
[g] & \text{Elements} & C_G(g) & \text{Irreps} \cr \hline 
[1] & 1 & D_4 & 1, 1_r, 1_s, 1_{rs},  E \cr  
[r^2] & r^2  & D_4 & 1, 1_r, 1_s, 1_{rs}, E \cr  
[r] & r, r^3 & \Z_4 & i^a\cr 
[s] & s, r^2s & \Z_2^2 & (\epsilon_1 ,\epsilon_2) \cr 
[rs] & rs, r^3s & \Z_2^2 & (\epsilon_1 ,\epsilon_2) \cr \hline
\end{array}
\ee
Where the irreps are labeled as follows: $1_r$ is the 1-dim irrep, in which $r$ acts trivially, and $s, rs$ act with a  sign etc. Furthermore, the irreps of $\Z_4$ and $\Z_2^2$ are labeled by $i^a,a\in\{0,1,2,3\}$ and $\epsilon_k\in\{+,-\},k\in\{1,2\}$. 

\vspace{1mm}
For $D_{4N}$, one has:
 \be
 \begin{array}{|c|c|c|c|}
 \hline
 [g] & \text{Elements} & C_G(g) & \text{Irreps} \cr \hline 
 [1] & 1 & D_{4N} & 1, 1_r, 1_s, 1_{rs},  E_\ell \cr  
 [r^{2N}] & r^{2N}  & D_{4N} & 1, 1_r, 1_s, 1_{rs}, E_\ell \cr  
 [r^a] & r^a, r^{-a} & \Z_{4N} & \zeta_{4N}^b \cr 
 [s] & s, r^2s, \cdots, r^{4N-2}s & \Z_2^2 & (\epsilon_1 ,\epsilon_2) \cr 
 [rs] & rs, r^3s, \cdots, r^{4N-1}s & \Z_2^2 & (\epsilon_1 ,\epsilon_2) \cr 
 \hline
 \end{array}
 \ee
 with the $D_{4N}$ irrep operators reviewed in Sec. \ref{sec:CliffHierStab}. Here, $\ell,a,b\in\{1,\cdots,2N-1\}$, $\zeta_{4N}=e^{i\pi/2N}$ and $\epsilon_k\in\{+,-\},k\in\{1,2\}$. 

\subsection{Gapped Boundary Conditions}
\label{sec:GappedBC}

Gapped boundary conditions are in one-to-one correspondence with Lagrangian algebras in $D(G)$. These are in turn classified by a subgroup $K\subseteq G$ and a cocycle
\be
(K, \gamma) \,,\qquad \gamma \in H^2 (K, U(1)) \,,
\ee
{up to conjugation}.
We can think of these as arising from the Dirichlet boundary condition that realizes $G$ by stacking the SPT $\gamma$ and gauging $K$. 
In turn one can characterize them in terms of a collection of anyons, that form an algebra (the anyons alone will not fix the algebra in general fully -- but in the current examples they do, so we will refrain from elaborating on this here, see \cite{Gai:2026hjk}).
We will write these as 
\be
\cL_{(K, \gamma)} = \bigoplus_{n_{ ([g], \bm{R})}} n_{ ([g], \bm{R})}([g], \bm{R}) \,,
\ee
with $n_{ ([g], \bm{R})}$ non-negative integers.

\subsection{Group 2-Cocycles} 
\label{app:2coc}
In this appendix, we review the theory of group 2-cocycles. A \emph{group 2-cocycle} is a function $\alpha:G\times G\to U(1)$ that satisfies the equation:
\be \label{eq:alpha_2-coc_cond}
    \alpha(g,h)\alpha(gh,k)=\alpha(g,hk)\alpha(h,k)\,,
\ee
which is known as the 2-cocycle condition. Two 2-cocycles $\alpha$ and $\alpha'$ are \emph{equivalent} if there exists a function $\kappa:G\to U(1)$ (known as a group 1-cochain) such that
\be
    \alpha(g,h)=\frac{\kappa(g)\kappa(h)}{\kappa(gh)}\alpha'(g,h)\,,\quad \forall\,g,h\in G\,.
\ee
The above equation can be written more concisely as $\alpha=(\delta\kappa)\,\alpha'$ and characterizes a \emph{cohomology class} in $H^2(G,U(1))$. Any 2-cocycle in that class is called a \emph{representative} of the cohomology class. 

Let us briefly review how a representative 
\be
    \alpha \in H^2(G,U(1))
\ee
can be computed. A \emph{central extension} 
\be
1 \to A \to \Gamma \to G \to 1 
\ee
of a group $G$ is a group $\Gamma$ together with a group homomorphism
\be \label{eq:p_Gamma_G}
    p:\; \Gamma \to G 
\ee
such that there is an Abelian group 
\be
A=\ker(p)\subseteq Z(\Gamma)=\{h\in\Gamma\;|\;hk=kh\;\forall\;k\in \Gamma \}\,.
\ee
For each $g\in G$ let $\ell(g)\in \Gamma$ be a lift of $g$, i.e. $p(\ell(g))=g$ and we will always chose $\ell(\id)=\id$. Define 
\be
\varphi:\; G \times G \to A
\ee
by 
\be
    \ell(g)\ell(h)=\varphi(g,h)\ell(gh)\,.
\ee
Then
\be\ba
    \ell(g)\ell(h)\ell(k)&=\varphi(g,h)\ell(gh)\ell(k)=\varphi(g,h)\varphi(gh,k)\ell(ghk)\\
    \ell(g)\ell(h)\ell(k)&=\ell(g)\varphi(h,k)\ell(hk)=\varphi(g,hk)\varphi(h,k)\ell(ghk)
\ea\ee
hence $\varphi$ obeys the 2-cocycle condition
\be \label{eq:alpha2coc}
\varphi(g,h)\varphi(gh,k)=\varphi(g,hk)\varphi(h,k)\,,
\ee
(as a consequence of associativity of multiplication in $\Gamma$).

Let $\lambda:\;A\to U(1)$
be a group homomorphism, then 
\be
    \alpha':=\lambda \circ \varphi \in H^2(G,U(1))
\ee
follows from (\ref{eq:alpha2coc}) and the fact that $\lambda$ is a homomorphism. We then define 
\be
\alpha:=(\delta\kappa)\alpha'
\ee
for a 1-cochain $\kappa$, chosen such that $\alpha$ satisfies \cite[Lemma 5.1]{Beigi:2010htr}:
\be\ba \label{eq:lemma5.1}
    |\alpha(g,h)|&=1\,,
    &\alpha(\id,g)&=\alpha(g,\id)=1\,, \\
    \alpha(g,g^{-1})&=1\,,\;\;
    &\alpha(h^{-1},g^{-1})&=\alpha(g,h)^{-1}\,,
\ea\ee
$\forall \,g,h\in G$. The reason we require these properties is for compatibility with the action of bulk vertex operators $A_v^{g}$ on the lattice, as we illustrate in detail in App. \ref{app:proof_logical_action_indep}.

Note that when we solve $\alpha|_K=\delta\beta$, with $K\subset G$ a subgroup labeling a boundary condition, if a solution $\beta:G\to U(1)$ exists, we can chose it such that:
\be \label{eq:betagginv}
    \beta(g)\beta(g^{-1})=1\,,\quad \beta(\id)=1\,.
\ee
This is possible as a consequence of property \eqref{eq:lemma5.1} of $\alpha$, in particular:
\be
1=\alpha(g,g^{-1})=\frac{\beta(g)\beta(g^{-1})}{\beta(\id)} \,.
\ee

\vspace{1mm}
\noindent\textbf{2-cocycles for $D_{4N}$.}
To compute 
\be
    \alpha_N\in H^2(D_{4N},U(1))\,,
\ee
we will take a non-trivial central extension of $D_{4N}$ to be specified by the group
\be 
    \qquad\Gamma=D_{8N}=\langle r,s \,|\, r^{8N}=s^2=\id\,,\; srs=r^{-1}\rangle\,,
\ee
and the homomorphism 
\be\ba
    p:\;D_{8N}&\to D_{4N}\\ 
    r^{4N} &\mapsto\id\,,
\ea\ee
with $A=\ker(p)=\langle r^{4N}\rangle\cong \Z_2$
and $\lambda (r^{4N}) =- 1$.
In $D_{8N}$, we have
\be
    r^a s^j r^b s^k  =  r^{a+(1-2j)b} s^{j+k}\,.
\ee
To write a non-trivial 2-cocycle on $D_{4N}$, we will identify the group element $r^{a}s^j\in D_{4N}$ for $i\in\{0,1\}$ and $a\in\{0,\cdots ,4N-1\}$ as its lift in $D_{8N}$ with the same values of $a,j$, i.e. $\ell(r^as^j)=r^as^j$. Following the procedure summarized above, we obtain the following non-trivial 2-cocycle on $D_{4N}$:
\be
    \alpha_N'(r^a s^j, r^b s^k)=
    \begin{cases}
        -1 & \text{if } (a+(1-2j)b) \,\text{mod}\,8N \geqslant 4N\\
        +1 &\text{otherwise} \\
    \end{cases}
\ee
where the $-1$ is the phase comes from $\lambda(r^{4N})=-1$ since $r^{4N}\in D_{8N}$ appears in the $D_{8N}$ multiplication when $(a+(1-2j)b) \,\text{mod}\, 8N \geqslant 4N$. 

Define $\beta'_N$ to be such that $\alpha'_N\big|_{\langle r\rangle}=\delta\beta'_N$. For example:
     \be\ba
        -1&=\alpha'_N(r^{2N},r^{2N})=\frac{\beta'_N(r^{2N})^2}{\beta(\id)} &&\Rightarrow\; \beta'_N(r^{2N})=e^{i\pi/2} 
     \ea\ee
Recursively, a solution for $\beta'_N$ can be computed to be:
\be
     \beta'_N(r^a)=e^{i\pi a/(4N)}\,.
\ee

Note however, that $\alpha_N'$ does not satisfy \eqref{eq:lemma5.1}: in particular $\alpha_N'(r^{2N},r^{2N})=-1$ and $\alpha_N'(r^a,r^{-a})=-1$ for $a \neq 2N$. We therefore define a 1-cochain $\kappa_N:D_{4N}\to U(1)$ as follows:
\be
    \kappa_N(r^a s^j)=
    \begin{cases}
        -i & \text{if } a = 2N \,\text{and}\,j=0\\
        -1 & \text{if } a > 2N \,\text{and}\,j=0\\
        +1 &\text{otherwise} \\
    \end{cases}
\ee
for $a\in\{0,1,2,3\}$ and $j\in\{0,1\}$. This ensures that
\be
    \alpha_N:=\alpha_N'\delta \kappa
\ee
satisfies \eqref{eq:lemma5.1}. A 1-cochain $\beta_N$ such that $\alpha_N\big|_{\langle r\rangle}=\delta\beta_N$ is given by:
\be
    \beta_N:=\beta'_N\kappa_N\,,
\ee
whose explicit values are
\be
    \beta_N(r^a)=
    \begin{cases}
        +e^{i\pi a/(4N)} & \text{if } a<2N \\
        +1 & \text{if } a = 2N \\
        -e^{i\pi a/(4N)} & \text{if } a>2N\,. \\
    \end{cases}
\ee

\subsection{\prx{Proof that $U_{\alpha,\beta}$ commutes with $A_v^h$ operators}} 
\label{app:proof_logical_action_indep}

{
In this Appendix, we prove that $U_{\alpha,\beta}$ commutes with the $A_v^h$ operators: it therefore realizes a well-defined logical action, i.e. it acts identically on all physical states in the superpositions \eqref{eq:idid_state_lattice} and \eqref{eq:g1g2_state_lattice}. 

\vspace*{2mm}
\noindent\textbf{Proof that bulk vertex terms commute with $U_{\alpha,\beta}$.}
Consider a state on the four plaquettes around a bulk vertex $v$:
\begin{equation} \label{eq:plaquette_before}
\begin{split}
\begin{tikzpicture}[scale=1.5]
\node at (1.1,0.9) {$_v$};
\draw[->-] (0,1) -- node[left] {$g_1$} (0,2);
\draw[->-] (0,2) -- node[above] {$g_2$} (1,2);
\draw[->-] (1,1) -- node[left] {$g_3$} (1,2);
\draw[->-] (0,1) -- node[below] {$g_4$} (1,1);
\draw[->-] (1,2) -- node[above] {$g_5$} (2,2);
\draw[->-] (2,1) -- node[right] {$g_6$} (2,2);
\draw[->-] (1,1) -- node[below] {$g_7$} (2,1);
\draw[->-] (2,0) -- node[right] {$g_8$} (2,1);
\draw[->-] (1,0) -- node[below] {$g_9$} (2,0);
\draw[->-] (1,0) -- node[left] {$g_{10}$} (1,1);
\draw[->-] (0,0) -- node[below] {$g_{11}$} (1,0);
\draw[->-] (0,0) -- node[left] {$g_{12}$} (0,1);
\end{tikzpicture}
\end{split}
\end{equation}
We require that the state is stabilized by $\prod_pB_p$ in order for it to appear in a ground state of the Hamiltonian \eqref{eq:Hamiltonian}, i.e.
\be\ba \label{eq:state_Bp_cond}
    g_{1}g_{2}&=g_{4}g_{3}\,, & g_{3}g_{5}&=g_{7}g_{6} \,,\\
    g_{12}g_{4}&=g_{11}g_{10}\,, & g_{10}g_{7}&=g_{9}g_{8} \,.\\
\ea\ee
Recalling \eqref{eq:Malpha_general}, the phase produced by $\prod_pM_p^\alpha$ acting on the state \eqref{eq:plaquette_before} is:
\be \label{eq:phi}
\phi:=\frac{\alpha(g_{1},g_{2})}{\alpha(g_{4},g_{3})}\cdot
\frac{\alpha(g_{3},g_{5})}{\alpha(g_{7},g_{6})}\cdot
\frac{\alpha(g_{12},g_{4})}{\alpha(g_{11},g_{10})}\cdot
\frac{\alpha(g_{10},g_{7})}{\alpha(g_{9},g_{8})}\,.
\ee
After acting with $A_v^{\green{h}}$, for any $\green{h}\in G$, the state \eqref{eq:plaquette_before} becomes:
\begin{equation} \label{eq:plaquette_after}
\begin{split}
\begin{tikzpicture}[scale=1.5]
\node at (1.1,0.9) {$_v$};
\draw[->-] (0,1) -- node[left] {$g_1$} (0,2);
\draw[->-] (0,2) -- node[above] {$g_2$} (1,2);
\draw[->-] (1,1) -- node[left] {${\green{h}}g_3$} (1,2);
\draw[->-] (0,1) -- node[below] {$g_4{\green{h^{-1}}}$} (1,1);
\draw[->-] (1,2) -- node[above] {$g_5$} (2,2);
\draw[->-] (2,1) -- node[right] {$g_6$} (2,2);
\draw[->-] (1,1) -- node[below] {${\green{h}}g_7$} (2,1);
\draw[->-] (2,0) -- node[right] {$g_8$} (2,1);
\draw[->-] (1,0) -- node[below] {$g_9$} (2,0);
\draw[->-] (1,0) -- node[left] {$g_{10}{\green{h^{-1}}}$} (1,1);
\draw[->-] (0,0) -- node[below] {$g_{11}$} (1,0);
\draw[->-] (0,0) -- node[left] {$g_{12}$} (0,1);
\end{tikzpicture}
\end{split}
\end{equation}
The phase produced by $\prod_pM_p^\alpha$ on the state \eqref{eq:plaquette_after} is:
\be\ba
\phi^{\green{h}}:=&\frac{\alpha(g_{1},g_{2})}{\alpha(g_{4}\green{h^{-1}},\green{h}g_{3})}\cdot
\frac{\alpha(\green{h}g_{3},g_{5})}{\alpha(\green{h}g_{7},g_{6})}\\\cdot&\frac{\alpha(g_{12},g_{4}\green{h^{-1}})}{\alpha(g_{11},g_{10}\green{h^{-1}})}\cdot
\frac{\alpha(g_{10}\green{h^{-1}},\green{h}g_{7})}{\alpha(g_{9},g_{8})}
\ea\ee
We will now show that $\phi^{\green{h}}=\phi$. Using the 2-cocycle condition \eqref{eq:alpha_2-coc_cond} and the property that the state minimizes the $B_p$ operators, i.e. \eqref{eq:state_Bp_cond}, we can re-write each term in $\phi^{\green{h}}$ as follows:
\be\ba \label{eq:phi_h_lines}
    \frac{\alpha(g_{1},g_{2})}{\alpha(g_{4}\green{h^{-1}},\green{h}g_{3})}&=
    \frac{\alpha(g_{1},g_{2})}{\alpha(g_{4},g_{3})}\cdot 
    \frac{\alpha(g_{4},\green{h^{-1}})}{\alpha(\green{h^{-1}},\green{h}g_{3})}\,,
    \\
   \frac{\alpha(\green{h}g_{3},g_{5})}{\alpha(\green{h}g_{7},g_{6})}&=
   \frac{\alpha(g_{3},g_{5})}{\alpha(g_{7},g_{6})}\cdot
   \frac{\cancel{\alpha(\green{h},g_{3}g_{5})}}{\cancel{\alpha(\green{h},g_{7}g_{6})}}\cdot 
   \frac{\alpha(\green{h},g_{7})}{\alpha(\green{h},g_{3})} \,,\\
    \frac{\alpha(g_{12},g_{4}\green{h^{-1}})}{\alpha(g_{11},g_{10}\green{h^{-1}})}&=
    \frac{\alpha(g_{12},g_{4})}{\alpha(g_{11},g_{10})}\cdot
    \frac{\cancel{\alpha(g_{12}g_{4},\green{h^{-1}})}}{\cancel{\alpha(g_{11}g_{10},\green{h^{-1}})}}\cdot
    \frac{\alpha(g_{10},\green{h^{-1}})}{\alpha(g_{4},\green{h^{-1}})} \,,\\
    \frac{\alpha(g_{10}\green{h^{-1}},\green{h}g_{7})}{\alpha(g_{9},g_{8})}&=
    \frac{\alpha(g_{10},g_{7})}{\alpha(g_{9},g_{8})}\cdot
    \frac{\alpha(\green{h^{-1}},\green{h}g_{7})}{\alpha(g_{10},\green{h^{-1}})} \,.
\ea\ee
In $\phi^{\green{h}}$, the four lines above are multiplied together: in the product,  $\alpha(g_{4},\green{h^{-1}})$ cancels since it appears in a numerator and a denominator, and similarly for $\alpha(g_{10},\green{h^{-1}})$. Furthermore, using \eqref{eq:alpha_2-coc_cond} and \eqref{eq:lemma5.1}:
\be\ba \label{eq:hg3_simpl}
    \alpha(\green{h^{-1}},\green{h}g_{3})\alpha(\green{h},g_{3})&=\alpha(\green{h^{-1}},\green{h})\alpha(\id,g_{3})=1\,,\\
    \alpha(\green{h},g_{7})\alpha(\green{h^{-1}},\green{h}g_{7})&=\alpha(\green{h^{-1}},\green{h})\alpha(\id,g_{7})=1\,,
\ea\ee
so in the product of the four lines of \eqref{eq:phi_h_lines} all terms containing $\green{h}$ or $\green{h^{-1}}$ cancel and we are left precisely with the product of the four factors of $\phi$, \eqref{eq:phi}. We have proven that
\be
    \phi^{\green{h}}=\phi\,.
\ee
This shows that any bulk vertex term $A_v^{\green{h}}$ commutes with $\prod_pM^\alpha_p$, and therefore with the $U_{\alpha,\beta}$ circuit \eqref{eq:U_alpha_lattice}, when acting on the states that minimize the $B_p$ terms in \eqref{eq:Hamiltonian}. 

\vspace*{2mm}
\noindent\textbf{Proof that boundary vertex terms commute with $U_{\alpha,\beta}$.}
Consider a state on the four plaquettes around a vertex $v$ on the bottom boundary (the cases where $v$ is on a different boundary can be checked similarly):
\begin{equation} \label{eq:plaquette_boundary_before}
\begin{split}
\begin{tikzpicture}[scale=1.5]
\node at (1,0.9) {$_v$};
\draw[->-] (0,1) -- node[left] {$g_1$} (0,2);
\draw[->-] (0,2) -- node[above] {$g_2$} (1,2);
\draw[->-] (1,1) -- node[left] {$g_3$} (1,2);
\draw[->-] (0,1) -- node[below] {$g_4$} (1,1);
\draw[->-] (1,2) -- node[above] {$g_5$} (2,2);
\draw[->-] (2,1) -- node[right] {$g_6$} (2,2);
\draw[->-] (1,1) -- node[below] {$g_7$} (2,1);
\end{tikzpicture}
\end{split}
\end{equation}
The $U_{\alpha,\beta}$ circuit \eqref{eq:U_alpha_lattice} acting on the boundary state \eqref{eq:plaquette_boundary_before} produces the following phase, in which we denote by $\beta$ the boundary 1-cochain such that $\alpha|_K=\delta\beta$, where $K$ is the subgroup labeling the boundary:
\be\ba
    \phi:=\frac{\alpha(g_{1},g_{2})}{\alpha(g_{4},g_{3})}\cdot
\frac{\alpha(g_{3},g_{5})}{\alpha(g_{7},g_{6})}\cdot \beta(g_{4})\beta(g_{7})\,.
\ea\ee
After acting with $A_v^{\green{h}}$ for $\green{h}\in K$, the state \eqref{eq:plaquette_boundary_before} becomes:
\begin{equation} \label{eq:plaquette_boundary_after}
\begin{split}
\begin{tikzpicture}[scale=1.5]
\node at (1,0.9) {$_v$};
\draw[->-] (0,1) -- node[left] {$g_1$} (0,2);
\draw[->-] (0,2) -- node[above] {$g_2$} (1,2);
\draw[->-] (1,1) -- node[left] {${\green{h}}g_3$} (1,2);
\draw[->-] (0,1) -- node[below] {$g_4{\green{h^{-1}}}$} (1,1);
\draw[->-] (1,2) -- node[above] {$g_5$} (2,2);
\draw[->-] (2,1) -- node[right] {$g_6$} (2,2);
\draw[->-] (1,1) -- node[below] {${\green{h}}g_7$} (2,1);
\end{tikzpicture}
\end{split}
\end{equation}
on which the phase produced by $U_{\alpha,\beta}$ is:
\be\ba \label{eq:phy_bdr_h}
    \phi^{\green{h}}:=\frac{\alpha(g_{1},g_{2})}{\alpha(g_{4}\green{h^{-1}},\green{h}g_{3})}\cdot
\frac{\alpha(\green{h}g_{3},g_{5})}{\alpha(\green{h}g_{7},g_{6})}\cdot
\beta(g_{4}\green{h^{-1}})\beta(\green{h}g_{7})
\ea\ee
Similarly to the first two equations in \eqref{eq:phi_h_lines}, we can use \eqref{eq:alpha_2-coc_cond} and \eqref{eq:state_Bp_cond} to write:
\be\ba  \label{eq:bdr_calc_1}
    \frac{\alpha(g_{1},g_{2})}{\alpha(g_{4}\green{h^{-1}},\green{h}g_{3})}\cdot \frac{\alpha(\green{h}g_{3},g_{5})}{\alpha(\green{h}g_{7},g_{6})}&=
    \frac{\alpha(g_{1},g_{2})}{\alpha(g_{4},g_{3})}\cdot \frac{\alpha(g_{3},g_{5})}{\alpha(g_{7},g_{6})} \\
    &\cdot
    \frac{\alpha(g_{4},\green{h^{-1}})}{\alpha(\green{h^{-1}},\green{h}g_{3})} \cdot \frac{\alpha(\green{h},g_{7})}{\alpha(\green{h},g_{3})}\,.
\ea\ee
We now use $\alpha|_K=\delta\beta$ to write:
\be
    \alpha(g_{4},\green{h^{-1}})\cdot \alpha(\green{h},g_{7})=\frac{\beta(g_{4})\beta(\green{h}^{-1})}{\beta(g_{4}\green{h^{-1}})}\cdot
    \frac{\beta(\green{h})\beta(g_{7})}{\beta(\green{h}g_{7})}
\ee
and recall from \eqref{eq:betagginv} that
\be \label{eq:bdr_calc_2}
    \beta(\green{h^{-1}})\beta(\green{h})=1\,.
\ee
Replacing  \eqref{eq:bdr_calc_1}-\eqref{eq:bdr_calc_2} and the first line of \eqref{eq:hg3_simpl} into \eqref{eq:phy_bdr_h}, we conclude that:
\be
    \phi^{\green{h}}=\phi\,.
\ee
The calculation for the other boundaries is analogous. Therefore, any boundary vertex term $A_v^{\green{h}}$ commutes with the $U_{\alpha,\beta}$ circuit \eqref{eq:U_alpha_lattice}, when acting on the states that minimize the $B_p$ terms in \eqref{eq:Hamiltonian}.

\vspace*{2mm}
\noindent\textbf{Proof that corner vertex terms commute with $U_{\alpha,\beta}$.}
In general, Hamiltonians can have corner vertex terms, $A_v^{\green{h}}$ \eqref{eq:A_corner}, at vertices $v=s_{ij}$ between boundaries $i$ and $j$ for ${\green{h}}\in K_i\cap K_j$, although in our concrete examples they are all trivial since $ K_i\cap K_j=\{\id\}$. We will show that corner terms, when present, also commute with the operators in the $U_{\alpha,\beta}$ circuit \eqref{eq:U_alpha_lattice}.
Consider a state on a plaquette near a  corner vertex $v$, which we take to be in the bottom-left, i.e. between boundaries 1 and 3 (other cases are analogous): 
\begin{equation} \label{eq:plaquette_corner_before}
\begin{split}
\begin{tikzpicture}[scale=1.5]
\node at (0,0.9) {$_v$};
\draw[->-] (0,1) -- node[left] {$g_1$} (0,2);
\draw[->-] (0,2) -- node[above] {$g_2$} (1,2);
\draw[->-] (1,1) -- node[right] {$g_3$} (1,2);
\draw[->-] (0,1) -- node[below] {$g_4$} (1,1);
\end{tikzpicture}
\end{split}
\end{equation}
$U_{\alpha,\beta}$ \eqref{eq:U_alpha_lattice} acting on  \eqref{eq:plaquette_corner_before} produces the phase
\be
    \phi:=\frac{\alpha(g_1,g_2)}{\alpha(g_4,g_3)}\cdot\frac{\beta^{(3)}(g_4)}{\beta^{(1)}(g_1)}\,.
\ee
Applying a corner vertex term $A_v^{\green{h}}$ \eqref{eq:A_corner} at $v$, the state becomes
\begin{equation} \label{eq:plaquette_corner_after}
\begin{split}
\begin{tikzpicture}[scale=1.5]
\node at (0,0.9) {$_v$};
\draw[->-] (0,1) -- node[left] {$\green{h}g_1$} (0,2);
\draw[->-] (0,2) -- node[above] {$g_2$} (1,2);
\draw[->-] (1,1) -- node[right] {$g_3$} (1,2);
\draw[->-] (0,1) -- node[below] {$\green{h}g_4$} (1,1);
\end{tikzpicture}
\end{split}
\end{equation}
on which the phase produced by the $U_{\alpha,\beta}$ circuit \eqref{eq:U_alpha_lattice} is:
\be \label{eq:phi_corner_h}
    \phi^{\green{h}}:=\frac{\alpha(\green{h}g_1,g_2)}{\alpha(\green{h}g_4,g_3)}\cdot\frac{\beta^{(3)}(\green{h}g_4)}{\beta^{(1)}(\green{h}g_1)}\,.
\ee
Using the 2-cocycle condition \eqref{eq:alpha_2-coc_cond}, and the property $g_1g_2=g_4g_3$ that ensures that the state minimizes the plaquette term, we can write
\be\ba
    \frac{\alpha(\green{h}g_1,g_2)}{\alpha(\green{h}g_4,g_3)}&=\frac{\alpha(g_1,g_2)}{\alpha(g_4,g_3)}\cdot\frac{\cancel{\alpha(\green{h},g_1g_2)}}{\cancel{\alpha(\green{h},g_4g_3)}}\cdot
    \frac{\alpha(\green{h},g_4)}{\alpha(\green{h},g_1)}\,.
\ea\ee
Now, we use the property that $\beta^{(i)}$ trivializes $\alpha|_{K_i}$ \eqref{eq:alphaK_delta_beta} and the matching $\beta^{(i)}(g)=\beta^{(j)}(g)$ \eqref{eq:beta_i_beta_j}, to write
\be\ba
    \frac{\alpha(\green{h},g_4)}{\alpha(\green{h},g_1)}&=\frac{\cancel{\beta^{(3)}(\green{h})}\beta^{(3)}(g_4)}{\cancel{\beta^{(1)}(\green{h})}\beta^{(1)}(g_1)}\cdot\frac{\beta^{(1)}(\green{h}g_1)}{\beta^{(3)}(\green{h}g_4)}
\ea\ee
therefore, combining the above equations, we have shown that:
\be
    \phi^{\green{h}}=\phi\,,
\ee
proving that the corner $A_v^\green{h}$ terms also commute with the $U_{\alpha,\beta}$ circuit \eqref{eq:U_alpha_lattice}.

\vspace*{2mm}
\noindent\textbf{Conclusion.}
In summary, we have shown that any vertex term $A_v^{\green{h}}$ commutes with the $U_{\alpha,\beta}$ circuit \eqref{eq:U_alpha_lattice}, when acting on the states that minimize the $B_p$ terms in \eqref{eq:Hamiltonian}. This implies that all states that minimize the $B_p$ terms in \eqref{eq:Hamiltonian} and that are related by $A_v$ terms, in particular all the physical states that comprise each logical state \eqref{eq:idid_state_lattice} and \eqref{eq:g1g2_state_lattice}, will carry the same phase when acted upon by $U_{\alpha,\beta}$, whose logical action is therefore well-defined. 

}

\section{Non-Abelian Stabilizer Code for $D(D_{4N})$} 
\label{app:StabilizerComms}

\subsection{$D(D_{4N})$ stabilizer group}
In this appendix we will prove that a stabilizer group for the $D(D_{4N})$ surface code is given by (\ref{StabGroupsD4N}), i.e. 
\be \label{eq:stab_group_app}
\cS_{D(D_{4N})} = \left\langle A_v^r \,,\, A_v^s\,,\, S_p^r \,,\, S_p^s  \right\rangle \,,
\ee
which satisfy the non-trivial commutation relations (\ref{D4NStabComms}).

\vspace{1mm}
\noindent\textbf{$D_{4N}$ stabilizer group generators.}
The stabilizer group is comprised of vertex and plaquette unitary operators, for which the ground states of the quantum double Hamiltonian \eqref{eq:Hamiltonian} are +1 eigenstates (i.e. they are stabilized). We will also require that all elementary excited states (as defined in \cite{Kitaev:1997wr}) of \eqref{eq:Hamiltonian} are uniquely distinguished by the eigenvalues of the stabilizers.

The vertex stabilizer operators are generated by the quantum double unitaries $A^r_v$ and $A^s_v$ (since $r,s$ are generators of $D_{4N}$ and the other vertex operators can be obtained from them via the first equation in \eqref{eq:non-comm_A_B}). Explicitly,
inserting the $D_{4N}$ left and right multiplication operator expressions \eqref{eq:D4N_LR} into \eqref{eq:AvBp}, they take the form:
\be \label{eq:A_stab}
\ba
A_v^{r} &= 
\begin{tikzpicture}[baseline]
\begin{scope}[shift={(0,0.1)}]
\draw[thick, ->-] (-1,0) to (0,0);
\draw[thick, ->-] (0,-1) to (0,0);
\draw[thick, ->-] (0,0) to (0,1);
\draw[thick, ->-] (0,0) to (1,0);
\draw[fill=black] (0,0) ellipse (0.05 and 0.05);
\node[] at (0.15, 0.15) {$_v$};
\node[above] at (-0.6,0.05) {$\mathcal{X}^{-Z}$};
\node[right] at (0.04,-0.7) {$\mathcal{X}^{-Z}$};
\node[right] at (0.04,0.8) {$\mathcal{X}$};
\node[above] at (0.75,0.05) {$\mathcal{X}$};
\draw[fill=black] (0,0) ellipse (0.05 and 0.05);
\end{scope}
\end{tikzpicture} \quad
A_v^{s} = 
\begin{tikzpicture}[baseline]
\begin{scope}[shift={(0,0.1)}]
\draw[thick, ->-] (-1,0) to (0,0);
\draw[thick, ->-] (0,-1) to (0,0);
\draw[thick, ->-] (0,0) to (0,1);
\draw[thick, ->-] (0,0) to (1,0);
\draw[fill=black] (0,0) ellipse (0.05 and 0.05);
\node[] at (0.15, 0.15) {$_v$};
\node[above] at (-0.7,0.05) {$X$};
\node[right] at (0.04,-0.7) {$X$};
\node[right] at (0.04,0.8) {$\mathcal{C}X$};
\node[above] at (0.75,0.05) {$\mathcal{C}X$};
\draw[fill=black] (0,0) ellipse (0.05 and 0.05);
\end{scope}
\end{tikzpicture} \,,
\ea\ee
where we recall that we denote the $\Z_{4N}$ qudit operators by $\cX, \cZ,\cC$ and the qubit Paulis by $X, Z$. 

The quantum double plaquette operators $B_p^g$, which we recall, act as projectors onto the plaquettes with $g$ flux as in \eqref{eq:AvBp}, are instead non-unitary. We will therefore define unitaries $S_p^r$ and $S_p^s$ as the following linear combinations of the $B_p^g$'s, with $\zeta_{4N}:=e^{2\pi i/(4N)}$:
\be\ba \label{eq:UrUs_def} 
    S_p^r&:=\sum_{\alpha=0}^{4N-1} \sum_{\beta=0}^{1} \,\zeta_{4N}^{\alpha}\, B_p^{r^\alpha s^\beta} \,,\cr
    S_p^s&:=\sum_{\alpha=0}^{4N-1} \sum_{\beta=0}^{1} \,\zeta_{4N}^{\beta}\, B_p^{r^\alpha s^\beta} \,. 
    \ea
    \ee
Note that\footnote{We include $S_p^s$ (rather than only $(S_p^s)^{2N}$) for two reasons: the former will be necessary for the boundary stabilizers and will appear (with power of $-2$) as the result of the commutator $ \left[ A_{v}^{r}, S_{p_{NE}}^r \right]$. }
\be
(S_p^s)^{2N}=\sum_{\alpha=0}^{4N-1} \sum_{\beta=0}^{1} \,(-1)^{\beta}\, B_p^{r^\alpha s^\beta} \,.
\ee
The eigenvalues of $S_p^r,S_p^s$ depend only on the magnetic flux enclosed by the plaquette $p$ and uniquely identify it. Specifically, if the eigenvalues of $(S_p^r,(S_p^s)^{2N})$  are $(\zeta_{4N}^\alpha,(-1)^\beta)$ then the flux is necessarily $r^\alpha s^\beta$. The quantum double $B_p^g$ operators can conversely be expressed as linear combinations of $S_p^r,(S_p^s)^{2N}$: 
\be\ba
    B^{r^\alpha s^\beta}=\frac{1}{8N}\sum_{\gamma=0}^{4N-1} \sum_{\delta=0}^{1} \,\zeta_{4N}^{-\alpha\gamma}\, (S_p^r)^\gamma \,(-1)^{\beta\delta} (S_p^s)^{2N\delta}\,.
\ea\ee

\noindent
{\bf Boundary Stabilizer Group.}
The Hamiltonian \eqref{eq:Hamiltonian}, also contains boundary terms. Let use therefore introduce boundary stabilizers. Let $K$ be the subgroup labeling a boundary condition.\footnote{We do not consider boundaries with non-trivial 2-cocycles in this work.} 
For each $k\in K$ and $v$ a vertex on the boundary, let the 
truncated vertex operator acting on the three edges around $v$ be denoted by $A_{v}^{k}$. 
The boundary plaquette operators in the Hamiltonian, $B_{p}^k$ act on a single boundary edge (also denoted as $s_i$) and enforce that its group element lies in $K$ (since anyons with flux in $K$ can condense at the boundary). 
Therefore, at the boundary, the operators in \eqref{eq:UrUs_def} also act on a single edge, with $B_p^h$  projecting onto the group element $h$ at that boundary edge. Since all fluxes with group elements in $K$ are admissible at the boundary, we define the boundary stabilizer subgroup $S_p^{\fB_{K_i}}$ as the subgroup of the operators $\langle S_p^r, S_p^s\rangle\big|_{\fB_i}$ that have $+1$ eigenvalue on all group elements in $K$. Explicitly, for our choices of boundary conditions $\mathfrak{B}_{\langle g \rangle}$ defined by (\ref{D4NLags}), we have: 
\be\ba
    S^{\fB_{\langle rs \rangle}}_p&=\langle (S_p^r)(S_p^s)^{-1} \rangle\big|_{\fB_{\langle rs \rangle}}\,,\\
    S^{\fB_{\langle s \rangle}}_p&=\langle S_p^r \rangle\big|_{\fB_{\langle s \rangle}} \,, \\
    S^{\fB_{\langle r \rangle}}_p&=\langle S_p^s \rangle\big|_{\fB_{\langle r \rangle}}\,.
\ea\ee
The boundary stabilizer group for boundary $\fB_{K_{i}}$ labeled by  the subgroup $K_i$, is:
\be \label{eq:boundary_stabilizers}
    \cS_{D(D_{4N})}^{\fB_{K_i}}=\langle A_v^k\,,k\in K\,;\, S_p^{\fB_{K_i}}\rangle\big|_{\fB_{K_i}}\,.
\ee

\noindent
{\bf Stabilizer Property.}
Let us prove the equivalence between the set of ground states of \eqref{eq:Hamiltonian} and the set of states stabilized by \eqref{eq:stab_group_app} and \eqref{eq:boundary_stabilizers}. First, consider the vertex operators. If $A_v^g\ket{\psi}=\ket{\psi}$ for all $g\in G$ then $A_v\ket{\psi}=\ket{\psi}$, as required for $\ket{\psi}$ to be a ground state of \eqref{eq:Hamiltonian}. Conversely, if $A_v\ket{\psi}=\ket{\psi}$, then, noting that $A_v^gA_v=A_v$ for all $g\in G$, one has $\bra{\psi}A_v^g\ket{\psi}=\bra{\psi}A_v^gA_v\ket{\psi}=\bra{\psi}A_v\ket{\psi}=1$. Therefore, $A_v^g\ket{\psi}=\ket{\psi}$ for all $g\in G$. 

Next, we turn to the plaquette operators. From their definitions in \eqref{eq:UrUs_def}, it is clear that all eigenvalues of the bulk $S_p^r,S_p^s$ operators are $+1$ if and only if the flux enclosed by the plaquette is the identity group element, id. Requiring $S_p^r\ket{\psi}=S_p^s\ket{\psi}=\ket{\psi}$ is therefore equivalent to the condition $B_p\ket{\psi}=\ket{\psi}$. On the boundaries, by definition, the condition that the state is stabilized by $\cS_{D(D_{4N})}^{\fB_{K_i}}$ is equivalent to it being a $+1$ eigenstate of the $B_{s_i}^{K_i}$ terms in \eqref{eq:Hamiltonian}. Stabilized states are therefore all and only all ground states of the quantum double Hamiltonian \eqref{eq:Hamiltonian}.

\vspace{1mm}
\noindent\textbf{Lattice Realization of Stabilizers.}
To derive the lattice expression of $S_p^r$ and $S_p^s$, consider a plaquette with group elements denoted as follows:
\be \label{eq:p_g1g2g3g4}
\begin{tikzpicture}[baseline]
\begin{scope}[shift={(0,-0.4)}, scale=1.2]
    \draw[thick, ->-] (-1,0) to (-1,1);
    \node at (-2,0.5) {$g_1=r^as^i$};
    \draw[thick, ->-] (-1,1) to (0,1);
    \node at (-0.5,1.4) {$g_2=r^bs^j$};
    \draw[thick, ->-] (0,0) to (0,1);
    \node at (1,0.5) {$g_3=r^cs^k$};
    \draw[thick, ->-] (-1,0) to (0,0);
    \node at (-0.5,-0.35) {$g_4=r^ds^\ell$};
    \node at (-0.5,0.5) {$_p$};
\end{scope}     
\end{tikzpicture} 
\ee
for $a,b,c,d\in\{0,1,\cdots,4N-1\}$ and $i,j,k,\ell\in\{0,1\}$. Then, from $D_{4N}$ multiplication,
\be\ba
    g_1g_2&=r^as^ir^bs^j=r^{a+(1-2i)b}s^{i+j}\,,\cr
    g_3^{-1}g_4^{-1}&=s^kr^{-c} s^\ell r^{-d}=s^{k+\ell} r^{-(1-2\ell)c-d}\,,
\ea\ee
and the magnetic flux enclosed by the plaquette is:
\be\ba
    &g_1g_2g_3^{-1}g_4^{-1}=\cr
    =\,&r^{a+(1-2i)b-[1-2(i+j+k+\ell \text{ mod }2)][(1-2\ell)c+d]}\,s^{i+j+k+\ell}\,,
\ea\ee
from which we derive the lattice realization of the operators \eqref{eq:UrUs_def}:
\be\ba
S_p^r &=
\begin{tikzpicture}[baseline]
\begin{scope}[shift={(0,-0.4)}]
    \draw[thick, ->-] (-1,0) to (-1,1);
    \node at (-1.4,0.5) {$\mathcal{Z}_{1}$};
    \draw[thick, ->-] (-1,1) to (0,1);
    \node at (-0.5,1.4) {$\mathcal{Z}_{2}^{Z_1}$};
    \draw[thick, ->-] (0,0) to (0,1);
    \node[right] at (0.2,0.5) {$\mathcal{Z}_{3}^{-Z_1Z_2Z_3}$};
    \draw[thick, ->-] (-1,0) to (0,0);
    \node at (0.2,-0.35) {$\mathcal{Z}^{-Z_1Z_2Z_3Z_4}_{4}$};
    \node at (-0.5,0.5) {$_p$};
\end{scope}     
\end{tikzpicture} \cr 
S_p^s&=\zeta_{4N}^{(\bbI-Z_1Z_2Z_3Z_4)/2}\cr
&=\tfrac{1}{2}(\bbI+Z_1Z_2Z_3Z_4)+\tfrac{1}{2}(\bbI-Z_1Z_2Z_3Z_4)\zeta_{4N}\cr
(S_p^s)^{2N} &=
\begin{tikzpicture}[baseline]
\begin{scope}[shift={(0,-0.4)}]
    \draw[thick, ->-] (-1,0) to (-1,1);
    \node[left] at (-1,0.5) {$Z_1$};
    \draw[thick, ->-] (-1,1) to (0,1);
    \node at (-0.5,1.35) {$Z_2$};
    \draw[thick, ->-] (0,0) to (0,1);
    \node[right] at (0.1,0.5) {$Z_3$};
    \draw[thick, ->-] (-1,0) to (0,0);
    \node at (-0.5,-0.3) {$Z_4$};
     \node at (-0.5,0.5) {$_p$};
\end{scope}    
\end{tikzpicture} \,. 
\ea
\ee

\vspace{1mm}
\noindent\textbf{Clifford-hierarchy level of stabilizers.}
We now proceed to compute the level in the Clifford hierarchy of the $D(D_{4N})$ stabilizers \eqref{eq:stabilizer_ops}. Recall that, denoting by $\cP_n$ the $n$-qubit Pauli group defined as level 1 of the Clifford hierarchy, the $k$-th level of the Clifford hierarchy is defined as \cite{Gottesman:1999tea,Zeng:2008zow,Cui:2016bxt}
\be \label{eq:Cliff_k_def}
    \cC^{(k)}_n:=\{ U\in U(2^n)\,|\, UPU^\dagger \in \cC^{(k-1)}_n\!\!,\,\forall P\in \cP_n\}.
\ee
Note that since $\cP_n$ is a group, it is sufficient to check the above condition for its generators, it is furthermore trivially satisfied for operators proportional to identity.

If $4N$ is not a power of two, the stabilizer group lies beyond the Clifford hierarchy, since the Hilbert space does not admit a purely-qubit description. 

For $4N=2^{n-1}$, instead, we can realize the gates on a Hilbert space of $n$ physical qubits on each lattice edge, as explained in Sec. \ref{sec:Qubits}: in this case, we computed  with GAP \cite{GAP4} {for $n=3,4,5$} that the stabilizer group is in level $n-1$ of the Clifford hierarchy, as summarized in Table \ref{tab:Levels}, and conjecture that this holds for all $n$. Note that the level shown is that of the full stabilizer group, some of its operators can be in lower levels.

\begin{table}
$$
\begin{array}{|c|c|c|c|c|}
\hline
N & n &  D_{4N} & \makecell[c]{\text{Level of} \\ \cS_{D(D_4)}} & \makecell[c]{\;\,\text{Level of $T^{1/N}$} \\ \text{Logical Gate}}  \\
\hline
1 & 3 &  D_4=\Z_{4}\rtimes\Z_2 & 2 & 3 \\
2 & 4 &   D_{8}=\Z_{8}\rtimes\Z_2 & 3 & 4  \\
4 & 5 &  D_{16}=\Z_{16}\rtimes\Z_2 & 4 & 5 \\
2^{n-3} & n & D_{2^{n-1}} = \Z_{2^{n-1}} \rtimes\Z_2 & n-1 & n \cr 
\hline
\end{array}
$$
\caption{To complement Table \ref{tab:Resources}, we provide the level in the Clifford hierarchy of the stabilizer group of $D(D_{4N})$ and of the logical $T^{1/N}$ gate constructed in this work, in terms of $n$, the number of physical qubits on each lattice edge. For $n=3,4,5$ we computed the level of the stabilizer group with GAP \cite{GAP4}, while the last line (for general $n$) is a conjecture. \label{tab:Levels}}
\end{table}

\subsection{Commutation Relations} \label{app:comm_rels} 

The stabilizer group is non-abelian: its non-trivial commutators as defined in \eqref{commutdef} are given by 
\be
\ba
    \left[ A_{v}^{r}, A_{v}^{s} \right] 
    &= (A_{v}^{r})^{-2} \cr 
    \left[ A_{v}^{r}, S_{p_{NE}}^r \right] 
    &= (S_{p_{NE}}^s)^{-2} \cr     
    \left[ A_{v}^{s}, S_{p_{NE}}^r \right] 
    &= (S_{p_{NE}}^r)^2
\ea\ee
where $p_{NE}$ denotes the plaquette in the North-East direct with respect to the vertex $v$.
We will now prove the above commutation relations. 

The commutators between the vertex terms are computed as follows
\be\ba
&[A^r, A^s] = (A^{r^{-1}} A^{s^{-1}}) (A^r A^s) \cr 
&\begin{tikzpicture}[baseline]
\begin{scope}[shift={(0,0)}]
\node at (-1.5, 0) {$=$};
\draw[thick, ->-] (0,0) to (1,0);
\draw[thick, ->-] (-1,0) to (0,0);
\draw[thick, ->-] (0,0) to (0,1);
\draw[thick, ->-] (0,-1) to (0,0);
\draw[fill=black] (0,0) ellipse (0.05 and 0.05);
\node[] at (0.15, 0.15) {$_v$};
\node[above] at (-0.8,0) {$\mathcal{X}^{Z}X$};
\node[right] at (0,-0.8) {$\mathcal{X}^{Z}X$};
\node[right] at (0,1) {$\mathcal{X}^{-1}\cC X$};
\node[above] at (1,0) {$\mathcal{X}^{-1}\cC X$};
\draw[fill=black] (0,0) ellipse (0.05 and 0.05);
\end{scope}
\begin{scope}[shift={(3.5,0)}]
\draw[thick, ->-] (0,0) to (1,0);
\draw[thick, ->-] (-1,0) to (0,0);
\draw[thick, ->-] (0,0) to (0,1);
\draw[thick, ->-] (0,-1) to (0,0);
\draw[fill=black] (0,0) ellipse (0.05 and 0.05);
\node[] at (0.15, 0.15) {$_v$};
\node[above] at (-0.8,0) {$\mathcal{X}^{-Z}X$};
\node[right] at (0,-0.8) {$\mathcal{X}^{-Z}X$};
\node[right] at (0,1) {$\mathcal{X}\cC X$};
\node[above] at (1,0) {$\mathcal{X}\cC X$};
\draw[fill=black] (0,0) ellipse (0.05 and 0.05);
\end{scope}
\begin{scope}[shift={(0,-3)}]
\node at (-1.5, 0) {$=$};
\node at (2.5,0) {$= \ A_v^{r^{-2}}=(A_v^r)^{-2}\,.$};
\draw[thick, ->-] (0,0) to (1,0);
\draw[thick, ->-] (-1,0) to (0,0);
\draw[thick, ->-] (0,0) to (0,1);
\draw[thick, ->-] (0,-1) to (0,0);
\draw[fill=black] (0,0) ellipse (0.05 and 0.05);
\node[] at (0.15, 0.15) {$_v$};
\node[above] at (-0.8,0) {$\mathcal{X}^{2Z}$};
\node[right] at (0,-0.8) {$\mathcal{X}^{2Z}$};
\node[right] at (0,1) {$\mathcal{X}^{-2}$};
\node[above] at (1,0) {$\mathcal{X}^{-2}$};
\draw[fill=black] (0,0) ellipse (0.05 and 0.05);
\end{scope}
\end{tikzpicture} 
\ea
\ee

Let us now turn to the commutators between $A_v^h$ and $S_p^g$ for $h,g\in\{r,s\}$. From the definition,
\be \label{eq:A_U_comm_general}
    [A_v^h,S_p^g]:=A_v^{h^{-1}} (S_p^g)^{-1}\,A_v^hS_p^g\,,
\ee
and the fact that the eigenvalue of $S_p^g$ depends only the magnetic flux, i.e. the (oriented) product of group elements along the plaquette, we infer that the commutator can be non-trivial only if $A_v^h$ changes said flux. Recall from \eqref{eq:AvBp} the expression of $A_v^h$, in which we now also label the plaquettes according to their direction with respect to $v$:
\be
A_v^{h}= 
\begin{tikzpicture}[baseline]
\begin{scope}[shift={(0,0)}]
\draw[thick, ->-] (-1,0) to (0,0);
\draw[thick, ->-] (0,0) to (1,0);
\draw[thick, ->-] (0,-1) to (0,0);
\draw[thick, ->-] (0,0) to (0,1);
\node[above] at (0.2, 0) {$_v$};
\node[above] at (-0.7,0.05) {$R^h$};
\node[above] at (0.7,0.05) {$L^h$};
\node[right] at (0.05,0.7) {$L^h$};
\node[right] at (0.8,0.7) {$_{p_{NE}}$};
\node[left] at (-0.5,0.7) {$_{p_{NW}}$};
\node[right] at (0.05,-0.7) {$R^h$};
\node[right] at (0.8,-0.7) {$_{p_{SE}}$};
\node[left] at (-0.5,-0.7) {$_{p_{SW}}$};
\draw[fill=black] (0,0) ellipse (0.05 and 0.05);
\end{scope}
\end{tikzpicture}
\ee
and we denote the group elements along each plaquette as $g_1,g_2,g_3,g_4$ as in \eqref{eq:p_g1g2g3g4}. We summarize below the action of $A_v^h$ on the magnetic flux 
\be
    f:=g_1g_2g_3^{-1}g_4^{-1}\,,
\ee
for each plaquette around $v$:
\be\ba
    p_{NE}:& \quad f \xmapsto{A_v^h} hg_1g_2g_3^{-1}(hg_4)^{-1}=hfh^{-1}\,,\cr
    p_{SE}:& \quad f \xmapsto{A_v^h} g_1h^{-1}hg_2g_3^{-1}g_4^{-1}=f\,,\cr
    p_{SW}:& \quad f \xmapsto{A_v^h} g_1g_2h^{-1}(g_3h^{-1})^{-1}g_4^{-1}=f\,,\cr
    p_{NW}:& \quad f \xmapsto{A_v^h} g_1g_2(hg_3)^{-1}(g_4h^{-1})^{-1}=f\,,
\ea\ee
from which 
infer that non-trivial commutators \eqref{eq:A_U_comm_general} can only occur for $p_{NE}$. The stabilizer group generators for the $p_{NE}$ plaquette 
\be
\begin{tikzpicture}
\begin{scope}[shift={(0,0)}]
    \draw[thick, ->-] (0,0) to (1,0);
    \node[below]at  (0.5,0) {$4$};
    \draw[thick, ->-] (0,0) to (0,1);
    \node[left] at (0,0.5) {$1$};
    \node at (0.6,0.5) {$p_{NE}$} ;
    \draw[fill=black] (0,0) ellipse (0.05 and 0.05);
    \node[] at (-0.15, -0.15) {$_v$};
\end{scope}    
\end{tikzpicture} 
\ee
take the following form
\be \label{eq:NE_ops}
\begin{tikzpicture}[baseline]
\begin{scope}[shift={(0,0)}]
\node at (-1, 0.5) {$A_v^{r} = $};
    \draw[thick, ->-] (0,0) to (1,0);
    \node[below]at  (0.5,0) {$\cX_4$};
    \draw[thick, ->-] (0,0) to (0,1);
    \node[left] at (0,0.5) {$\cX_1$};
    \draw[fill=black] (0,0) ellipse (0.05 and 0.05);
    \node[] at (-0.15, -0.15) {$_v$};
\end{scope}    
\begin{scope}[shift={(4,0)}]
\node at (-1.5, 0.5) {$A_v^{s} = $};
    \draw[thick, ->-] (0,0) to (1,0);
    \node[below]at  (0.5,0) {$\cC_4X_4$};
    \draw[thick, ->-] (0,0) to (0,1);
    \node[left] at (0,0.5) {$\cC_1X_1$};
    \draw[fill=black] (0,0) ellipse (0.05 and 0.05);
    \node[] at (-0.15, -0.15) {$_v$};
\end{scope} 
\begin{scope}[shift={(0,-2.5)}]
\node at (-1, 0.5) {$S_{p}^{r}  = $};
  \draw[thick, ->-] (0,0) to (1,0);
    \node[below]at  (1,0) {$\mathcal{Z}^{-Z_1Z_2Z_3Z_4}_{4}$};
    \draw[thick, ->-] (0,0) to (0,1);
    \node[left] at (0,0.5) {$\cZ_1$};
    \draw[fill=black] (0,0) ellipse (0.05 and 0.05);
    \node[] at (-0.15, -0.15) {$_v$};
\end{scope}    
\begin{scope}[shift={(4,-2.5)}]
\node at (-0.5,0.5) {$S_p^s=\zeta_{4N}^{(\bbI-Z_1Z_2Z_3Z_4)/2}$};
\end{scope}    
\end{tikzpicture} 
\ee
Let us first compute $[A_v^r,S_{p_{NE}}^r]$, denoting by $\ket{r^\alpha s^\beta}$ the flux enclosed by $p_{NE}$, and using $r(r^\alpha s^\beta)r^{-1}=r^{\alpha+2\beta}s^\beta$:
\be\ba
   [A_v^r,S_{p_{NE}}^r]\ket{r^\alpha s^\beta}&=A_v^{r^{-1}} (S_p^r)^{-1}\,A_v^rS_p^r\ket{r^\alpha s^\beta}\cr
   &=A_v^{r^{-1}} (S_p^r)^{-1}\,A_v^r \zeta_{4N}^\alpha \ket{r^\alpha s^\beta}\cr
   &=A_v^{r^{-1}} (S_p^r)^{-1}\,\zeta_{4N}^\alpha \ket{r^{\alpha+2\beta} s^\beta}\cr
   &=\zeta_{4N}^{-2\beta}\ket{r^\alpha s^\beta}=(S_{p_{NE}}^s)^{-2}\ket{r^\alpha s^\beta}\,.
\ea\ee
This is consistent with the computation using the operator expressions \eqref{eq:NE_ops}, since:
\be\ba
[\cX_i,\cZ_i] &= \zeta_{4N}^{-1} \bbI\,,\\
\cZ_4^{-Z_1Z_2Z_3Z_4} &=\tfrac{1}{2}(\bbI+Z_1Z_2Z_3Z_4)\cZ_4^{-1} \\
&+\tfrac{1}{2}(\bbI-Z_1Z_2Z_3Z_4)\cZ_4 \,,\\
[\cX_4,\cZ_4^{-Z_1Z_2Z_3Z_4}]&=\tfrac{1}{2}(\bbI+Z_1Z_2Z_3Z_4)\zeta_{4N} \\
&+\tfrac{1}{2}(\bbI-Z_1Z_2Z_3Z_4)\zeta_{4N}^{-1} \\
[\cX_1,\cZ_1][\cX_4,\cZ_4^{-Z_1Z_2Z_3Z_4}]&=\tfrac{1}{2}(\bbI+Z_1Z_2Z_3Z_4) \\
&+\tfrac{1}{2}(\bbI-Z_1Z_2Z_3Z_4)\zeta_{2N}^{-1}\,,\\
\ea\ee
therefore
\be
[A_v^r,S_{p_{NE}}^r]=(S_{p_{NE}}^s)^{-2}\,.
\ee

Since conjugation by $r$ does not change the exponent of $s$, $[A_v^r,S^s_{p_{NE}}]=\bbI$, as can also be checked by observing that $A_v^r$ contains operators that all commute with the ones in $S_{p_{NE}}^s$.

We now compute $[A_v^s,S_{p_{NE}}^r]$, denoting by $\ket{r^\alpha s^\beta}$ the flux enclosed by $p_{NE}$, and using $s(r^\alpha s^\beta)s^{-1}=r^{-\alpha}s^\beta$:
\be\ba
   [A_v^s,S_p^r]\ket{r^\alpha s^\beta}&=A_v^{s^{-1}} (S_p^r)^{-1}\,A_v^s S_p^r\ket{r^\alpha s^\beta}\cr
   &=A_v^{r^{-1}} (S_p^r)^{-1}\,A_v^s \zeta_{4N}^\alpha \ket{r^\alpha s^\beta}\cr
   &=A_v^{r^{-1}} (S_p^r)^{-1}\,\zeta_{4N}^\alpha \ket{r^{-\alpha} s^\beta}\cr
   &=\zeta_{4N}^{2\alpha}\ket{r^\alpha s^\beta}=(S_{p_{NE}}^r)^2\ket{r^\alpha s^\beta}\,.
\ea\ee
Indeed, from the operator commutators, we also have that
\be\ba
    [\cC_1X_1,\cZ_1]&=\cZ_1^2\,,\cr
    [\cC_1X_1\cC_4 X_4, \mathcal{Z}^{-Z_1Z_2Z_3Z_4}_{4}]&=\mathcal{Z}^{-2Z_1Z_2Z_3Z_4}_{4}\,,
\ea\ee
therefore
\be
[A_v^s,S_p^r]=(S_{p_{NE}}^r)^2\,.
\ee
Since conjugation by $s$ does not change the exponent of $s$, $[A_v^s,S^s_{p_{NE}}]=\bbI$, and indeed $X_1X_4$ commutes with $Z_1Z_2Z_3Z_4$.

As shown above, for the other plaquettes, i.e. $p_{NW},p_{SE},p_{SW}$ the commutators are trivial since the flux enclosed by these plaquettes is left unchanged by the vertex operators. Finally, $S_p^r$ and $S_p^s$ commute since they are both diagonal.

\vspace{1mm}
\noindent\textbf{Comments on error detection and correction.}
The stabilizers can distinguish all elementary local errors. Here, two-particle elementary excitation, as defined in \cite{Kitaev:1997wr} refers to a state for which the non-minimized terms of the Hamiltonian \eqref{eq:Hamiltonian} are, at most:
\be\ba \label{eq:AB_violations}
    A_s-\bbI &\neq 0\,, & B_s-\bbI &\neq 0\,,\\ 
    A_{s'}-\bbI &\neq 0\,, & B_{s'}-\bbI &\neq0 \,,
\ea\ee
at the two sites $s=(v,p)$ and $s'=(v',p')$ (recall that a site is comprised of a vertex and a plaquette adjacent to it). Physically, this corresponds to an anyon stretched between $s$ and $s'$. Each anyon is mathematically described by an irreducible representation of the quantum double $D(G)$, labeled by $([g],\bm{R})$, as we reviewed in App. \ref{app:anyons_review}. The group elements in $[g]$ can be detected by applying the $\{B^h\,:\,h\in G\}$ operators, while $\bm{R}$ can be inferred from the action of $\{A_v^k\,:\,k\in G\}$, as explained in detail in \cite{Kitaev:1997wr}.\footnote{An alternative Hamiltonian with projectors onto the $[g]$ and $\bm{R}$ sectors was proposed in \cite{Komar:2017uuc}.} We can similarly determine $\bm{R}$ from the action of the $A_v^k$ stabilizers and the elements in $[g]$ from the $S_p^r$ and $(S_p^s)^{2N}$ stabilizers, as we discussed above. Since stabilizers are local operators (they act on four $G$-qudits), they can only detect errors shorter than the code distance.

For a full QEC analysis we expect to use the just-in-time decoder \cite{Bombin:2018wjx, Brown:2020xxo, Scruby:2020pvw, Davydova:2025ylx} that was applied recently to non-Abelian surface codes in \cite{Huang:2025ump}. We will leave an in depth {discussion} of this for future work.

\section{Review of Constant-Depth $T$-gate from $D^\omega(\Z_2^3)$} \label{app:T_from_rgb}

We summarize here the alternative description of $D(D_4)$ in terms of $D^\omega (\Z_2^3)$, as it was presented in \cite{Kobayashi:2025cfh}. Here, $\omega$ is the type-III 3-cocycle
\be \label{eq:omIII_2}
    \omega(\r^{i_\r}\g^{i_\g}\b^{i_\b}, \r^{j_\r} \g^{j_\g} \b^{j_\b}, \r^{k_\r} \g^{k_\g}\b^{k_\b})=(-1)^{i_\r j_\g k_\b}\,.
\ee 
Note that the generalization to $D(D_{4N})$ does not admit such a description as a quantum double for abelian groups with cocycle, which required us to develop a general approach instead, applicable for any quantum double $D(G)$. 
The protocol in \cite{Kobayashi:2025cfh}, which we review below, realizes the logical $T^\dagger$ gate with a constant-depth circuit using both a group automorphism, and an SPT stacking, whereas our construction only requires SPT stacking. The two descriptions map into each other by the following isomorphism between the topological orders $D(D_4)$ and $D^\omega (\Z_2^3)$:
\be\ba
1 & \ \longleftrightarrow \ 1 \,,&\quad
1_r & \ \longleftrightarrow \ e_{\r\b} \,,\cr
1_s & \ \longleftrightarrow \  e_\b \,,&\quad
1_{rs} & \ \longleftrightarrow \  e_\r \,,\cr
E & \ \longleftrightarrow \  m_\g \,,&\quad
[r^2] & \ \longleftrightarrow \  e_\g \,,\cr
[r^2]1_r & \ \longleftrightarrow \  e_{\r\g\b} \,,&\quad
[r^2]1_s & \ \longleftrightarrow \  e_{\g\b} \,,\cr
[r^2]1_{rs} & \ \longleftrightarrow \  e_{\r\g} \,,&\quad
[r^2]E & \ \longleftrightarrow \  f_\g \,,\cr
[s]_{++} & \ \longleftrightarrow \  m_\r \,,&\quad
[s]_{-+} & \ \longleftrightarrow \  f_\r \,,\cr
[s]_{+-} & \ \longleftrightarrow \  m_{\r\g} \,,&\quad
[s]_{--} & \ \longleftrightarrow \  f_{\r\g} \,,\cr
[rs]_{++} & \ \longleftrightarrow \  m_\b \,,&\quad
[rs]_{-+} & \ \longleftrightarrow \  f_\b \,,\cr
[rs]_{+-} & \ \longleftrightarrow \  m_{\g\b} \,,&\quad
[rs]_{--} & \ \longleftrightarrow \  f_{\g\b} \,,\cr
[r] & \ \longleftrightarrow \ m_{\r\b} \,,&\quad
[r]_{i} & \ \longleftrightarrow \ s_{\r\b\g} \,,\cr
[r]_{-1} & \ \longleftrightarrow \ f_{\r\b} \,,&\quad
[r]_{-i} & \ \longleftrightarrow \ \bar{s}_{\r\b\g}\,.
\ea\ee

A $T$-gate protocol was proposed in \cite{Davydova:2025ylx} from corners between gapped boundaries. In \cite{Kobayashi:2025cfh} a constant-depth $T^\dagger$-gate was obtained from a bulk automorphism in the presence of boundaries and corners. The analysis of \cite{Kobayashi:2025cfh} is as follows. Consider the three gapped boundaries (first computed in \cite{Bhardwaj:2024qrf}):
\be\ba
    \cL_\r=\cL_{\langle \g,\b\rangle}&=e_\r\oplus m_\b \oplus m_\g \oplus m_{\g\b} \cr
   \cL_\b=\cL_{\langle \r,\g\rangle}&=e_\b\oplus m_\g \oplus m_\r \oplus m_{\r\g} \cr
    \cL_{\r\b}=\cL_{\langle \r\b\rangle}&=e_\g\oplus e_{\r\b} \oplus e_{\r\g\b} \oplus 2m_{\r\b}\,,
\ea\ee
arranged to form in a spatial triangle.

Performing the group automorphism
\be
    f:\;\; (\r,\g,\b) \mapsto (\r\g,\g,\g\b)\,,
\ee
the 3-cocycle \ref{eq:omIII_2} becomes
\be\ba
    &\omega(f(\r^{i_\r}\g^{i_\g}\b^{i_\b}), f(\r^{j_\r} \g^{j_\g} \b^{j_\b}), f(\r^{k_\r} \g^{k_\g}\b^{k_\b}))=\\
    &=(-1)^{i_\r j_\g k_\b+i_\r j_\b k_\b+i_\r j_\r k_\b}=\\
    &=(-1)^{i_\r j_\g k_\b} +\delta\alpha
\ea\ee
with
\begin{align}
    &\alpha(\r^{i_\r}\g^{i_\g}\b^{i_\b}, \r^{j_\r} \g^{j_\g} \b^{j_\b})=\\
     &=\begin{cases}
         e^{i\pi/2} & \text{if } (i_\r, j_\b)=(1,1)\\
         1 & \text{otherwise}\,.
     \end{cases}
\end{align}
$\alpha$ restricted to the subgroups $\langle \g, \b\rangle$ or $\langle \g, \r\rangle$ is identically 1, while
\be
    \alpha|_{\langle \r \b\rangle}=\delta\beta
\ee
for the 1-cochain
\be
    \beta(\id)=1\,,\quad \beta(\r\b)=e^{i\pi/4}\,,
\ee
In combination, they act as the non-Clifford $T^\dagger$ gate on the logical qubit.

\twocolumngrid
\bibliographystyle{ytphys}
\small 
\baselineskip=.7\baselineskip
\let\bbb\bibitem\def\bibitem{\itemsep3.3pt\bbb}
\bibliography{ref}

\end{document}